\documentclass[11pt]{article}
\usepackage{graphics,amsmath,amssymb,amsfonts}
\def \Tr#1{{\rm Tr}(#1)}
\def \ui#1{^{(#1)}}
\def \W#1{\widehat{#1}}

\def \hh{\vskip0.5\baselineskip \hbox to \hsize}

\def \ds{\displaystyle\displaystyle}

\newtheorem{Proposition}{Proposition}[section]
\numberwithin{equation}{section}
\title{Notes on solutions in Wronskian form\\
to soliton equations: KdV-type }

\author{ Da-jun Zhang\footnote{E-mail: djzhang@staff.shu.edu.cn}
\\
{\small\it Department of Mathematics,
 Shanghai University, Shanghai 200444,  P.R. China}}

\begin{document}

\maketitle

\begin{abstract}

This paper can be an overview on solutions in Wronskian/Casoratian form to
soliton equations with KdV-type bilinear forms.
We first investigate properties of  matrices commuting with a Jordan block,
by which we derive explicit general solutions to equations satisfied by
Wronskian/Casoratian entry vectors, which we call  {\it condition equations}.
These solutions are given according to
the coefficient matrix in the condition equations
taking diagonal or Jordan block form.
Limit relations between theses different solutions
are described.  We take  the KdV equation and the Toda lattice
to serve as two examples for solutions in Wronskian form and Casoratian form, respectively.
We also discuss Wronskian solutions for the KP equation.
Finally, we formulate the Wronskian technique as four steps.

\end{abstract}

\tableofcontents

\newpage
\vskip 20pt

\section {\normalsize Introduction}

Soliton solutions can be represented in terms of Wronskian\cite{Wadati,Satsuma},
and this has been realized through the Darboux transformation\cite{Matveev-book},
Sato theory\cite{Sato,Sato-elementary} where $\tau$-functions expressed in terms of Wronskian
are governed in common by the Pl\"{u}cker relations,
and Wronskian technique\cite{Freeman-Nimmo-KP}-\cite{Freeman-IMA}
where bilinear equations are reduced to the Laplace expression of some zero determinants.
Among them, Wronskian technique provides direct verifications of solutions to bilinear equations
by taking the advantage that special structure of a Wronskian contributes
simple forms of its derivatives.
And this technique, together with the Hirota method\cite{Hirota-1971,Hirota-book},
is considered as one of efficient direct approaches to
deriving soliton solutions to nonlinear evolution equations possessing bilinear forms.

Besides solitons, many other kinds solutions can also be expressed in terms of Wronskian, such as
rational solutions, positons, negatons, complexitons and mixed solutions.
Rational solutions in Wronskian form were first given for the KdV equation in 1983
by Nimmo and Freeman\cite{rational-1}, based on the idea of long-wave limit proposed by
Ablowitz and Satsuma\cite{Ablowitz-Satsuma}.
In 1988 Sirianunpiboon, Howard and Roy (SHR)\cite{Sirianunpiboon-1988}
generalized the conditions satisfied by Wronskian entries,
and still through the standard Wronskian procedure, derived more solutions to the KdV equation.
They obtained Wronskian entries for their generalized conditions by
Taylor-expanding the original entries and the solutions obtained can include
positons, negatons, rational solutions and mixed solutions.
The name of positons was first introduced by Matveev\cite{Matveev-positon-kdv1,Matveev-positon-kdv2}
in 1992. He obtained positons of the KdV equation from the results of Darboux transformation
by considering the Taylor expansions of some entries in Wronskian, and analyzed the property
of slowly decaying of positons. The name of positons for the KdV
equation comes from the fact that these solutions correspond to
positive eigenvalues of the stationary Schr\"{o}dinger equation (Schr\"{o}dinger spectral problem)
with zero potential.  Similarly,
negatons correspond to the negative eigenvalues.
Complexitons of KdV equation was first named by Ma\cite{Ma-complexiton} in 2002,
which correspond to complex eigenvalues (appearing in conjugate pair)
of the stationary Schr\"{o}dinger equation,
and essentially are breathers\cite{breather-KdV} and high-order breathers.

Many developments on Wronskian technique have been occurred recently. For example,
Rational solutions in Casoratian form to the Toda lattice were obtained\cite{Zhang-Toda rational} by
using Nimmo and Freeman's procedure proposed in \cite{rational-1}.
(Note: It was mentioned in \cite{rational-1} that:
``It might be hoped that
the rational solutions of other equations of the type whose
solutions take Wronskian form may be obtained in a similar way,
however we have found this not to be possible.'')
Besides, a new identity was given to get Wronskian solutions for some
soliton equations with self-consistent sources\cite{Zhang-mKdVscs}-\cite{DDKPscs}.
Many nonisospectral evolution equations have been shown to possess solutions
in Wronskian form\cite{Zhang-nonisos-KdV}-\cite{Zhang-nonisos-NLSE}.
Recently, Ma and You\cite{Ma-You-KdV}, based on their previous discussions\cite{Ma-wronskian,Ma-complexiton},
reviewed solutions of the KdV equation
from the viewpoint of Wronskian form.
They naturally considered the coefficient matrix in the Wronskian condition equations
to be its canonical form, i.e., a diagonal or Jordan form;
and particularly, they answered an important question about Wronskian technique,
i.e., how to obtain all solutions to the Wronskian condition equations
when the coefficient matrix taking a Jordan form.
By solving the so-called  representative systems through the variation of parameters approach,
they worked out a set of recursive formulae by which all the Wronskian entries for an $N$-order Jordan-block
solution can be determined one by one. Similar results for the Toda lattice can be found in
Refs.\cite{Ma-You-toda}-\cite{Ma-Maruno-PA}.

Our paper aims to give explicit general solutions to the Wronskian/Casoratian condition equations
and investigate the relations between Jordan-block solutions and diagonal cases for
those soliton equations with KdV-type bilinear forms\cite{Hietarinta-KdV,Hietarinta-rev}.
To achieve that, we first discuss properties of  matrices commuting with a Jordan block,
i.e., the lower triangular Toeplitz matrices.
Then we take the KdV equation and the Toda lattice
to serve as two examples for solutions in Wronskian form and Casoratian form, respectively.
In each case, we first study the conditions satisfied by
Wronskian/Casiratian entry vectors, which we call  {\it condition equations}.
These condition equations are more general than the known ones.
Then we give explicit forms of these entry vectors according to
the coefficient matrix in the condition equations
taking diagonal or Jordan block form.
The corresponding solutions in Wronskians/Casiratians form are called diagonal or Jordan-block solutions.
For Jordan-block case, we give explicit forms and effective forms
for general solutions to the  condition equations.
The effective form means the number of arbitrary parameters in $N$-Jordan block solutions
has been reduced to the least,
which will be helpful when we discuss the parameter effects and dynamics of solutions.
For the high-order complexitons of the KdV equation and Toda lattice,
we give several different choices for Wronskian/Casoratian entries
which generate same solutions.
Besides, we propose a limit procedure to describe the relations between Jordan-block solutions and diagonal cases.
Those properties of  lower triangular Toeplitz matrices will
play important roles in our paper. They enable us to easily obtain
general Jordan-block solutions from the known special ones
and  to write out their effective forms.
Some properties also help us to explain the exact relations between
Jordan-block and diagonal solutions.
Besides the KdV equation and the Toda lattice, as an (1+2)-dimensional example,
the Wronskian solutions for the KP equation will also be investigated.
We prove that some generalizations are trivial for generating
new solutions.
Finally, we formulate the Wronskian technique as four steps.

The paper is arranged  as follows.
In Sec.2, we investigate the properties of  matrices commuting with a Jordan block.
Then in Sec.3, 4 we respectively take
the KdV equation and the Toda lattice  as two different examples
to discuss their solutions in Wronskian form and Casoratian form.
Wronskian solutions for the KP equation will  be investigated in Sec.5.

\section {\normalsize Lower triangular Toeplitz matrices --- matrices commuting with a Jordan block}

In this section, we collect some properties of  lower triangular Toeplitz matrices
which are also the matrices commuting with a Jordan block.
These properties will play important roles in the following sections. We list them
through several propositions.
Some of them have been discussed in Ref.\cite{Zhang-Hietarinta}.

\begin{Proposition}
\label{Prop 2.1}
 {\it Let
\begin{equation}
J_S=J_{S}(k,\kappa)=\left(\begin{array}{cccccc}
k & 0    & 0   & \cdots & 0   & 0 \\
\kappa & k  & 0   & \cdots & 0   & 0 \\
\cdots &\cdots &\cdots &\cdots &\cdots &\cdots \\
0   & 0    & 0   & \cdots &  \kappa   & k
\end{array}\right)_{S\times S}
\label{Jordan block}
\end{equation}
be a Jordan block, where $k$  and $\kappa$ are arbitrary non-zero
complex numbers, $\mathcal{A}$ is
an $S\times S$ complex matrix. Then
\begin{equation}
\mathcal{A} J_S=J_S \mathcal{A} \label{commute}
\end{equation}
if and only if $\mathcal{A}$ is  a lower triangular Toeplitz matrix,
i.e., $\mathcal{A}$ is in the following form
\begin{equation}
\mathcal{A}=\left(\begin{array}{cccccc}
a_0 & 0    & 0   & \cdots & 0   & 0 \\
a_1 & a_0  & 0   & \cdots & 0   & 0 \\
a_2 & a_1  & a_0 & \cdots & 0   & 0 \\
\cdots &\cdots &\cdots &\cdots &\cdots &\cdots \\
a_{S-1} & a_{S-2} & a_{S-3}  & \cdots &  a_1   & a_0
\end{array}\right)_{S\times S}
\label{A}
\end{equation}
where $\{a_j\}^{S-1}_{j=1}$ are arbitrary complex
numbers.\hfill$\Box$}
\end{Proposition}

%
%

\begin{Proposition}
\label{Prop 2.2}
{\it Suppose that $J_S$ is given as
\eqref{Jordan block}, and let
\begin{subequations}
\begin{eqnarray}
\widetilde{G}_S=\bigl \{\mathcal{A}_{S\times S}\bigl
|~\bigr. \mathcal{A} J_S=J_S \mathcal{A}\bigr\} =\{
~S\hbox{\rm -order lower triangular Toeplitz matrices }\},
\label{G-tilde}\\
G_S=\bigl \{\mathcal{A} \bigl |~\bigr. \mathcal{A}\in \widetilde{G}_S,~|\mathcal{A}|\neq 0 \bigr\},
\label {G}
\end{eqnarray}
\end{subequations}
then, $G_S$ forms an Abelian group with respect to matrix multiplication
and inverse, and $\widetilde{G}_S$ is an Abelian semigroup with identity.}\hfill $\Box$
\end{Proposition}

\vskip 8pt We can also verify the following proposition.

\begin{Proposition}
\label{Prop 2.3}
{\it Suppose that $\mathcal{A}$ is a lower triangular Toeplitz matrix defined as
\eqref{A}, $\varphi(k)$ and $\alpha(k)$ are complex functions
arbitrarily differential with respect to $k$, and
\begin{equation*}
\mathcal{Q}_0=(\mathcal{Q}_{0,0},\mathcal{Q}_{0,1},\cdots,\mathcal{Q}_{0,S-1})^T,~~
\widetilde{\mathcal{Q}}_0=(\widetilde{\mathcal{Q}}_{0,0},\widetilde{\mathcal{Q}}_{0,1},\cdots,
\widetilde{\mathcal{Q}}_{0,S-1})^T
\end{equation*}
with
\begin{equation*}
\mathcal{Q}_{0,j}=\frac{\kappa^j}{j!}\partial^{j}_{k}
\varphi(k),~~
\widetilde{\mathcal{Q}}_{0,j}=\frac{\kappa^j}{j!}\partial^{j}_{k}
(\alpha(k)\varphi(k)),~~j=0,1,\cdots,S-1,
\end{equation*}
where $\kappa$ is some nonzero complex number. Then we have
\begin{equation*}
\widetilde{\mathcal{Q}}_0=\mathcal{A}\mathcal{Q}_0,~~~\mathcal{A}\in \widetilde{G}_S,
\end{equation*}
where
\begin{equation}
a_j=\frac{\kappa^j}{j!}\partial^{j}_{k}
\alpha(k),~~j=0,1,\cdots,S-1.
\end{equation}
}\hfill$\Box$
\end{Proposition}

\vskip 8pt

\begin{Proposition}
\label{Prop 2.4}
{\it Suppose that $\mathcal{A}\in \widetilde{G}_S$ is
defined as \eqref{A}, $\alpha(z)$ is a complex polynomial defined as
\begin{equation}
\alpha(z)=\alpha_0 z^{S-1} + \alpha_1 z^{S-2} + \cdots
+\alpha_{S-2} z +\alpha_{S-1}, \label{alpha-poly}
\end{equation}
satisfying
\begin{equation}
\partial^{j}_{z} \alpha(z)|_{z=k}=\frac{j!}{\kappa^j}a_j,~~j=0,1,\cdots,S-1,
\label{alpha-poly-cond}
\end{equation}
where $\kappa$ is some nonzero complex number. Let
\begin{equation*}
\bar{\alpha}=(a_0,\frac{1}{\kappa} a_1,\frac{2!}{\kappa^2}
a_2,\cdots,\frac{(S-1)!}{\kappa^{S-1}} a_{S-1})^T,~~
\tilde{\alpha}=(\alpha_0,\alpha_1,\cdots,\alpha_{S-1})^T.
\end{equation*}
Then the linear equations
\begin{equation*}
Z \tilde{\alpha}= \bar{\alpha}
\end{equation*}
have a unique solution vector $\tilde{\alpha}=Z^{-1} \bar{\alpha}$, where
\begin{equation*}
Z=\left(\begin{array}{cccccc}
z^{S-1} & z^{S-2}      & \cdots & z^{2} & z   & 1 \\
\partial_z z^{S-1} & \partial_z z^{S-2}      & \cdots & 2z & 1   & 0 \\
\partial^2_z z^{S-1} & \partial^2_z z^{S-2}      & \cdots & 2 & 0   & 0 \\
\cdots &\cdots &\cdots &\cdots &\cdots &\cdots \\
\partial^{S-1}_z z^{S-1} & 0      & \cdots & 0 & 0   & 0 \\
\end{array}\right)_{z=k}.
\end{equation*}
}
\end{Proposition}

{\it Proof:} This proposition holds by noting that
\begin{equation*}
|Z|=(-1)^{\sum^{S-1}_{j=0}j}\prod^{S-1}_{j=0}j!\neq 0.
\end{equation*}
\hfill$\Box$

\vskip 8pt
\begin{Proposition}
\label{Prop 2.5}
{\it Given $\mathcal{A}\in G_S$
we can uniquely determine  a complex polynomial $\alpha(z)$
\eqref{alpha-poly} by imposing the condition
\eqref{alpha-poly-cond} where $\kappa$ is nonzero. Suppose that
\begin{equation}
b_j=\frac{\kappa^j}{j!}\partial^{j}_{z}
\frac{1}{\alpha(z)}|_{z=k},~~j=0,1,\cdots,S-1,
\label{alpha-invers-cond}
\end{equation}
and
\begin{equation}
\mathcal{B}=\left(\begin{array}{cccccc}
b_0 & 0    & 0   & \cdots & 0   & 0 \\
b_1 & b_0  & 0   & \cdots & 0   & 0 \\
b_2 & b_1  & b_0 & \cdots & 0   & 0 \\
\cdots &\cdots &\cdots &\cdots &\cdots &\cdots \\
b_{S-1} & b_{S-2} & b_{S-3}  & \cdots &  b_1   & b_0
\end{array}\right),
\label{B}
\end{equation}
then we have $\mathcal{B}=\mathcal{A}^{-1}\in G_S$.}
\end{Proposition}

{\it Proof:} Let
$\mathcal{AB}=\mathcal{D}=(\mathcal{D}_{i,j})_{S\times S}$ and we
prove $\mathcal{D}=I$. In fact, for $i\leq j$, one can find from
\eqref{alpha-poly-cond} and \eqref{alpha-invers-cond} that
\begin{equation*}
\mathcal{D}_{i,j}=\sum^{i-j}_{l=0}a_{i-j-l}b_{l}
=\frac{\kappa^{i-j}}{(i-j)!}\partial_z^{i-j}\Bigl[\alpha(z)\cdot
\frac{1}{\alpha(z)}\Bigr]_{z=k}.
\end{equation*}
So we conclude $\mathcal{D}=I$.\hfill$\Box$

\vskip 8pt
\begin{Proposition}
{\it For any given
$\mathcal{A}\in G_S$, there exist  $\pm \mathcal{B}\in G_S$ such
that $\mathcal{B}^2=\mathcal{A}$.}
\label{Prop 2.6}
\end{Proposition}

{\it Proof:} Let $\mathcal{B}$ be given as \eqref{B}. From
$\mathcal{B}^2=\mathcal{A}$ we can uniquely obtain that
\begin{equation}
\begin{array}{l}
\ds b_0=\pm \sqrt{a_0},~~~b_1=\frac{a_1}{2b_0},\\
\ds b_j=\frac{1}{2b_0}\Bigl(a_j-\sum^{j-1}_{l=1}b_l
b_{j-l}\Bigr),~~~ j=2,3,\cdots, S-1.
\end{array}
\end{equation}
\hfill$\Box$

Consider the following complex block Jordan block
\begin{equation}
J^B_{2S}=\left(\begin{array}{cccccc}
K & 0    & 0   & \cdots & 0   & 0 \\
I_1 & K & 0   & \cdots & 0   & 0 \\
\cdots &\cdots &\cdots &\cdots &\cdots &\cdots \\
0   & 0    & 0   & \cdots &  I_1  & K
\end{array}\right)_{2S\times 2S}
\label{B-Jordan block}
\end{equation}
where
$K=\biggl(\begin{array}{cc}
k_1 & 0   \\
0 & k_2
\end{array}\biggr)$ and
$I_1=\biggl(\begin{array}{cc}
1 & 0   \\
0 & 1
\end{array}\biggr)$, then a $2S\times 2S$ matrix $\mathcal{A}^B$
commuting with $J^B_{2S}$, i.e., $\mathcal{A}^B J^B_{2S}=J^B_{2S} \mathcal{A}^B$,
if and only if $\mathcal{A}^B$ is a block lower triangular Toeplitz matrix, i.e.,
\begin{equation}
\mathcal{A}^B=\left(\begin{array}{cccccc}
A_0 & 0    & 0   & \cdots & 0   & 0 \\
A_1 & A_0  & 0   & \cdots & 0   & 0 \\
A_2 & A_1  & A_0 & \cdots & 0   & 0 \\
\cdots &\cdots &\cdots &\cdots &\cdots &\cdots \\
A_{S-1} & A_{S-2} & A_{S-3}  & \cdots &  A_1   & A_0
\end{array}\right)_{2S\times 2S},
\label{block-A-block}
\end{equation}
where
$A_j=\biggl(\begin{array}{cc}
a_{j1} & 0   \\
0 & a_{j2}
\end{array}\biggr)$
and $\{a_{js}\}$ are arbitrary complex numbers. So we have the following proposition.

\begin{Proposition}
\label{Prop 2.7}
 {\it The $2S\times 2S$ matrix set
\begin{equation}
\begin{array}{rl}
\widetilde{G}^B_{2S}&=\bigl \{\mathcal{A}^B \bigl |~\bigr. \mathcal{A}^B J^B_{2S}=J^B_{2S} \mathcal{A}^B\bigr\}\\
&=\{2S\hbox{\rm -order block lower triangular Toeplitz matrices
defined as \eqref{block-A-block}}\}
\end{array}
\label{B-G-tilde}
\end{equation}
is an Abelian semigroup with identity and
\begin{equation}
G^B_{2S}=\bigl \{\mathcal{A}^B \bigl |~\bigr. \mathcal{A}^B\in \widetilde{G}^B_S,~|\mathcal{A}^B|\neq 0 \bigr\}
\label {B-G}
\end{equation}
is an Abelian group.}\hfill $\Box$
\end{Proposition}

Consider the following real block Jordan block
\begin{equation}
J^{\widetilde{B}}_{2S}=\left(\begin{array}{cccccc}
\widetilde{K}& 0    & 0   & \cdots & 0   & 0 \\
I_1 & \widetilde{K} & 0   & \cdots & 0   & 0 \\
\cdots &\cdots &\cdots &\cdots &\cdots &\cdots \\
0   & 0    & 0   & \cdots &  I_1  & \widetilde{K}
\end{array}\right)_{2S\times 2S}
\end{equation}
where
$\widetilde{K}=\biggl(\begin{array}{cc}
a & -b   \\
b & a
\end{array}\biggr)$ and $b\neq 0$, then a $2S\times 2S$ matrix $\mathcal{A}^{\widetilde{B}}$
commuting with $J^{\widetilde{B}}_{2S}$, i.e.,
$\mathcal{A}^{\widetilde{B}} J^{\widetilde{B}}_{2S}=J^{\widetilde{B}}_{2S} \mathcal{A}^{\widetilde{B}}$,
if and only if $\mathcal{A}^{\widetilde{B}}$ is   the following  block lower triangular Toeplitz matrix
\begin{equation}
\mathcal{A}^{\widetilde{B}}=\left(\begin{array}{cccccc}
\widetilde{A}_0 & 0    & 0   & \cdots & 0   & 0 \\
\widetilde{A}_1 & \widetilde{A}_0  & 0   & \cdots & 0   & 0 \\
\widetilde{A}_2 & \widetilde{A}_1  & \widetilde{A}_0 & \cdots & 0   & 0 \\
\cdots &\cdots &\cdots &\cdots &\cdots &\cdots \\
\widetilde{A}_{S-1} & \widetilde{A}_{S-2} & \widetilde{A}_{S-3}  & \cdots &  \widetilde{A}_1   & \widetilde{A}_0
\end{array}\right)_{2S\times 2S},
\label{block-A-block-1}
\end{equation}
where
$\widetilde{A}_j=\biggl(\begin{array}{cc}
\tilde{a}_{j1} & -\tilde{a}_{j2}  \\
\tilde{a}_{j2} & \widetilde{a}_{j1}
\end{array}\biggr)$
and $\{\tilde{a}_{js}\}$ are arbitrary real numbers. We also have the following proposition.

\begin{Proposition}
\label{Prop 2.8}
{\it The $2S\times 2S$ matrix set
\begin{equation}
\widetilde{G}^{\widetilde{B}}_{2S}=\bigl \{\mathcal{A}^{\widetilde{B}} \bigl |~\bigr.
\mathcal{A}^{\widetilde{B}} J^{\widetilde{B}}_{2S}=J^{\widetilde{B}}_{2S} \mathcal{A}^{\widetilde{B}}\bigr\}
=\{\hbox{\rm block lower triangular Toeplitz matrices
defined as \eqref{block-A-block-1}}\}
\label{B-G-tilde-1}
\end{equation}
is an Abelian semigroup with identity and
\begin{equation}
G^{\widetilde{B}}_{2S}=\bigl \{\mathcal{A}^{\widetilde{B}} \bigl |~\bigr.
\mathcal{A}^{\widetilde{B}}\in \widetilde{G}^{\widetilde{B}}_S ,~
|\mathcal{A}^{\widetilde{B}}|\neq 0\bigr\}
\label {B-G-1}
\end{equation}
is an Abelian group.}\hfill $\Box$
\end{Proposition}

The above propoties will
play important roles in the following sections. For example,
with the help of Proposition \ref{Prop 2.2}, \ref{Prop 2.7} and \ref{Prop 2.8}, we can easily get
general Jordan-block solutions from the known special ones.

\section {\normalsize Solutions in Wronskian form to the KdV equation}

In this section, we will first consider the Wronskian condition equations of the KdV equation.
Then we discuss general solutions to the condition equations according to
the coefficient matrix taking diagonal or Jordan block form.
Particularly, for Jordan-block solutions, we give explicit general forms and effective forms
of the Wronskian entries.
We will also investigate the relations between Jordan-block solutions and diagonal cases.

The well-known KdV equation is
\begin{equation}
u_t+6uu_x + u_{xxx}=0
\label{KdV}
\end{equation}
with Lax pair
\begin{equation}
-\varphi_{xx}=(\lambda+u)\varphi,
\label{Lax-KdV-a}
\end{equation}
\begin{equation}
\varphi_t=-4\varphi_{xxx}-3u_x\varphi -6u\varphi_x,
\label{Lax-KdV-b}
\end{equation}
where $\varphi$ is the wave function and $\lambda= -k^2$ is the spectral parameter.
Employing the transformation
\begin{equation}
u=2(\ln f)_{xx}=\frac{2(f_{xx}f -f^2_x)}{f^2},
\label{trans-KdV}
\end{equation}
Hirota\cite{Hirota-1971} first transformed \eqref{KdV} into its bilinear form
\begin{equation}
(D_{t}D_{x}+D^4_{x})f\cdot f=0,
\label{blinear-KdV}
\end{equation}
where $D$ is the well-known Hirota's bilinear operator defined by\cite{Hirota-1971,Hirota-book}
\begin{equation*}
D^m_{t}D^n_{x}a(t,x)\cdot b(t,x)
=\frac{\partial^m}{{\partial s}^m}\frac{\partial^n}{{\partial y}^n}
a(t+s,x+y)b(t-s,x-y)|_{s=0,y=0},~~m,n=0,1,2,\cdots.
\end{equation*}
Here we note that $2(\ln f)_{xx}$ in \eqref{trans-KdV} should be considered as a formal
expression where the function $f$ can have arbitrary values.

An $N\times N$ Wronskian is defined as
\begin{equation*}
W(\phi_1, \phi_2,\cdots,\phi_N)
       =\left | \begin{array}{llll}
                \phi_{1} & \phi_{1}\ui{1} & \cdots  & \phi_{1}\ui{N-1}\\
                \vdots   & \vdots  &  ~ & \vdots\\
                \phi_{N} & \phi_{N}\ui{1} & \cdots  & \phi_{N}\ui{N-1}
                \end{array}
        \right |
\end{equation*}
where $\phi_{j}\ui{l}=\partial^l \phi_j/{\partial x}^l$. It can be denoted by the following
compact form\cite{Freeman-Nimmo-KP}
\begin{equation}
W(\phi)=|\phi, \phi\ui{1},\cdots,\phi\ui{N-1}|=|0,1,\cdots,N-1|=|\widehat{N-1}|,
\label{wronskian}
\end{equation}
where $\phi=(\phi_1, \phi_2,\cdots,\phi_N)^T$ and $\widehat{N-j}$ indicates the set of consecutive
columns $0,1,\cdots,N-j$.

To smoothly discuss Wronskian solutions to the bilinear KdV equation, we start from the following
propositions.

\begin{Proposition}\cite{Zhang-Hietarinta}
\label{Prop 3.1}
{\it  Suppose that $\Xi$ is an
  $N\times N$ matrix with column vector set $\{\Xi_j\}$;  $\Omega$ is an
  $N\times N$ operator matrix with column vector set $\{\Omega_j\}$ and
  each entry $\Omega_{j,s}$ being an operator. Then
  we have
\begin{equation}
\sum^N_{j=1} |\Omega_j * \Xi|
=\sum^N_{j=1}|(\Omega^T)_{j} * \Xi^T|,
\label{2.4}
\end{equation}
 where for any $N$-order column vectors $A_j$ and $B_j$ we define
\begin{equation*}
A_j \circ B_j=(A_{1,j}B_{1,j},~A_{2,j}B_{2,j},\cdots, A_{N,j}B_{N,j})^T
\end{equation*}
 and
\begin{equation*}
|A_j * \Xi|=|\Xi_1,\cdots,\Xi_{j-1},~A_j \circ\Xi_j,~\Xi_{j+1},\cdots, \Xi_{N}|
\end{equation*}}
\hfill $\Box$
\end{Proposition}

This proposition is more general than  the similar results in
Ref.\cite{Sirianunpiboon-1988} and \cite{Zhang-Toda} in the sense that
$\Omega_{j,s}$ can be an arbitrary operator or a constant.
We can prove it through the
expansion rule of a determinant.

\begin{Proposition}
\label{Prop 3.2}
{\it Consider $N\times N$ Wronskian $f(\phi)=W(\phi)$ defined by \eqref{wronskian}
where the entry vector satisfies
\begin{equation}
\phi_t=B\phi,
\label{t-relation}
\end{equation}
 $B=(B_{ij})$ is an $N\times N$
matrix and each $B_{ij}$ can be a function of $t$ but independent of $x$.
Then we have
\begin{equation}
f_t={\rm Tr}(B) f,
\label{t-result}
\end{equation}
and consequently
\begin{equation*}
D_xD_t f\cdot f=\frac{1}{2}D_x f_t \cdot f=0.
\end{equation*}
em In addition, if $f\neq 0$, then
\begin{equation}
D^2_t f\cdot f=0
\end{equation}
if and only if ${\rm Tr}(B)_t=0$, i.e., ${\rm Tr}(B)$ is independent of $t$.}
\end{Proposition}

{\it Proof} ~
We only prove \eqref{t-result}. It can easily be derived by using \eqref{t-relation}
if we calculate $t$-derivative of $f$ row by row instead of usually column by column.
\hfill $\Box$

\begin{Proposition}
\label{Prop 3.3}
{\it Consider $N\times N$ Wronskians $f(\phi)=W(\phi)$ and $f(\psi)=W(\psi)$
where the entry vectors satisfy $\psi=P\phi$ and $P=(P_{ij})$ is an $N\times N$
non-singular constant matrix.
Then we have
\begin{equation*}
f(\psi)=|P|f(\phi),
\end{equation*}
and hence if $f(\phi)$ solves \eqref{blinear-KdV}, then
$f(\phi)$ and $f(\psi)$ lead to same solutions to the KdV equation through the transformation
\eqref{trans-KdV}, no matter $|P|$ is real or complex.} \hfill $\Box$
\end{Proposition}

\begin{Proposition}
\label{Prop 3.4}
{\it Suppose that the KdV-type bilinear equation\cite{Hietarinta-KdV, Hietarinta-rev}
\begin{equation}
B(D_t,D_x)f\cdot f=0,
\label{bilinear eq}
\end{equation}
has a Wronskian solution
\begin{equation*}
f=W(\phi_1, \phi_2,\cdots,\phi_N),
\end{equation*}
where each $\phi_j=\phi_1(\varrho_j,t,x)$ is infinitely differentiable with respect to parameter $\varrho_j$,
then we have
\begin{equation}
B(D_t,D_x)W_j(\phi_1, \phi_2,\cdots,\phi_N)\cdot W_j(\phi_1, \phi_2,\cdots,\phi_N)=0,
~~2\leq j\leq N,
\end{equation}
where
\begin{equation}
W_j(\phi_1, \phi_2,\cdots,\phi_N)=
W(\phi_1, \frac{\partial \phi_1}{\partial \varrho_1},
\frac{\partial^2 \phi_1}{{\partial \varrho_1}^2},\cdots,
\frac{\partial^{j-1} \phi_1}{{\partial \varrho_1}^{j-1}},
\phi_{j+1},\cdots,\phi_N),~~ 2\leq j\leq N.
\end{equation}
}
\end{Proposition}

{\it Proof:} ~First we replace $f$ by
\begin{equation}
f=\frac{W(\phi_1, \phi_2,\cdots,\phi_N)}{\prod^{N}_{j=2}(\varrho_j-\varrho_1)^{j-1}\frac{1}{(j-1)!}}
\label{wrons for bi-eq}
\end{equation}
which also solves \eqref{bilinear eq}.
Next, let  $\varrho_2 \rightarrow \varrho_1$ in \eqref{wrons for bi-eq}.
Using Lopita rule we get
\begin{equation*}
f\rightarrow
\frac{W(\phi_1, \frac{\partial}{\partial \varrho_1}\phi_1,\cdots,\phi_N)}
{\prod^{N}_{j=3}(\varrho_j-\varrho_1)^{j-1}\frac{1}{(j-1)!}},
\end{equation*}
which means
\begin{equation*}
B(D_t,D_x)W_2(\phi_1, \phi_2,\cdots,\phi_N)\cdot W_2(\phi_1, \phi_2,\cdots,\phi_N)=0.
\end{equation*}
Then, repeating the same procedure for $\varrho_j \rightarrow \varrho_1$ up to $j=N$,
we can find each $W_j(\phi_1, \phi_2,\cdots,\phi_N)$ solves \eqref{bilinear eq}.
Thus we have completed the proof. \hfill $\Box$

\subsection{Wronskian solutions}

\begin{Proposition}
\label{Prop 3.5}
{\it A Wronskian solution to the bilinear KdV equation \eqref{blinear-KdV} is given as
\begin{equation}
f=|\widehat{N-1}|,
\label{wrons-KdV}
\end{equation}
provided that its entries  satisfy
\begin{equation}
-\phi_{xx}=A(t)\phi,
\label{cond-a}
\end{equation}
\begin{equation}
\phi_{t}=-4\phi_{xxx}+B(t)\phi,
\label{cond-b}
\end{equation}
where $A(t)=(A_{ij}(t))_{N\times N}$ and $B(t)=(B_{ij}(t))_{N\times N}$
are two arbitrary $N\times N$  matrices of $t$ but
independent of $x$.
Considering that \eqref{cond-a} and \eqref{cond-b} should be solvable, $A(t)$ and $B(t)$ must satisfy
\begin{equation}
A_t(t)+[A(t),B(t)]=0,
\label{compat}
\end{equation}
where $[A(t),B(t)]=A(t)B(t)-B(t)A(t)$.}\hfill $\Box$
\end{Proposition}

The proof for the proposition is quite similar to the case that $A(t)$ is constant and
triangular\cite{Sirianunpiboon-1988}.
The key point is that,
by taking $\Xi=|\W{N-1}|$ and $\Omega_{js}\equiv \partial^2_x$ in Proposition \ref{Prop 3.1}  and noticing
\eqref{cond-a},
we can get
\begin{equation*}
-\Tr{A(t)} |\W{N-1}|=-|\W{N-3},N-1,N|
+|\W{N-2},N+1|,
\end{equation*}
which is the same as in the cases that $A(t)$ is  diagonal\cite{Freeman-Nimmo-KP}
and triangular\cite{Sirianunpiboon-1988}.
In addition, we also need to use Proposition \ref{Prop 3.2}.

Equation \eqref{compat} comes from the compatibility of \eqref{cond-a} and \eqref{cond-b},
which guarantees \eqref{cond-a} and \eqref{cond-b} solvable, like in Ref.\cite{Ma-You-KdV}.

Can we simplify equations \eqref{cond-a} and\eqref{cond-b}?
Can the arbitrariness of $B(t)$ in \eqref{cond-b} generate any new solutions to the KdV equation?

\begin{Proposition}
\label{Prop 3.6}
{\it Suppose that $\{B_{ij}(t)\}\in C[a,b]$ ($a$ and $b$ can be infinite).
Then, there exists an non-singular $N\times N$ t-dependent matrix $H(t)$ satisfying
\begin{equation}
H_t(t)=-H(t)B(t).
\label{p3.6.0}
\end{equation}}

{\it Proof:}~ Consider the following homogeneous linear ordinary differential equations
\begin{equation}
h_t(t)=-B^T(t)h(t),
\label{p3.6.1}
\end{equation}
where $h(t)$ is an $N$-order vector function of $t$.
For any given number set $(t_j, \tilde{h}_{j1},\tilde{h}_{j2},\cdots, \tilde{h}_{jN})$,
under the condition of the proposition, \eqref{p3.6.1}
has unique solution vector
\begin{equation*}
h_j(t)=(h_{j1}(t),h_{j2}(t),\cdots,h_{jN}(t))^T
\end{equation*}
satisfying $h_j(t_j)=(\tilde{h}_{j1},\tilde{h}_{j2},\cdots,\tilde{h}_{jN})^T$
for each $j=1,2,\cdots N$.
Taking $Det \{\tilde{h}_{js} \}_{N\times N} \neq 0$,
then $\{{h}_{j}(t) \}^N_{j=1}$ is a basic solution set, i.e.,
$\{{h}_{j}(t) \}^N_{j=1}$ is linearly independent.
Finally, it turns out that the matrix
\begin{equation*}
H(t)=(h_{1}(t),h_{2}(t),\cdots,h_{N}(t))^T
\end{equation*}
is a non-singular solution to \eqref{p3.6.0}.
\hfill $\Box$
\end{Proposition}

By virtue of this proposition, taking
$\psi=H(t)\phi$ where $H(t)$ satisfies \eqref{p3.6.0},
we then from \eqref{cond-a} and \eqref{cond-b} have
\begin{equation*}
-\psi_{xx}=H(t)A(t)H^{-1}(t) \psi,
\end{equation*}
\begin{equation*}
\psi_{t}=-4\psi_{xxx};
\end{equation*}
and using Proposition \ref{Prop 3.2} we also have the relation $f(\psi)=|H(t)|f(\phi)$
which means $\psi$ and $\phi$ lead to same solutions to the KdV equation through
the transformation \eqref{trans-KdV}.
Thus, in the following we always
%
simplify equations \eqref{cond-a} and \eqref{cond-b} to
\begin{equation}
-\phi_{xx}=A \phi,
\label{cond-a'}
\end{equation}
\begin{equation}
\phi_{t}=-4\phi_{xxx},
\label{cond-b'}
\end{equation}
where $A$ is arbitrary but has to be  constant due to \eqref{compat}.
Further, as done in \cite{Ma-You-KdV}, based on Proposition\ref{Prop 3.3}, we can replace $A$ by
any matrices which are similar to $A$,
i.e., we only need to discuss
\begin{equation}
-\phi_{xx}=\Gamma \phi,
\label{cond-KdV-a}
\end{equation}
\begin{equation}
\phi_{t}=-4\phi_{xxx},
\label{cond-KdV-b}
\end{equation}
where $\Gamma=T^{-1}AT$ and $|T|\neq 0$.
In the paper we call \eqref{cond-a'} and \eqref{cond-b'} or \eqref{cond-KdV-a} and \eqref{cond-KdV-b}
{\it Wronskian condition equations} of the KdV equation.
For the solutions to \eqref{cond-a} and \eqref{cond-KdV-a}, we have the following proposition.

\begin{Proposition}
\label{Prop 3.7}
{\it If $\Gamma=T^{-1}AT$, $|T|\neq 0$
and $\{\Phi_1, \Phi_2, \cdots, \Phi_m \}$, $(m\leq N)$, are $m$ linearly independent
solution vectors of equation \eqref{cond-KdV-a}, then
$\{T\Phi_1, T\Phi_2, \cdots, T\Phi_m \}$ are $m$ linearly independent
solution vectors of equation \eqref{cond-a'} as well.}\hfill $\Box$
\end{Proposition}

In general, we take $\Gamma$ in \eqref{cond-KdV-a} to be the canonical form of $A$ for simplicity.

\subsection{Solutions related to $\Gamma$ }

In the following we give explicit expressions of general solutions to
the condition equations \eqref{cond-KdV-a} and \eqref{cond-KdV-b}
where $\Gamma$ takes different canonical forms of the $N\times N$ constant matrix $A$,
which corresponds to $A$ having different kinds of eigenvalues.
In addition, we describe relations between different kinds of solutions.

{\it Case 1}
\begin{equation}
\Gamma=D^{-}_{N}[\lambda_1,\lambda_2,\cdots,\lambda_N]
={\rm Diag}(-k^2_1,-k^2_2,\cdots,-k^2_N),
\label{nor-matrix-Case1}
\end{equation}
where  $\{ -k^2_j= \lambda_j \}$ are distinct negative numbers.
$\{ k_j\}$ are positive without loss of generality.
In this case, $\phi$ in the condition equations \eqref{cond-KdV-a} and \eqref{cond-KdV-b} is given as
\begin{equation}
\phi=\phi^{-}_{N}[\lambda_1,\lambda_2,\cdots,\lambda_N]
=(\phi^{-}_1,\phi^{-}_2,\cdots,\phi^{-}_N)^T,
\label{entry-I}
\end{equation}
in which
\begin{equation}
\phi^{-}_j=a^{+}_{j}\cosh {\xi_j} + a^{-}_{j}\sinh {\xi_j},
\label{entry-Case1}
\end{equation}
or
\begin{equation}
\phi^{-}_j=b^{+}_{j}e^{\xi_j} + b^{-}_{j}e^{-\xi_j},
\label{entry-Case1-1}
\end{equation}
where
\begin{equation}
\xi_j=k_j x -4k^3_j t +\xi^{(0)}_j,~~j=1,2,\cdots N,
\label{xi}
\end{equation}
$a^{\pm}_{j}$, $b^{\pm}_{j}$ and $\xi^{(0)}_j$  are all real constants.
Obviously, each $\phi_j$ is a solution of \eqref{Lax-KdV-a} and \eqref{Lax-KdV-b} when $u=0$ and the spectral
parameter $\lambda=-k_j^2$ is negative.
If we take $\phi$ as \eqref{entry-Case1} with $a^{\pm}_{j}=[1\mp(-1)^{j}]/2$ and $0<k_1 <k_2 < \cdots < k_N$
then the corresponding Wronskian generates a normal $N$-soliton
solution. In fact, such a Wronskian can  alternatively be written as
\begin{equation}
f=
\biggl (\prod^{N}_{j=1}e^{-\xi_j}\biggr)
        \biggl(\prod_{1\leq j<l\leq N}(k_{l}-k_{j})\biggr)
    \sum_{\mu=0, 1}\exp \biggl\{\sum_{j=1}^{N}2\mu_{j}\eta_{j}+\sum_{1\leq j<l\leq N}
   \mu_{j}\mu_{l}A_{jl} \biggr\},
\label{Hirota-KdV}
\end{equation}
where the sum over $\mu=0,1$ refers to each of $\mu_j=0,1$ for $j=0,1,\cdots, N$, and
\begin{equation*}
\eta_{j}=\xi_{j}-\frac{1}{4}\sum_{l=1,l\neq j}^{N}A_{jl},
~~e^{A_{jl}}=\biggl(\frac{k_l-k_j}{k_l+k_j}\biggr)^2.
\end{equation*}
This is nothing but Hirota's expression for the $N$-soliton solution in terms of polynomials of exponential
which admits classical $N$-soliton scattering.
A similar proof for the above statement can be found in Ref.\cite{Zhang-sinescs}.

Besides, we have the following result.

\begin{Proposition}
\label{Prop 3.8}
{\it
Suppose that $D^{-}_{N}[\lambda_1,\lambda_2,\cdots,\lambda_N]=T^{-1}AT$. Then
$T\phi^{-}_{N}[\lambda_1,\lambda_2,\cdots,\lambda_N]$
provides a general solution to \eqref{cond-a'} and \eqref{cond-b'}}.
\end{Proposition}

{\it Proof:} ~ Let $\phi^{-}_j$ be given by \eqref{entry-Case1-1}.
We can have
\begin{equation}
T\phi^{-}_{N}[\lambda_1,\lambda_2,\cdots,\lambda_N]=\sum^{N-1}_{j=0}b^{+}_j T\Phi^{+}_{j}+
\sum^{N-1}_{j=0}b^{-}_j T\Phi^{-}_{j},
\label{TPhi-KdV-1}
\end{equation}
where the vector set $\Phi^{\pm}_{j}$ are linearly independent and defined by
\begin{equation}
\Phi^{\pm}_{j}=(\Phi^{\pm}_{j,1},\Phi^{\pm}_{j,2},\cdots,\Phi^{\pm}_{j,N})^T,
~~~\Phi^{\pm}_{j,s}=\delta_{j,s}e^{\pm\xi_s}.
\end{equation}
If setting $t$ in $\Phi^{\pm}_{j}$ to be constant, then, by virtue of Proposition \ref{Prop 3.7},
\eqref{TPhi-KdV-1} is a general  solution to \eqref{cond-a'},
where $b^{\pm}_j$ should be considered as arbitrary functions of $t$.
Now substituting \eqref{TPhi-KdV-1} into \eqref{cond-b'},
one can find all the $b^{\pm}_j$ are arbitrary constant.
Thus the proposition holds.
\hfill $\Box$

{\it Case 2}
\begin{equation}
\Gamma=D^{+}_{N}[\lambda_1,\lambda_2,\cdots,\lambda_N]
={\rm Diag}(k^2_1,k^2_2,\cdots,k^2_N),
\label{nor-matrix-Case2}
\end{equation}
where  $\{ k^2_j= \lambda_j \}$ are distinct positive  numbers.
$\{ k_j\}$ are positive without loss of generality.
$\phi$ in the condition equations \eqref{cond-KdV-a} and \eqref{cond-KdV-b} is given as
\begin{equation}
\phi=\phi^{+}_{N}[\lambda_1,\lambda_2,\cdots,\lambda_N]
=(\phi^{+}_1,\phi^{+}_2,\cdots,\phi^{+}_N)^T,
\label{entry-II}
\end{equation}
in which
\begin{equation}
\phi^{+}_j=a^{+}_{j}\cos {\theta_j} + a^{-}_{j}\sin {\theta_j},
\label{entry-Case2}
\end{equation}
or
\begin{equation}
\phi^{+}_j=b^{+}_{j}e^{i\theta_j} + b^{-}_{j}e^{-i\theta_j},
\end{equation}
where
\begin{equation}
\theta_j=k_j x +4k^3_j t+ \theta^{(0)}_j ,~~j=1,2,\cdots N,
\label{theta}
\end{equation}
$a^{\pm}_{j}$, $b^{\pm}_{j}$ and $\theta^{(0)}_j$ are all real constant.
Such a  $\phi_j$ is also a solution of \eqref{Lax-KdV-a} and \eqref{Lax-KdV-b} when $u=0$
and the spectral parameter
$\lambda=k_j^2$ is positive.

{\it Case 3}
\begin{equation}
\Gamma=J^{-}_{N}[\lambda_1]
=\left(\begin{array}{cccccc}
-k^2_1 & 0    & 0   & \cdots & 0   & 0 \\
1   & -k^2_1  & 0   & \cdots & 0   & 0 \\
\cdots &\cdots &\cdots &\cdots &\cdots &\cdots \\
0   & 0    & 0   & \cdots & 1   & -k^2_1
\end{array}\right)_N,
\label{nor-matrix-Case3}
\end{equation}
where $ -k^2_1=\lambda_1$ is a positive  number and $k_1$ also positive.

In the following we will first give explicit general solutions to the condition
equations \eqref{cond-KdV-a} and \eqref{cond-KdV-b} with \eqref{nor-matrix-Case3},
and then  discuss relations between the solutions obtained in Case 3 and Case 1.

{\it a).~~ General solutions to the condition
 equations \eqref{cond-KdV-a} and \eqref{cond-KdV-b} with \eqref{nor-matrix-Case3} }

A special solution in this case is easily given by
\begin{equation}
\phi=\mathcal{Q}^{+}_0+\mathcal{Q}^{-}_0
\label{special-sol}
\end{equation}
with
\begin{equation}
\mathcal{Q}^{\pm}_0=(\mathcal{Q}^{\pm}_{0,0},\mathcal{Q}^{\pm}_{0,1},\cdots,\mathcal{Q}^{\pm}_{0,N-1})^T,
\end{equation}
and
\begin{equation}
\mathcal{Q}^{\pm}_{0,j}=\frac{(-1)^j}{j!}\partial^{j}_{\lambda_1} b^{\pm}_1e^{\pm \xi_1},
\end{equation}
where

Now we define
\begin{equation}
\phi^{J^{-}}_N[\lambda_1]=\mathcal{A}\mathcal{Q}^{+}_{0}+ \mathcal{B} \mathcal{Q}^{-}_{0},
\label{gen-sol-KdV}
\end{equation}
where
$\mathcal{A}$ and $\mathcal{B}$ are arbitrary $N$-order lower triangular
Toeplitz matrices defined as \eqref{A} and \eqref{B},
i.e., arbitrary elements
of semigroup $\widetilde{G}_N$ defined by \eqref{G-tilde}.
Then we have

\begin{Proposition}
\label{Prop 3.9}
{\it
 $\phi=\phi^{J^{-}}_N[\lambda_1]$ defined by \eqref{gen-sol-KdV}
gives  an explicit expression for all solutions to
the condition equations \eqref{cond-KdV-a} and \eqref{cond-KdV-b} when $\Gamma=J^{-}_{N}[\lambda_1]$.}
\end{Proposition}

{\it Proof:} ~
In fact, first, \eqref{gen-sol-KdV} is still a solution to \eqref{cond-KdV-a} and \eqref{cond-KdV-b}
due to $\mathcal{A}$ and $\mathcal{B}$ commuting with the Jordan block $J^{-}_{N}[\lambda_1]$.
Secondly, we alternatively write \eqref{gen-sol-KdV} as
\begin{equation}
\phi^{J^{-}}_N[\lambda_1]=\sum^{N-1}_{j=0}a_j \mathcal{Q}^{+}_{j}+
\sum^{N-1}_{j=0}b_j \mathcal{Q}^{-}_{j},
\label{gen-sol-KdV-3}
\end{equation}
where
\begin{equation}
\mathcal{Q}^{\pm}_{j}=(\overbrace{0,0,\cdots,0}^{j},
\mathcal{Q}^{\pm}_{0,0},\mathcal{Q}^{\pm}_{0,1},\cdots,\mathcal{Q}^{\pm}_{0,N-j-1})^T,
~~~(j=0,1,\cdots, N-1).
\label{Q-pm}
\end{equation}
$\{\mathcal{Q}^{\pm}_{j} \}_{j=0}^{N-1}$
are nothing but  $2N$ linearly independent solution vectors of
the condition equations \eqref{cond-KdV-a} and \eqref{cond-KdV-b}.
We argue that \eqref{gen-sol-KdV-3} is a general solution to  \eqref{cond-KdV-a} and \eqref{cond-KdV-b}
and we explain this fact in the following.
First, since there is no $t$ involved in \eqref{cond-KdV-a}, so, taking $t$ to be constant
in $\{\mathcal{Q}^{\pm}_{j} \}$,
\eqref{gen-sol-KdV-3} provides a general solution to  \eqref{cond-KdV-a}, where
$\{ a_j\}$ and $\{b_j\} $ can be considered as arbitrary functions of $t$.
Then, to determine $\{ a_j\}$ and $\{b_j\} $, we substitute \eqref{gen-sol-KdV-3} into \eqref{cond-KdV-b}
and it is easy to show that all the  $\{ a_j\}$ and $\{b_j\} $ are  arbitrary constants.
Thus we finish the proof.
\hfill $\Box$

Employing a similar proof for Proposition \ref{Prop 3.9} and by virtue of Proposition \ref{Prop 3.7},
we can get general solutions to \eqref{cond-a'} and \eqref{cond-b'} with any
$A$ which is similar to $J^{-}_{N}[\lambda_1]$.

\begin{Proposition}
\label{Prop 3.10}
{\it
Suppose that $J^{-}_{N}[\lambda_1]=T^{-1}AT$. Then,
\begin{equation}
T\phi^{J^{-}}_N[\lambda_1]=\sum^{N-1}_{j=0}a_j T\mathcal{Q}^{+}_{j}+
\sum^{N-1}_{j=0}b_j T\mathcal{Q}^{-}_{j}
\end{equation}
provides a general solution to \eqref{cond-a'} and \eqref{cond-b'}}.
\hfill $\Box$
\end{Proposition}

{\it b).~~ Effective form of the general solution \eqref{gen-sol-KdV} }

For the parameters in $\mathcal{A}$ and $\mathcal{B}$ in \eqref{gen-sol-KdV}, we have the following proposition.

\begin{Proposition}
\label{Prop 3.11} {\it
The effective form of \eqref{gen-sol-KdV} is
\begin{equation}
\phi=\mathcal{A}\mathcal{Q}^{+}_{0}+ \mathcal{Q}^{-}_{0},
\label{egen-sol-KdV}
\end{equation}
where $\mathcal{A}\in \widetilde{G}_N$, i.e.,
\eqref{egen-sol-KdV} and \eqref{gen-sol-KdV} lead to same solutions to the KdV equation.}
\end{Proposition}

{\it Proof:}~~ First, one of matrices $\mathcal{A}$
and $\mathcal{B}$ in \eqref{gen-sol-KdV} must be in the group $G_N$, otherwise, \eqref{gen-sol-KdV} generates a zero
Wronskian. We take $\mathcal{B}\in G_N$ without loss of generality.
Next, on the basis of Proposition \ref{Prop 3.3}, \eqref{gen-sol-KdV} and
$\mathcal{B}^{-1}\mathcal{A}\mathcal{Q}^{+}_{0}+ \mathcal{Q}^{-}_{0}$
lead to same solutions to the KdV equation.
In addition, making use of
Proposition \ref{Prop 2.2}, we can further substitute $\mathcal{A}$ for $\mathcal{B}^{-1}\mathcal{A}$
and then get the  effective form  \eqref{egen-sol-KdV}.
Thus, the number of effective parameters in $N$-order Wronskian is essentially $N$ not $2N$
and we have completed the proof.
\hfill $\Box$

{\it c).~~ Obtaining the general solution  \eqref{gen-sol-KdV} through a limit procedure}

As \eqref{gen-sol-KdV} is a general solution to \eqref{cond-KdV-a} and \eqref{cond-KdV-b} when $\Gamma$ is
the Jordan block \eqref{nor-matrix-Case3}, we call the Wronskian $f(\phi^{J^{-}}_N[\lambda_1])$ a
Jordan block solution to the bilinear KdV equation \eqref{blinear-KdV}.
However, such a solution can also be obtained from a limit of some Wronskian obtained in Case 1.
To achieve that, let us recall the proof for Proposition \ref{Prop 3.4}. We start from
the following Wronskian
\begin{equation}
\frac{W(\phi^{-}_1, \phi^{-}_2,\cdots,\phi^{-}_N)}{\prod^{N}_{j=2}(\lambda_1-\lambda_j)^{j-1}}
\label{limit}
\end{equation}
which is a solution to the bilinear KdV equation \eqref{blinear-KdV} corresponding to Case 1,
where we take
\begin{equation*}
\phi^{-}_1=\phi^{-}_1(\lambda_1, t,x)=b^{+}_{1}e^{\xi_1} + b^{-}_{1}e^{-\xi_1},
\end{equation*}
and $\phi^{-}_j=\phi^{-}_j(\lambda_1, t,x)$.
Then, following the limit procedure in the proof for Proposition \ref{Prop 3.4},
and taking $\{\lambda_j\}_{j=2}^{N}\rightarrow \lambda_1$ successively,
we can get the general solution \eqref{gen-sol-KdV} in Case 3,
where the arbitrary matrices in semigroup $\widetilde{G}_N$, by virtue Proposition \ref{Prop 2.4},
will come from considering $b^{+}_{1}$ and $ b^{-}_{1}$
as some polynomials of $\lambda_1$.

{\it d).~~ General solutions to equations \eqref{cond-KdV-a} and
\eqref{cond-KdV-b} with $\Gamma=\widehat{\Gamma}^{-}_N[k_1]$ \eqref{Gamma-k-KdV}}

Let us recall the results in \cite{Sirianunpiboon-1988}.
\begin{equation}
\widehat{\phi}=\widehat{\mathcal{Q}}^{+}_0+\widehat{\mathcal{Q}}^{-}_0,
\end{equation}
with
\begin{equation}
\widehat{\mathcal{Q}}^{\pm}_0=(\widehat{\mathcal{Q}}^{\pm}_{0,0},\widehat{\mathcal{Q}}^{\pm}_{0,1},\cdots,
\widehat{\mathcal{Q}}^{\pm}_{0,N-1})^T,~~~
\widehat{\mathcal{Q}}^{\pm}_{0,j}=\frac{1}{j!}  \partial^{j}_{k_1}b^{\pm}_1 e^{\pm \xi_1},
\label{vector-k-KdV-3}
\end{equation}
is a special  solution to the condition equations \eqref{cond-KdV-a} and \eqref{cond-KdV-b} but with
\begin{equation}
\Gamma=\widehat{\Gamma}^{-}_N[k_1]=\left(\begin{array}{ccccccc}
-k^2_1 & 0    & 0   & \cdots & 0   & 0 & 0 \\
-2k_1   & -k^2_1  & 0   & \cdots & 0   & 0 & 0\\
-1   & -2k_1    & -k^2_1 & \cdots & 0   & 0 & 0\\
\cdots &\cdots &\cdots &\cdots &\cdots &\cdots &\cdots \\
0   & 0    & 0   & \cdots & -1 & -2k_1  & -k^2_1
\label{Gamma-k-KdV}
\end{array}\right)_N.
\end{equation}
$\widehat{\phi}$ is first given  in \cite{Sirianunpiboon-1988} and leads to the same solution to the KdV equation
as $\phi$ \eqref{special-sol} does. In fact, we can find the following relation
\begin{equation}
\phi=\mathcal{Q}^{+}_0+\mathcal{Q}^{-}_0=M\widehat{\phi},
\end{equation}
where $M=(M_{js})_{0\leq j,s\leq N-1}$ is an $N\times N$ lower triangular matrix and $\{M_{js}\}$ come from
\begin{equation}
\Bigl(\frac{1}{2k_1}\partial_{k_1}\Bigr)^{j}=\sum^{j}_{s=0}M_{js}\partial^s_{k_1},~~~(j=0,1,\cdots, N-1).
\end{equation}

Obviously, to calculate derivatives of $b^{\pm}_1 e^{\pm \xi_1}$ with respect to $k_1$
is much easier than with respect to $\lambda_1$. In what follows we give an explicit general solution
to the condition equations \eqref{cond-KdV-a} and
\eqref{cond-KdV-b} with $\Gamma=\widehat{\Gamma}^{-}_N[k_1]$
by virtue of Proposition \ref{Prop 2.2}.

\begin{Proposition}
\label{Prop 3.12}
{\it
\begin{equation}
\widehat{\phi}^{J^{-}}_N[k_1]=\mathcal{A}\widehat{\mathcal{Q}}^{+}_{0}+ \mathcal{B} \widehat{\mathcal{Q}}^{-}_{0},
~~~~\mathcal{A},~\mathcal{B} \in \widetilde{G}_N
\label{gen-sol-KdV-k-3}
\end{equation}
provides a general solution to equations \eqref{cond-KdV-a} and
\eqref{cond-KdV-b} with $\Gamma=\widehat{\Gamma}^{-}_N[k_1]$.}
\hfill $\Box$
\end{Proposition}

In fact, by noting that $\widehat{\Gamma}^{-}_N[k_1]\in \widetilde{G}_N$
and $\widetilde{G}_N$ is an Abelian semigroup, we have
\begin{equation*}
\mathcal{A}\widehat{\Gamma}^{-}_N[k_1]=\widehat{\Gamma}^{-}_N[k_1]\mathcal{A},~~~
\forall \mathcal{A}\in \widetilde{G}_N,
\end{equation*}
which means \eqref{gen-sol-KdV-k-3} is still a solution to \eqref{cond-KdV-a} and
\eqref{cond-KdV-b}
Then, similar to the proof for Proposition \ref{Prop 3.10},
\eqref{gen-sol-KdV-k-3}
can alternatively be expressed as
\begin{equation*}
\widehat{\phi}^{J^{-}}_N[k_1]=\sum^{N-1}_{j=0}a_j \widehat{\mathcal{Q}}^{+}_{j}+
\sum^{N-1}_{j=0}b_j \widehat{\mathcal{Q}}^{-}_{j},
\end{equation*}
where
\begin{equation*}
\widehat{\mathcal{Q}}^{\pm}_{j}=(\overbrace{0,0,\cdots,0}^{j},
\widehat{\mathcal{Q}}^{\pm}_{0,0},\widehat{\mathcal{Q}}^{\pm}_{0,1},\cdots,\widehat{\mathcal{Q}}^{\pm}_{0,N-j-1})^T,
~~~(j=0,1,\cdots, N-1),
\label{Q-pm-k}
\end{equation*}
and $\{\widehat{\mathcal{Q}}^{\pm}_{j} \}_{j=0}^{N-1}$
 are just $2N$ linearly independent solution vectors of the condition equations
\eqref{cond-KdV-a} and \eqref{cond-KdV-b} with $\Gamma=\widehat{\Gamma}^{-}_N[k_1]$.
That means \eqref{gen-sol-KdV-k-3} is a general solution to \eqref{cond-KdV-a} and \eqref{cond-KdV-b}.

For the links with the solutions obtained in Case 1,
\eqref{gen-sol-KdV-k-3} can be obtained by substituting
$\prod^{N}_{j=2}(k_1-k_j)^{j-1}$ for the denominator of
\eqref{limit} and considering successively the limit $k_j\rightarrow k_1$.

The effective form of \eqref{gen-sol-KdV-k-3} is
\begin{equation}
\widehat{\phi}=\mathcal{A}\widehat{\mathcal{Q}}^{+}_{0}+ \widehat{\mathcal{Q}}^{-}_{0},
~~\mathcal{A}\in \widetilde{G}_N.
\label{e-gen-sol-KdV-k-3}
\end{equation}
Obviously, \eqref{gen-sol-KdV-k-3} and \eqref{e-gen-sol-KdV-k-3} are preferable.

Thus, for Case 3, we have given the explicit general  solutions to the condition equations
\eqref{cond-KdV-a} and \eqref{cond-KdV-b} when $\Gamma$ is respectively \eqref{nor-matrix-Case3}
and \eqref{Gamma-k-KdV},
and further given the effective forms of them, where parameter number has been reduced half.
We also investigated the relationship of solutions between Case 3 and Case 1, and
explained the Jordan block solution as a limit one. In fact, such a kind of  limit solutions can also
be obtained through the IST
as  multi-pole solutions\cite{Wadati-82,Wadati-84}, or through a limit procedure
in Darboux transformation\cite{Matveev-positon-kdv1,Matveev-positon-kdv2},
or through a generalized Hirota's procedure
in Refs.\cite{Chen-02-JPSJ1,Chen-02-JPSJ2}.

{\it Case 4}
\begin{equation}
\Gamma=J^{+}_{N}[\lambda_1]
=\left(\begin{array}{cccccc}
k^2_1 & 0    & 0   & \cdots & 0   & 0 \\
1   & k^2_1  & 0   & \cdots & 0   & 0 \\
\cdots &\cdots &\cdots &\cdots &\cdots &\cdots \\
0   & 0    & 0   & \cdots & 1   & k^2_1
\end{array}\right)_N,
\label{nor-matrix-Case4}
\end{equation}
where $ k^2_1=\lambda_1$ is a positive number and $k_1$ also positive.

Similar to the relationship between Case 3 and Case 1, Wronskian solutions in this case
can be considered as a limit result from those in Case 2. The explicit general solution
to equations \eqref{cond-KdV-a} and \eqref{cond-KdV-b} in this case can be expressed as
\begin{equation}
\phi^{J^{+}}_{\hbox{\tiny{\it N}}}[\lambda_1]
=\mathcal{A}\mathcal{P}^{+}_{0}+ \mathcal{B} \mathcal{P}^{-}_{0},
~~\mathcal{A},\mathcal{B}\in \widetilde{G}_N,
\label{gen-sol-KdV-4}
\end{equation}
and its effective form is
\begin{equation}
\phi=\mathcal{A}\mathcal{P}^{+}_{0}+  \mathcal{P}^{-}_{0},
~~\mathcal{A}\in \widetilde{G}_N,
\label{egen-sol-KdV-4}
\end{equation}
where
\begin{equation}
\mathcal{P}^{\pm}_{0}=(\mathcal{P}^{\pm}_{0,0},\mathcal{P}^{\pm}_{0,1},\cdots,\mathcal{P}^{\pm}_{0,N-1})^T,
~~~\mathcal{P}^{\pm}_{0,j}=\frac{(-1)^j}{j!}\partial^{j}_{\lambda_1}
b^{\pm}_1 e^{\pm i \theta_1},
\end{equation}
$\lambda_1=k^2_1$ and $\theta_1$ is defined by \eqref{theta}.

The general solution to the condition equations \eqref{cond-KdV-a} and \eqref{cond-KdV-b} with
\begin{equation}
\Gamma=\widehat{\Gamma}^{+}_N[k_1]=\left(\begin{array}{ccccccc}
k^2_1 & 0    & 0   & \cdots & 0   & 0 & 0 \\
2k_1   & k^2_1  & 0   & \cdots & 0   & 0 & 0\\
1   & 2k_1    & k^2_1 & \cdots & 0   & 0 & 0\\
\cdots &\cdots &\cdots &\cdots &\cdots &\cdots &\cdots \\
0   & 0    & 0   & \cdots & 1 & 2k_1  & k^2_1
\label{Gamma-k-KdV-4}
\end{array}\right)_N.
\end{equation}
is
\begin{equation}
\widehat{\phi}^{J^{+}}_{\hbox{\tiny{\it N}}}[k_1]
=\mathcal{A}\widehat{\mathcal{P}}^{+}_{0}+ \mathcal{B} \widehat{\mathcal{P}}^{-}_{0},
~~~\mathcal{A}, \mathcal{B}\in \widetilde{G}_N,
\end{equation}
and its effective form is
\begin{equation}
\widehat{\phi}=\mathcal{A}\widehat{\mathcal{P}}^{+}_{0}+ \widehat{ \mathcal{P}}^{-}_{0},
~~~ \mathcal{A} \in \widetilde{G}_N,
\end{equation}
where
\begin{equation}
\widehat{\mathcal{P}}^{\pm}_{0}
=(\widehat{\mathcal{P}}^{\pm}_{0,0},\widehat{\mathcal{P}}^{\pm}_{0,1},\cdots,\widehat{\mathcal{P}}^{\pm}_{0,N-1})^T,
~~~\widehat{\mathcal{P}}^{\pm}_{0,j}=\frac{1}{j!}\partial^{j}_{k_1}
b^{\pm}_1 e^{\pm i \theta_1}.
\end{equation}

{\it Case 5} ~
\begin{equation}
\Gamma=J^{0}_{N}
=\left(\begin{array}{cccccc}
0 & 0    & 0   & \cdots & 0   & 0 \\
1   & 0  & 0   & \cdots & 0   & 0 \\
0   & 1  & 0   & \cdots & 0   & 0 \\
\cdots &\cdots &\cdots &\cdots &\cdots &\cdots \\
0   & 0    & 0   & \cdots & 1   & 0
\end{array}\right)_N.
\label{nor-matrix-Case5}
\end{equation}
In this case, we can get rational solutions to the KdV equation from the corresponding Wronskians.

To give a general form of the solutions to the condition
equations \eqref{cond-KdV-a} and \eqref{cond-KdV-b} with $J^{0}_{N}$
\eqref{nor-matrix-Case5}, we first construct $2N$ explicit linearly independent solution vectors.
To achieve that, we consider
\begin{equation}
\widetilde{\phi}^{+}_{1}=\cosh \xi_1,~~\xi_1=k_1 x -4k_1^3t,
\end{equation}
where we have taken $\xi^{(0)}_1=0$ in \eqref{xi}.
Expanding $\widetilde{\phi}^{+}_{1}$ with respect to $k_1$ yields
\begin{equation}
\widetilde{\phi}^{+}_{1}=\sum^{\infty}_{j=0}(-1)^j R^{+}_{0,j}\lambda_1^{j},
\end{equation}
where $\lambda_1=-k_1^2$ and
\begin{equation}
R^{+}_{0,j}=\frac{1}{(2j)!}\Bigl [\frac{\partial^{2j}}{{\partial k_1}^{2j}}\cosh \xi_1\Bigr ]_{k_1=0}
\end{equation}
which is independent of $\lambda_1$.
Noting that
\begin{equation*}
\widetilde{\phi}^{+}_{1,xx}=-\lambda_1 \widetilde{\phi}^{+}_{1},~~
\widetilde{\phi}^{+}_{1,t}=-4\widetilde{\phi}^{+}_{1,xxx}
\end{equation*}
we have
\begin{equation*}
\sum^{\infty}_{j=0} (R^{+}_{0,j})_{xx} (-\lambda_1)^{j}
=\sum^{\infty}_{l=0} (R^{+}_{0,l})(- \lambda_1)^{l+1},~~~
\sum^{\infty}_{j=0}(-1)^j (R^{+}_{0,j})_{t} \lambda_1^{j}
=-4 \sum^{\infty}_{j=0}(-1)^j (R^{+}_{0,j})_{xxx} \lambda_1^{j}.
\end{equation*}
Then, taking
\begin{equation}
\mathcal{R}^{+}_0=(\mathcal{R}^{+}_{0,0},\mathcal{R}^{+}_{0,1},\cdots,\mathcal{R}^{+}_{0,N-1})^T
\end{equation}
and considering $\lambda_1$ as an arbitrary real number,
we get
\begin{equation}
\mathcal{R}^{+}_{0,xx}=- J^{0}_{N} \mathcal{R}^{+}_0, ~~
\mathcal{R}^{+}_{0,t}= -4 \mathcal{R}^{+}_{0,xxx},
\end{equation}
i.e., $\mathcal{R}^{+}_0$ is a special solution vector
to \eqref{cond-KdV-a} and \eqref{cond-KdV-b} with $J^{0}_{N}$.

Another special solution vector to \eqref{cond-KdV-a} and \eqref{cond-KdV-b} with $J^{0}_{N}$
is
\begin{equation}
\mathcal{R}^{-}_0=(\mathcal{R}^{-}_{0,0},\mathcal{R}^{-}_{0,1},\cdots,\mathcal{R}^{-}_{0,N-1})^T,~~
R^{-}_{0,j}=\frac{1}{(2j+1)!}\Bigl [\frac{\partial^{2j+1}}{{\partial k_1}^{2j+1}}\sinh \xi_1\Bigr ]_{k_1=0},
\end{equation}
which is derived by expanding
\begin{equation}
\widetilde{\phi}^{-}_{1}=\frac{\sinh \xi_1}{k_1}=\sum^{\infty}_{j=0}(-1)^j R^{-}_{0,j}\lambda_1^{j}.
\end{equation}

$ R^{+}_{0}$ and $ R^{-}_{0}$ are linearly independent. This is because
each $R^{+}_{0,j}$ is even with respect to $x^h t^s$, i.e., $h+s$ is even,
while $R^{-}_{0,j}$ is odd.

Then, as \eqref{Q-pm} in Case 3, we can easily construct $2N$ linearly independent solution vectors,
and a general solution for this case can be
\begin{equation}
\phi^{0}_N=\mathcal{A}\mathcal{R}^{+}_{0}+ \mathcal{B} \mathcal{R}^{-}_{0},
~~\mathcal{A},\mathcal{B}\in \widetilde{G}_N.
\label{gen-sol-KdV-5}
\end{equation}
Further, the effective form of \eqref{gen-sol-KdV-5} is
\begin{equation}
\phi=\mathcal{A}\mathcal{R}^{+}_{0}+ \mathcal{R}^{-}_{0},
~~\mathcal{A}\in \widetilde{G}_N.
\label{egen-sol-KdV-5}
\end{equation}

We note that, similar to the previous two cases, \eqref{gen-sol-KdV-5} can also be obtained by successively taking
$k_j\rightarrow 0, ~( j=2,3,\cdots N)$ in
\begin{equation}
\frac{W(\phi_1, \phi_2,\cdots,\phi_N)}{\prod^{N}_{j=2} (-k_j^2)^{j-1}},
\end{equation}
where
\begin{equation*}
\phi_1=\phi_1(k_1, t,x)=a^{+}_{1}\cosh {\xi_1} + a^{-}_{1}\frac{\sinh {\xi_1}}{k_1},
\end{equation*}
and $\phi_j=\phi_j(k_1, t,x)$.
$\mathcal{A}$ and $\mathcal{B}$ in \eqref{gen-sol-KdV-5} can also be obtained in this limit procedure
by considering $a^{+}_{1}$ and $ a^{-}_{1}$
to be  some differential functions of $k_1$, i.e., $a^{\pm}_{1}(k_1)$,
which can be proved through a similar proof for Proposition \ref{Prop 2.4}.
This fact consists with the way to generate rational solutions by considering long-wave limits
proposed by Ablowitz, Satsuma\cite{Ablowitz-Satsuma} and Nimmo, Freeman\cite{rational-1}.

We also note that $\sin \theta$ and $\cos \theta$ do not generate any
new rational solutions to the KdV equation due to
\begin{equation*}
\cosh k = \cos ik,~~i \sinh k = \sin ik.
\end{equation*}

{\it Case 6} ~
\begin{equation}
\Gamma=D^{\hbox{\it\tiny C}}_{\hbox{\tiny 2\it M}}[\lambda_1,\lambda_2,\cdots,\lambda_M]
={\rm Diag}(-k^2_1,-k^{*2}_1,-k^2_2,-k^{*2}_2,\cdots,-k^2_M,-k^{*2}_M),
\label{nor-matrix-Case6}
\end{equation}
where $\{-k^2_j=\lambda_j\}$ are $M$ distinct complex numbers, and $*$ means complex conjugate.
If we consider $\Gamma$ to be a canonical form of $A$ in \eqref{cond-a'}, then this case corresponds to
the real matrix $A$ having $N=2M$ distinct complex eigenvalues which appear in conjugate pairs.

In this case, just similar to Case 1, we can take Wronskian entry vector as
\begin{equation}
\phi^{\hbox{\it\tiny C}}_{\hbox{\tiny 2\it M}}[\lambda_1,\lambda_2,\cdots,\lambda_{M}]
=(\phi^{\hbox{\it\tiny C}}_1, \phi^{\hbox{\it\tiny C}*}_1,
\phi^{{\hbox{\it\tiny C}}}_2, \phi^{{\hbox{\it\tiny C}}*}_2,\cdots,
\phi^{\hbox{\it\tiny C}}_{\hbox{\it\tiny M}}, \phi^{\hbox{\it\tiny C}*}_{\hbox{\it\tiny M}})^T,
\label{entry-Case6-0}
\end{equation}
where
\begin{equation}
\phi^{\hbox{\it\tiny C}}_j=a^{+}_{j}\cosh {\xi_j} + a^{-}_{j}\sinh {\xi_j},
\end{equation}
or
\begin{equation}
\phi^{\hbox{\it\tiny C}}_j=b^{+}_{j}e^{\xi_j} + b^{-}_{j}e^{-\xi_j},
\label{entry-Case6-1}
\end{equation}
with
\begin{equation}
\xi_j=k_j x -4k^3_j t +\xi^{(0)}_j,~~j=1,2,\cdots M,
\label{xi-6}
\end{equation}
while, different from in Case 1, here
$a^{\pm}_{j}$, $b^{\pm}_{j}$ and $\xi^{(0)}_j$  are all complex constants.

Here we do not need to care whether $\Gamma$ is real or complex,
so long as it can generate a real solution to the KdV equation.
Obviously, \eqref{entry-Case6-0} implies the Wronskian
$f(\phi^{\hbox{\it\tiny C}}_{\hbox{\tiny 2\it M}}[\lambda_1,\lambda_2,\cdots,\lambda_{M}])$ satisfies
\begin{equation*}
f(\phi^{\hbox{\it\tiny C}}_{\hbox{\tiny 2\it M}}[\lambda_1,\lambda_2,\cdots,\lambda_{M}])
=(-1)^M f^*(\phi^{\hbox{\it\tiny C}}_{\hbox{\tiny 2\it M}}[\lambda_1,\lambda_2,\cdots,\lambda_{M}]).
\end{equation*}
That means $f(\phi^{\hbox{\it\tiny C}}_{\hbox{\tiny 2\it M}}[\lambda_1,\lambda_2,\cdots,\lambda_{M}])$,
which is either real or pure imaginary, always generates a real solution
to the KdV equation through the transformation \eqref{trans-KdV}.

An alternative form of $\Gamma=D^{c}_{2M}[\lambda_1,\lambda_2,\cdots,\lambda_M]$ is
\begin{equation}
\Gamma={\rm Diag}(\widetilde{\Lambda}_1,\widetilde{\Lambda}_2,\cdots,\widetilde{\Lambda}_M),
~~\widetilde{\Lambda}_j=\biggl(\begin{matrix}\lambda_{j1} & -\lambda_{j2}\\
                            \lambda_{j2} & \lambda_{j1}
              \end{matrix}\biggr),~~j=1,2,\cdots M,
\label{(2.48)}
\end{equation}
where
\begin{equation}
\lambda_{j1}+ i \lambda_{j2}=\lambda_{j},
\end{equation}
and $i$ is the imaginary unit.
\eqref{(2.48)} is the real version of \eqref{nor-matrix-Case6} and these two forms are
connected through
\begin{equation}
{\rm Diag}(\widetilde{\Lambda}_1,\widetilde{\Lambda}_2,\cdots,\widetilde{\Lambda}_M)=
U^{-1}D^{c}_{2M}[\lambda_1,\lambda_2,\cdots,\lambda_M]U
\end{equation}
where $U$ is the following $2M\times 2M$ block-diagonal matrix
\begin{equation}
U={\rm Diag}(U_1,U_1,\cdots,U_1),
~~U_1= \biggl(\begin{matrix}1 & i\\
                            1 & -i \end{matrix}\biggr).
\label{(2.50)}
\end{equation}
Following the relation $\lambda_j=-k^2_j$,
we can separate $k_j$ by
\begin{equation*}
k_j=k_{j1}+ i k_{j2}
\end{equation*}
where
\begin{equation}
k_{j1}=\varepsilon_1 \sqrt{\frac{\sqrt{\lambda_{j1}^2+\lambda_{j2}^2}-\lambda_{j1}}{2}},
~~k_{j2}=\varepsilon_2 \sqrt{\frac{\sqrt{\lambda_{j1}^2+\lambda_{j2}^2}+\lambda_{j1}}{2}},
\label{r-c parts of k_j}
\end{equation}
$\{\varepsilon_s\}=\{\pm 1\}$ and satisfy $\varepsilon_1 \varepsilon_2=-{\rm{sgn}}(\lambda_{j2})$.

Noting that $U^{-1}_1=\frac{1}{2}\biggl(\begin{matrix}1 & 1\\
                            -i & i \end{matrix}\biggr)$
can separate real and imaginary parts of a complex number, i.e.,
\begin{equation*}
U^{-1}_1 \biggl(\begin{array}{c}
A+iB\\
A-iB
\end{array}\biggr)
=\biggl(\begin{array}{c}
A\\
B
\end{array}\biggr),
\end{equation*}
and taking
\begin{equation*}
\phi^{\hbox{\it\tiny C}}_j=\phi_{j1}+ i \phi_{j2},~~~
b^{\pm}_{j}=b^{\pm}_{j1}+ ib^{\pm}_{j2},~~~
\xi^{(0)}_{j}=\xi^{(0)}_{j1}+ i \xi^{(0)}_{j2},
\end{equation*}
we can easily write out the following solution of the condition equations \eqref{cond-KdV-a} and \eqref{cond-KdV-b}
with \eqref{(2.48)}\cite{Ma-complexiton},
\begin{equation}
\phi^{\hbox{\it\tiny R}}_{\hbox{\tiny 2\it M}}[\lambda_1,\lambda_2,\cdots,\lambda_{M}]
=(\phi_{11},\phi_{12},\phi_{21},\phi_{22},\cdots,\phi_{\hbox{\tiny {\it M}1}},\phi_{\hbox{\tiny{\it M}2}})^T
=U^{-1}\phi^{\hbox{\it\tiny C}}_{\hbox{\tiny 2\it M}}[\lambda_1,\lambda_2,\cdots,\lambda_{M}]
\label{phi-R-C}
\end{equation}
where, if $\phi^{\hbox{\it\tiny C}}_j$ is defined by \eqref{entry-Case6-1},
\begin{equation}
\begin{array}{rl}
\phi_{j1}= & \Bigl\{ b^{+}_{j1}\cos [k_{j2}(x-\nu_j t)+\xi^{(0)}_{j2}]
-b^{+}_{j2} \sin [k_{j2}(x-\nu_j t)+\xi^{(0)}_{j2}]\Bigr\}
e^{[k_{j1}(x-\mu_j t)+\xi^{(0)}_{j1}]}\\
~& +\Bigl\{ b^{-}_{j1}\cos [k_{j2}(x-\nu_j t)+\xi^{(0)}_{j2}]
+b^{-}_{j2} \sin [k_{j2}(x-\nu_j t)+\xi^{(0)}_{j2}]\Bigr\}
e^{-[k_{j1}(x-\mu_j t)+\xi^{(0)}_{j1}]},
\end{array}
\label{entry-Case6-R-a}
\end{equation}
\begin{equation}
\begin{array}{rl}
\phi_{j2} = & \Bigl\{ b^{+}_{j2}\cos [k_{j2}(x-\nu_j t)+\xi^{(0)}_{j2}]
+b^{+}_{j1} \sin [k_{j2}(x-\nu_j t)+\xi^{(0)}_{j2}]\Bigr\}
e^{[k_{j1}(x-\mu_j t)+\xi^{(0)}_{j1}]}\\
~& +\Bigl\{ b^{-}_{j2}\cos [k_{j2}(x-\nu_j t)+\xi^{(0)}_{j2}]
-b^{-}_{j1} \sin [k_{j2}(x-\nu_j t)+\xi^{(0)}_{j2}]\Bigr\}
e^{-[k_{j1}(x-\mu_j t)+\xi^{(0)}_{j1}]},
\end{array}
\label{entry-Case6-R-b}
\end{equation}
and
\begin{equation*}
\mu_j=4(k_{j1}^2-3k_{j2}^2),~~ \nu_j=4(3k_{j1}^2-k_{j2}^2).
\end{equation*}

Obviously, by virtue of \eqref{phi-R-C},
$\phi^{\hbox{\it\tiny R}}_{\hbox{\tiny 2\it M}}[\lambda_1,\lambda_2,\cdots,\lambda_{M}]$ and
$\phi^{\hbox{\it\tiny C}}_{\hbox{\tiny 2\it M}}[\lambda_1,\lambda_2,\cdots,\lambda_{M}]$
lead to  same solutions to the KdV equation.

{\it Case 7} ~
\begin{equation}
\Gamma= J^{\hbox{\it\tiny C}}_{2M}[\lambda_1]=\biggl(\begin{matrix}\Lambda & 0\\
                                 0  & \Lambda^* \end{matrix}\biggr),
~~\Lambda
=\left(\begin{array}{ccccc}
-k^2_1 & 0    &  \cdots & 0   & 0 \\
1   & -k^2_1  &  \cdots & 0   & 0 \\
\cdots &\cdots &\cdots &\cdots  &\cdots \\
0   & 0    &  \cdots & 1   & -k^2_1
\end{array}\right)_{2M},
\label{(2.56)}
\end{equation}
i.e.,  $A$ has $M=N/2$ same complex conjugate eigenvalue pairs, where, still, $-k^2_1=\lambda_1$.

{\it a).~~ General solution to equations \eqref{cond-KdV-a} and \eqref{cond-KdV-b} with \eqref{(2.56)} }

Similar to  Case 3, the general solution
to the condition equations \eqref{cond-KdV-a} and \eqref{cond-KdV-b} in this case can be expressed as
\begin{equation}
\phi
=\Biggl(\begin{array}{cc}
\mathcal{A}& 0\\
0& \mathcal{A}^*
\end{array}\Biggr)\Biggl(
\begin{array}{c}
\mathcal{Q}^{+}_{0}\\
{\mathcal{Q}^{+}_{0}}^*
\end{array}\Biggr)
+\Biggl(\begin{array}{cc}
\mathcal{B}& 0\\
0& \mathcal{B}^*
\end{array}\Biggr)\Biggl(
\begin{array}{c}
\mathcal{Q}^{-}_{0}\\
{\mathcal{Q}^{-}_{0}}^*
\end{array} \Biggr),
~~ \mathcal{A}, \mathcal{B} \in \widetilde{G}_M,
\label{gen-sol-KdV-c}
\end{equation}
where
\begin{equation}
\mathcal{Q}^{\pm}_0=(\mathcal{Q}^{\pm}_{0,0},\mathcal{Q}^{\pm}_{0,1},\cdots,\mathcal{Q}^{\pm}_{0,M-1})^T,
~~
\mathcal{Q}^{\pm}_{0,j}=\frac{(-1)^j}{j!}\partial^{j}_{\lambda_1} b^{\pm}_1e^{\pm \xi_1}.
\end{equation}
The Wronskian $f(\phi)=|\widehat{N-1}|$ with \eqref{gen-sol-KdV-c} always generates a real solution to the KdV equation,
and the effective form of \eqref{gen-sol-KdV-c} can be taken as
\begin{equation}
\phi
=\Biggl(\begin{array}{cc}
\mathcal{A}& 0\\
0& \mathcal{A}^*
\end{array}\Biggr)\Biggl(
\begin{array}{c}
\mathcal{Q}^{+}_{0}\\
{\mathcal{Q}^{+}_{0}}^*
\end{array}\Biggr)
+\Biggl(
\begin{array}{c}
\mathcal{Q}^{-}_{0}\\
{\mathcal{Q}^{-}_{0}}^*
\end{array} \Biggr),
~~ \mathcal{A} \in \widetilde{G}_M.
\end{equation}

We note that $\mathcal{A}^*$ and $\mathcal{B}^*$ in \eqref{gen-sol-KdV-c} can be substituted by arbitrary matrices
$\mathcal{C}$ and $\mathcal{D}$ in $\widetilde{G}_M$, but such a $\phi$ does not
guarantee to generate a real solution to the KdV equation.

{\it b).~~ Alternative expressions of \eqref{gen-sol-KdV-c}}

If we substitute $J^{\hbox{\it\tiny C}}_{2M}[\lambda_1]$ by
\begin{equation}
\Gamma=J^{\hbox{\it\tiny C}}_{2M}[\Lambda_1]
=\left(\begin{array}{ccccc}
\Lambda_1 & 0    &  \cdots & 0   & 0 \\
I_1   & \Lambda_1  &  \cdots & 0   & 0 \\
\cdots &\cdots &\cdots &\cdots  &\cdots \\
0   & 0    &  \cdots & I_1   & \Lambda_1
\end{array}\right)_{2M},
\label{(2.57a)}
\end{equation}
which is similar to $J^{\hbox{\it\tiny C}}_{2M}[\lambda_1]$
where
\begin{equation}
I_1 = \biggl(\begin{matrix}1 & 0\\
                                 0  & 1 \end{matrix}\biggr), ~~
\Lambda_1\equiv \biggl(\begin{matrix}-k^2_1 & 0\\
                                 0 & -k^{*2}_1 \end{matrix}\biggr),
\label{(2.57b)}
\end{equation}
then the general solution to the condition equations \eqref{cond-KdV-a} and \eqref{cond-KdV-b} can be taken as
\begin{equation}
\phi^{J_C}_{\hbox{\tiny 2\it M}}[\lambda_1]
=\mathcal{A}^B\widetilde{\mathcal{Q}}^{+}_{0}+
\mathcal{B}^B \widetilde{\mathcal{Q}}^{-}_{0},
\label{gen-sol-KdV-c-7}
\end{equation}
where
\begin{equation}
\widetilde{\mathcal{Q}}^{\pm}_0=(\mathcal{Q}^{\pm}_{0,0},{\mathcal{Q}^{\pm *}_{0,0}},
\mathcal{Q}^{\pm}_{0,1},{\mathcal{Q}^{\pm *}_{0,1}},\cdots,\mathcal{Q}^{\pm}_{0,M-1},
{\mathcal{Q}^{\pm *}_{0,M-1}})^T,
~~
\mathcal{Q}^{\pm}_{0,j}=\frac{(-1)^j}{j!}\partial^{j}_{\lambda_1} b^{\pm}_1e^{\pm \xi_1},
\label{Q-tilde}
\end{equation}
$\mathcal{A}^B$ and $\mathcal{B}^B$, block lower triangular Toeplitz matrices
defined as the form \eqref{block-A-block}, can be arbitrary elements
in semigroup $\widetilde{G}^{B}_{2M}$ defined by \eqref{B-G-tilde}.
However, in order to get real solutions to the KdV equation, we always take $A_j$ and $B_j$
in $\mathcal{A}^B$ and $\mathcal{B}^B$  as
\begin{equation}
A_j = \biggl(\begin{matrix} a_{j1} & 0\\
                                 0  & a^*_{j1} \end{matrix}\biggr),~~~
B_j = \biggl(\begin{matrix} b_{j1} & 0\\
                                 0  & b^*_{j1} \end{matrix}\biggr).
\end{equation}
Similarly, the effective form of \eqref{gen-sol-KdV-c-7} is
\begin{equation}
\phi=\mathcal{A}^B\widetilde{\mathcal{Q}}^{+}_{0}+
\widetilde{\mathcal{Q}}^{-}_{0},
~~\mathcal{A}^B \in \widetilde{G}^{B}_{2M}.
\end{equation}

Besides $J^{\hbox{\it\tiny C}}_{2M}[\Lambda_1]$, we can also consider $\Gamma$ given as
\begin{equation}
\Gamma=\widetilde{J}^{\hbox{\it\tiny C}}_{2M}[\widetilde{\Lambda}_1]
=\left(\begin{array}{ccccc}
\widetilde{\Lambda}_1 & 0    &  \cdots & 0   & 0 \\
I_1   & \widetilde{\Lambda}_1  &  \cdots & 0   & 0 \\
\cdots &\cdots &\cdots &\cdots  &\cdots \\
0   & 0    &  \cdots & I_1   & \widetilde{\Lambda}_1
\end{array}\right)_{2M},
\end{equation}
where $\widetilde{\Lambda}_1$ is defined by \eqref{(2.48)}, i.e.,
\begin{equation}
\widetilde{\Lambda}_1 \equiv \biggl(\begin{matrix}\lambda_{11} & -\lambda_{12}\\
                                 \lambda_{12} & \lambda_{11} \end{matrix}\biggr).
\end{equation}
By noting that
\begin{equation}
\widetilde{J}^{\hbox{\it\tiny C}}_{2M}[\widetilde{\Lambda}_1]=
U^{-1}J^{\hbox{\it\tiny C}}_{2M}[\Lambda_1]U
\end{equation}
where $U$ is given by \eqref{(2.50)},
the solution to \eqref{cond-KdV-a} and \eqref{cond-KdV-b} with
$\widetilde{J}^{\hbox{\it\tiny C}}_{2M}[\widetilde{\Lambda}_1]$
can easily be taken as
\begin{equation}
\phi^{J_R}_{\hbox{\tiny 2\it M}}[\lambda_1]
=U^{-1}\phi^{J_C}_{\hbox{\tiny 2\it M}}[\lambda_1].
\end{equation}

To give the explicit form of $\phi^{J_R}_{\hbox{\tiny 2\it M}}[\lambda_1]$,
we first define
\begin{equation}
{\rm Diag}_{2M}[P]={\rm Diag}\Bigl(I_1, \frac{(-1)^1}{1!}P,
\frac{(-1)^2}{2!}P^2,\cdots, \frac{(-1)^{M-1}}{(M-1)!}P^{M-1}\Bigr),
\label{diag-P}
\end{equation}
where $P$ is a $2\times 2$ matrix.
Then, rewrite $\widetilde{\mathcal{Q}}^{\pm}_0$ as
\begin{equation}
\widetilde{\mathcal{Q}}^{\pm}_0
={\rm Diag}_{2M}[{\mathcal{C}}_{1,\lambda_1}]\Phi^{\pm}_{\hbox{\tiny C}},
\label{(2.59a)}
\end{equation}
where
\begin{equation}
{\mathcal{C}}_{1,\lambda_1}=\biggl(\begin{matrix} \partial_{\lambda_1} & 0\\
                                 0 & \partial_{\lambda^*_1} \end{matrix}\biggr),
\label{(2.59b)}
\end{equation}
and $\Phi^{\pm}_{\hbox{\tiny C}}$ is the following $2M$-order column vector
\begin{equation}
\Phi^{\pm}_{\hbox{\tiny C}}=(\mathcal{Q}^{\pm}_{0,0},{\mathcal{Q}^{\pm *}_{0,0}},
\mathcal{Q}^{\pm}_{0,0},{\mathcal{Q}^{\pm *}_{0,0}},\cdots,
\mathcal{Q}^{\pm}_{0,0},{\mathcal{Q}^{\pm *}_{0,0}})^T.
\label{(2.59c)}
\end{equation}
We further separate real and imaginary parts of $\mathcal{Q}^{\pm}_{0,0}$ as
$\mathcal{Q}^{\pm}_{0,0}=\mathcal{Q}^{R\pm}_{0,0}+i \mathcal{Q}^{I\pm}_{0,0}$, i.e.,
\begin{equation*}
\biggl(\begin{array}{c}
\mathcal{Q}^{R\pm}_{0,0}\\
\mathcal{Q}^{I\pm}_{0,0}
\end{array}\biggr)
=U^{-1}_1 \biggl(\begin{array}{c}
\mathcal{Q}^{\pm}_{0,0}\\
\mathcal{Q}^{\pm*}_{0,0}
\end{array}\biggr),
\end{equation*}
where from \eqref{entry-Case6-R-a} and \eqref{entry-Case6-R-b} we find
\begin{equation*}
\begin{array}{rl}
\mathcal{Q}^{R\pm}_{0,0}= & \Big\{ b^{\pm}_{11}\cos [k_{12}(x-\nu_1 t)+\xi^{(0)}_{12}] \mp
b^{\pm}_{12} \sin [k_{12}(x-\nu_1 t)+\xi^{(0)}_{12}]\Bigr\}
e^{\pm [k_{11}(x-\mu_1 t)+\xi^{(0)}_{11}]},\\
\mathcal{Q}^{I\pm}_{0,0} = & \Big\{b^{\pm}_{12} \cos [k_{12}(x-\nu_1 t)+\xi^{(0)}_{12}]\pm
b^{\pm}_{11} \sin [k_{12}(x-\nu_1 t)+\xi^{(0)}_{12}]\Bigr\}e^{\pm [k_{11}(x-\mu_1 t)+\xi^{(0)}_{11}]},
\end{array}
\end{equation*}
and here $k_{1j}$, $\mu_1$ and $\nu_1$ are defined as in Case 6.
Then, taking
\begin{equation}
\mathcal{R}_{1,\lambda_1}=U_1^{-1}{\mathcal{C}}_{1,\lambda_1} U_1,
\label{R-C}
\end{equation}
\begin{equation}
\Phi^{\pm}_{\hbox{\tiny R}}=U^{-1}\Phi^{\pm}_{\hbox{\tiny C}}
=(\mathcal{Q}^{R\pm}_{0,0},{\mathcal{Q}^{I\pm }_{0,0}},
\mathcal{Q}^{R\pm}_{0,0},{\mathcal{Q}^{I\pm }_{0,0}},\cdots,
\mathcal{Q}^{R\pm}_{0,0},{\mathcal{Q}^{I\pm }_{0,0}})^T,
\label{Phi-R}
\end{equation}
we have
\begin{equation}
\bar{\mathcal{Q}}^{\pm}_0
=U^{-1}\widetilde{\mathcal{Q}}^{\pm}_0
=U^{-1}{\rm Diag}_{2M}[{\mathcal{C}}_{1,\lambda_1}]UU^{-1}\Phi^{\pm}_{\hbox{\tiny C}}
={\rm Diag}_{2M}[{\mathcal{R}}_{1,\lambda_1}]\Phi^{\pm}_{\hbox{\tiny R}}.
\label{entry-Case7-R-a}
\end{equation}
Noting that
$[\mathcal{Q}^{\pm}_{0,0}(\lambda_1)]^*\equiv \mathcal{Q}^{\pm}_{0,0}(\lambda_1^*)$,
we can replace ${\mathcal{C}}_{1,\lambda_1}$ in \eqref{(2.59a)} by
\begin{equation}
{\mathcal{C}}_{1,\lambda_1}=\biggl(\begin{matrix} \partial_{\lambda_{11}} & 0\\
                              0 & \partial_{\lambda_{11}} \end{matrix}\biggr),
\end{equation}
or
\begin{equation}
{\mathcal{C}}_{1,\lambda_1}=\biggl(\begin{matrix} -i \partial_{\lambda_{12}} & 0\\
                              0 & i \partial_{\lambda_{12}} \end{matrix}\biggr).
\end{equation}
Consequently, from \eqref{R-C} $\mathcal{R}_{1,\lambda_1}$ can be
\begin{equation}
\mathcal{R}_{1,\lambda_1}=
\mathcal{R}_{1,\lambda_{11}}\equiv \partial_{\lambda_{11}}I_1,
\end{equation}
or
\begin{equation}
\mathcal{R}_{1,\lambda_1}=
\sigma_1\mathcal{R}_{1,\lambda_{12}}\equiv\partial_{\lambda_{12}}\sigma_1, ~~
\mathcal{R}_{1,\lambda_{12}}\equiv \partial_{\lambda_{12}}I_1,~~
\sigma_1=\biggl(\begin{matrix} 0 & -1\\
                               1 & 0 \end{matrix}\biggr).
\end{equation}
Then we have the following result.

\begin{Proposition}
\label{Prop 3.13}
{\it
The following vectors
\begin{equation}
\phi^{\lambda_{11}}_{_{2M}}
={\rm Diag}_{2M}[{\mathcal{R}}_{1,\lambda_{11}}]
(\Phi^{+}_{\hbox{\tiny R}}+\Phi^{-}_{\hbox{\tiny R}}),
\end{equation}
\begin{equation}
\phi^{\lambda_{12}}_{_{2M}}
={\rm Diag}_{2M}[\sigma_1{\mathcal{R}}_{1,\lambda_{12}}]
(\Phi^{+}_{\hbox{\tiny R}}+\Phi^{-}_{\hbox{\tiny R}})
\end{equation}
and
\begin{equation}
\widetilde{\phi}^{\lambda_{12}}_{_{2M}}
={\rm Diag}_{2M}[{{\mathcal{R}}}_{1,\lambda_{12}}]
(\Phi^{+}_{\hbox{\tiny R}}+\Phi^{-}_{\hbox{\tiny R}}),
\end{equation}
as Wronskian entries,
lead to the same solutions to the KdV equation as $\phi=\mathcal{Q}^{+}_{0}+\mathcal{Q}^{-}_{0}$ does.
In addition, it is easy to find that the following Wronskian entry vector
\begin{equation}
\phi=\left((\phi^{\lambda_{11}}_{_{2M_1}})^T,
({\phi}^{\lambda_{12}}_{_{2M_2}})^T\right)^T
~~{\rm or}~~
\phi=\left((\phi^{\lambda_{11}}_{_{2M_1}})^T,
(\widetilde{\phi}^{\lambda_{12}}_{_{2M_2}})^T\right)^T,
~~M_1 +M_2=M
\end{equation}
always generates a zero solution to the KdV equation.}
\hfill $\Box$
\end{Proposition}

Obviously, the general solution to the condition equations \eqref{cond-KdV-a} and \eqref{cond-KdV-b}
with $\widetilde{J}^{\hbox{\it\tiny C}}_{2M}[\widetilde{\Lambda}_1]$
can be taken as
\begin{equation}
\phi^{J_R}_{\hbox{\tiny 2\it M}}[\lambda_1]
=\mathcal{A}^{\widetilde{B}} \bar{\mathcal{Q}}^{+}_{0} + \mathcal{B}^{\widetilde{B}}
\bar{\mathcal{Q}}^{-}_{0},
\label{entry-Case7-R-b}
\end{equation}
and the effective form of \eqref{entry-Case7-R-b} is
\begin{equation}
\phi
=\mathcal{A}^{\widetilde{B}}\bar{\mathcal{Q}}^{+}_{0} +
\bar{\mathcal{Q}}^{-}_{0},
\label{entry-Case7-R-c}
\end{equation}
where $\bar{\mathcal{Q}}^{\pm}_{0}$ is given as \eqref{entry-Case7-R-a},
$\mathcal{A}^{\widetilde{B}}=U^{-1}\mathcal{A}^{B}U$ and $\mathcal{B}^{\widetilde{B}}=U^{-1}\mathcal{B}^{B}U$
are arbitrary elements in
$\widetilde{G}_{2M}^{\widetilde{B}}$.

{\it c).~~ Relations of solutions in Case 6 and Case 7}

In what follows we discuss  relations between the Wronskians generated by
\eqref{entry-Case6-0} and \eqref{gen-sol-KdV-c}.

Consider the following Wronskian
\begin{equation}
f=\frac{W(\phi^{\hbox{\it\tiny C}}_1, \phi^{{\hbox{\it\tiny C}}*}_1,
\phi^{\hbox{\it\tiny C}}_2, \phi^{{\hbox{\it\tiny C}}*}_2,\cdots,
\phi^{\hbox{\it\tiny C}}_{\hbox{\it\tiny M}}, \phi^{{\hbox{\it\tiny C}}*}_{\hbox{\it\tiny M}})}
{\prod^{M}_{j=2} (\lambda_1-\lambda_j)^{j-1}(\lambda^*_1-\lambda^*_j)^{j-1}},
\end{equation}
where $\phi^{\hbox{\it\tiny C}}_1=\phi^{\hbox{\it\tiny C}}_1(\lambda_1,t,x)=b^{+}_1 e^{ \xi_1}+b^{-}_1 e^{- \xi_1}$,
$\xi_1=k_1 x -4k^3_1 t +\xi^{(0)}_1$, and
$\phi^{\hbox{\it\tiny C}}_j=\phi^{\hbox{\it\tiny C}}_1(\lambda_j,t,x)$ for $j=2,3,\cdots,M$.

Then, for $j=2,3,\cdots,M$, Taylor expanding $\phi_j$  at $\lambda_1$, i.e.,
\begin{equation*}
\phi^{\hbox{\it\tiny C}}_j(\lambda_j,t,x)
=\sum^{+\infty}_{s=0}\frac{1}{s!}
\partial^{s}_{\lambda_1}\phi^{\hbox{\it\tiny C}}_1(\lambda_1,t,x)(\lambda_j-\lambda_1)^s,
~~~j=2,3,\cdots,M,
\end{equation*}
and hence successively taking limit $\lambda_j \rightarrow \lambda_1$, $(j=2,3,\cdots,M)$,
we get
\begin{equation*}
f\rightarrow W(\mathcal{A}^B\widetilde{\mathcal{Q}}^{+}_{0}+
\mathcal{B}^B \widetilde{\mathcal{Q}}^{-}_{0}),
\end{equation*}
where $\widetilde{\mathcal{Q}}^{\pm}_{0}$ is defined as \eqref{Q-tilde},
and the arbitrary matrix $\mathcal{A}^B$ and $\mathcal{B}^B$ in $\widetilde{G}^{B }_{2M}$
can be obtained by selecting $b^{\pm}_1$ as $b^{\pm}_1(\lambda_1)$,
according to Proposition \ref{Prop 2.4}.

Thus, we have described the limit relation between solutions in Case 6 and Case 7.

{\it d).~~ General solutions to the condition equations \eqref{cond-KdV-a} and \eqref{cond-KdV-b}
with $\Gamma=\check{J}^{\hbox{\it\tiny C}}_{2M}[K_1]$ \eqref{J-KdV-7-3-a}
and $\widehat{J}^{\hbox{\it\tiny C}}_{2M}[\widehat{K}_1]$ \eqref{J-KdV-7-3-b}
}

If we replace $J^{\hbox{\it\tiny C}}_{2M}[\Lambda_1]$ by its following
similar form
\begin{equation}
\Gamma=\check{J}^{\hbox{\it\tiny C}}_{2M}[K_1]
=\left(\begin{array}{ccccccc}
-K^2_1 & 0   & 0  &  \cdots & 0   & 0 & 0 \\
-2K_1   & -K^2_1  & 0& \cdots & 0   & 0 & 0 \\
-I_1   & -2K_1  & -K^2_1 & \cdots & 0   & 0 & 0 \\
\cdots &\cdots &\cdots & \cdots &\cdots  &\cdots & \cdots\\
0   & 0  & 0  &  \cdots & -I_1 & -2K_1   & -K^2_1
\end{array}\right)_{2M},
\label{J-KdV-7-3-a}
\end{equation}
where
\begin{equation}
K_1=\biggl(\begin{matrix}k_1 & 0\\
                                 0  & k_1^* \end{matrix}\biggr),
\label{J-KdV-7-3-b}
\end{equation}
and $-K^2_1=\Lambda_1$, then we can get the following Wronskian entry vector
\begin{equation}
\check{\phi}^{J_C}_{\hbox{\tiny 2\it M}}[k_1]
=\mathcal{A}^B\check{\mathcal{Q}}^{+}_{0}+
\mathcal{B}^B \check{\mathcal{Q}}^{-}_{0},
\label{gen-sol-KdV-c-7-3}
\end{equation}
where
\begin{equation}
\check{\mathcal{Q}}^{\pm}_0=(\check{\mathcal{Q}}^{\pm}_{0,0},\check{\mathcal{Q}}^{\pm *}_{0,0},
\check{\mathcal{Q}}^{\pm}_{0,1},\check{\mathcal{Q}}^{\pm *}_{0,1},\cdots,
\check{\mathcal{Q}}^{\pm}_{0,M-1},\check{\mathcal{Q}}^{\pm *}_{0,M-1})^T,
~~
\check{\mathcal{Q}}^{\pm}_{0,j}=\frac{1}{j!}\partial^{j}_{k_1} b^{\pm}_1e^{\pm \xi_1},
\label{Q-check}
\end{equation}
$\mathcal{A}^B$ and $\mathcal{B}^B$ are arbitrary elements
in semigroup $\widetilde{G}^{B}_{2M}$ defined by \eqref{B-G-tilde},
and in order to get real solutions to the KdV equation, we always take entries $A_j$ and $B_j$
in $\mathcal{A}^B$ and $\mathcal{B}^B$  as
\begin{equation}
A_j = \biggl(\begin{matrix} a_{j1} & 0\\
                                 0  & a^*_{j1} \end{matrix}\biggr),~~~
B_j = \biggl(\begin{matrix} b_{j1} & 0\\
                                 0  & b^*_{j1} \end{matrix}\biggr).
\end{equation}
The effective form of \eqref{gen-sol-KdV-c-7-3} is
\begin{equation}
\check{\phi}=\mathcal{A}^B\check{\mathcal{Q}}^{+}_{0}+
\check{\mathcal{Q}}^{-}_{0},
~~\mathcal{A}^B \in \widetilde{G}^{B}_{2M}.
\end{equation}

Obviously, to calculate  $\check{\mathcal{Q}}^{\pm}_{0,j}$ is much easier than
$\widetilde{\mathcal{Q}}^{\pm}_{0}$, and they  lead to same solutions to the KdV equation.

Now we rewrite $\check{\mathcal{Q}}^{\pm}_0$ as
\begin{equation}
\check{\mathcal{Q}}^{\pm}_0
={\rm Diag}_{2M}[{\mathcal{C}}_{1,k_1}]\Phi^{\pm}_{\hbox{\tiny C}},
\end{equation}
where
\begin{equation}
{\mathcal{C}}_{1,k_1}=\biggl(\begin{matrix} \partial_{k_1} & 0\\
                                 0 & \partial_{k^*_1} \end{matrix}\biggr),
\end{equation}
and $\Phi^{\pm}_{\hbox{\tiny C}}$ is \eqref{(2.59c)}.
Then it is easy to find that $\check{\mathcal{Q}}^{\pm}_0$ can equivalently be expressed as
\begin{equation}
\check{\mathcal{Q}}^{\pm}_0
={\rm Diag}_{2M}[\biggl(\begin{matrix} \partial_{k_{11}} & 0\\
                                 0 & \partial_{k_{11}} \end{matrix}\biggr)]\Phi^{\pm}_{\hbox{\tiny C}}
\label{C-1}
\end{equation}
and
\begin{equation}
\check{\mathcal{Q}}^{\pm}_0
={\rm Diag}_{2M}[\biggl(\begin{matrix} -i \partial_{k_{12}} & 0\\
                                 0 & i \partial_{k_{12}} \end{matrix}\biggr)]\Phi^{\pm}_{\hbox{\tiny C}}.
\label{C-2}
\end{equation}
That means the following proposition holds.

\begin{Proposition}
\label{Prop 3.14}
{\it
As a Wronskian entry vector, ${\rm Diag}_{2M}[P](\Phi^{+}_{\hbox{\tiny C}}+\Phi^{-}_{\hbox{\tiny C}})$
generates same solutions to the KdV equation when $P$ is taken as ${\mathcal{C}}_{1,k_1}$,
$\partial_{k_{11}}I_1$ or $\partial_{k_{12}}I_1$.}
\hfill $\Box$
\end{Proposition}

We can also consider the real version of $\check{J}^{\hbox{\it\tiny C}}_{2M}[K_1]$, i.e.,
\begin{equation}
\Gamma=\widehat{J}^{\hbox{\it\tiny C}}_{2M}[\widehat{K}_1]
=U\check{J}^{\hbox{\it\tiny C}}_{2M}[K_1]U^{-1}
=\left(\begin{array}{ccccccc}
-\widehat{K}^2_1 & 0   & 0  &  \cdots & 0   & 0 & 0 \\
-2\widehat{K}_1   & -\widehat{K}^2_1  & 0& \cdots & 0   & 0 & 0 \\
-I_1   & -2\widehat{K}_1  & -\widehat{K}^2_1 & \cdots & 0   & 0 & 0 \\
\cdots &\cdots &\cdots & \cdots &\cdots  &\cdots & \cdots\\
0   & 0  & 0  &  \cdots & -I_1 & -2\widehat{K}_1   & -\widehat{K}^2_1
\end{array}\right)_{2M},
\label{J-KdV-7-4-a}
\end{equation}
where $U$ is defined by \eqref{(2.50)}, and
\begin{equation}
\widehat{K}_1\! =U^{-1}_1K_1U^{-1}_1 \!=\! \biggl(\begin{matrix}k_{11} & -k_{12}\\
                                 k_{12}  & k_{11} \end{matrix}\biggr),~~
-\widehat{K}_1^2\! =\widetilde{\Lambda}_1\equiv \biggl(\begin{matrix}\lambda_{11} & -\lambda_{12}\\
                                 \lambda_{12} & \lambda_{11} \end{matrix}\biggr)
                                 =\biggl(\begin{matrix}
                                   -(k^2_{11}-k^2_{12}) & 2k_{11}k_{12}\\
                                 -2k_{11}k_{12} & -(k^2_{11}-k^2_{12})
                                 \end{matrix}\biggr).
\label{J-KdV-7-4-b}
\end{equation}
In this case, the general solution to the condition
equations \eqref{cond-KdV-a} and \eqref{cond-KdV-b} can be taken as
\begin{equation}
\widehat{\phi}^{J_R}_{\hbox{\tiny 2\it M}}[k_1]
=U^{-1}\check{\phi}^{J_C}_{\hbox{\tiny 2\it M}}[k_1]
=\mathcal{A}^{\widetilde{B}}\widehat{\mathcal{Q}}^{+}_{0}+
\mathcal{B}^{\widetilde{B}} \widehat{\mathcal{Q}}^{-}_{0},
\label{gen-sol-KdV-c-7-4}
\end{equation}
$\mathcal{A}^{\widetilde{B}}=U^{-1}\mathcal{A}^{B}U$ and $\mathcal{B}^{\widetilde{B}}=U^{-1}\mathcal{B}^{B}U$
are arbitrary elements in
$\widetilde{G}_{2M}^{\widetilde{B}}$,
\begin{equation}
\widehat{\mathcal{Q}}^{\pm}_{0}=U^{-1}\check{\mathcal{Q}}^{\pm}_{0}
={\rm Diag}_{2M}[U_1^{-1}{\mathcal{C}}_{1,k_1}U_1]\Phi^{\pm}_{\hbox{\tiny R}},
\end{equation}
and $\Phi^{\pm}_{\hbox{\tiny R}}$ is \eqref{Phi-R}.
Noticing  \eqref{C-1} and \eqref{C-2}, we can take $\widehat{\mathcal{Q}}^{\pm}_{0}$
as
\begin{equation}
\widehat{\mathcal{Q}}^{\pm}_{0}
={\rm Diag}_{2M}[\mathcal{R}_{1,k_{11}}]\Phi^{\pm}_{\hbox{\tiny R}},~~~\mathcal{R}_{1,k_{11}}=\partial_{k_{11}}I_1
\end{equation}
or
\begin{equation}
\widehat{\mathcal{Q}}^{\pm}_{0}
={\rm Diag}_{2M}[\sigma_1 \mathcal{R}_{1,k_{12}}]\Phi^{\pm}_{\hbox{\tiny R}},
~~~\mathcal{R}_{1,k_{12}}=\partial_{k_{12}}I_1.
\end{equation}
Consequently, we have the following result similar to Proposition \ref{Prop 3.13}.

\begin{Proposition}
\label{Prop 3.15}
{\it
As a Wronskian entry vector, ${\rm Diag}_{2M}[P](\Phi^{+}_{\hbox{\tiny R}}+\Phi^{-}_{\hbox{\tiny R}})$
generates same solutions to the KdV equation no matter  $P$ is taken as
$\mathcal{R}_{1,k_{11}}$, $\mathcal{R}_{1,k_{12}}$ or $\sigma_1 \mathcal{R}_{1,k_{12}}$.
In addition, the following vector
\begin{equation}
\phi={\rm Diag}({\rm Diag}_{2M_1}[\mathcal{R}_{1,k_{11}}],
{\rm Diag}_{2M_2}[\mathcal{R}_{1,k_{12}}])(\Phi^{+}_{\hbox{\tiny R}}+\Phi^{-}_{\hbox{\tiny R}}),
~~M_1 +M_2=M
\end{equation}
always generates a zero solution to the KdV equation.}
\hfill $\Box$
\end{Proposition}

The effective form of \eqref{gen-sol-KdV-c-7-4} is
\begin{equation}
\widehat{\phi}^{J_R}_{\hbox{\tiny 2\it M}}[k_1]
=\mathcal{A}^{\widetilde{B}}\widehat{\mathcal{Q}}^{+}_{0}+
\widehat{\mathcal{Q}}^{-}_{0},
~~\mathcal{A}^{\widetilde{B}}\in \widetilde{G}_{2M}^{\widetilde{B}},
\end{equation}
and \eqref{gen-sol-KdV-c-7-4} can also be obtained through a limit procedure.

\vskip 10pt
We end this section by the following remarks.

{\it Remark 3.1} ~
An arbitrary Wronskian entry vector can be composed by arbitrarily picking up entries from
the above 7 cases. For example,
\begin{equation}
\phi_{mix}=\left((\phi^{0}_{\rho_1})^T,
(\phi^{-}_{\rho_2}[\lambda_{h_1},\cdots, \lambda_{h_{\rho_2}}])^T,
(\phi^{+}_{\rho_3}[\lambda_{l_1},\cdots, \lambda_{l_{\rho_3}}])^T,
(\phi^{J^{-}}_{\rho_4}[\lambda_{g_1}])^T ,(\phi^{J^{-}}_{\rho_5}[\lambda_{g_2}])^T,
\right)^T,
\label{mix-KdV}
\end{equation}
where $\sum^{5}_{s=1}\rho_s=N$,
$\phi^{0}_{\rho_1}$,
$\phi^{-}_{\rho_2}[\lambda_{h_1},\cdots, \lambda_{h_{\rho_2}}]$,
$\phi^{+}_{\rho_3}[\lambda_{l_1},\cdots, \lambda_{l_{\rho_3}}]$
and $\phi^{J^{-}}_{\rho_j}[\lambda_{g_s}]$
are respectively defined as \eqref{gen-sol-KdV-5}, \eqref{entry-I}, \eqref{entry-II} and \eqref{gen-sol-KdV}.
In order to get a nonzero solution to the KdV equation, we always let all the $\{\lambda_j\}$ be distinct.
\eqref{mix-KdV} is a solution of the condition equations \eqref{cond-KdV-a} and \eqref{cond-KdV-b} when
$\Gamma$ is the following block-diagonal matrix
\begin{equation}
D_{mix}={\rm Diag} (J^{0}_{\rho_1},
D^{-}_{\rho_2}[\lambda_{h_1},\cdots, \lambda_{h_{\rho_2}}],
D^{+}_{\rho_3}[\lambda_{l_1},\cdots, \lambda_{l_{\rho_3}}],
J^{-}_{\rho_4}[\lambda_{g_1}], J^{-}_{\rho_5}[\lambda_{g_2}]).
\label{D-mix}
\end{equation}
This case corresponds to $A$ in \eqref{cond-a'} having the following eigenvalues
\begin{equation}
\lambda_{h_1},\cdots, \lambda_{h_{\rho_2}},
\lambda_{l_1},\cdots, \lambda_{l_{\rho_3}},
\overbrace{\lambda_{g_1},\cdots,\lambda_{g_1}}^{\rho_4},
\overbrace{\lambda_{g_2},\cdots,\lambda_{g_2}}^{\rho_5},
\overbrace{0,\cdots,0}^{\rho_1},
\label{eigenvalue-mix}
\end{equation}
where $\sum^{5}_{s=1}\rho_s=N$.
In this case, we get the so-called mixed solutions.

{\it Remark 3.2} ~ We have given explicit forms for the general solutions to the condition
equations \eqref{cond-KdV-a} and \eqref{cond-KdV-b} when
$\Gamma$ is taken each carnonical form of $A$ in \eqref{cond-a'}, as we have discussed in the above
7 cases. In Case 3,4 and 7, for some Jordan-block solutions, we listed alternative Wronskian entry vectors which
maybe easy to calculate. Particularly, in Proposition \ref{Prop 3.13} and \ref{Prop 3.15}, we pointed out
the Wronskian entry vectors which only generate zero solutions.
Besides, we have given effective forms for Jordan-block solutions
in Case 3, 4, 5 and 7. This will be helpful when we investigate parameter effects of solutions.
We also explained the limit relations between Wronskians corresponding to Jordan block
and diagonal matrices.

{\it Remark 3.3} ~ To give explicit general solutions to the Wronskian condition equations
for Jordan-block solutions, we have made use of some algebraic properties of the lower
triangular Toeplitz matrices, i.e., the matrices
commuting with a Jordan block, which list in Sec.2. These properties
enable us to easily generate a general solution to the Wronskian condition equations
from a set of special solutions and further give the effective form of the  general solution.
Besides, Proposition \ref{Prop 2.4} and \ref{Prop 2.5}, taking Case 3 as an example,
make connections between the matrices $\mathcal{A},\mathcal{B}\in \widetilde{G}_N$ and
the coefficients $b^{\pm}_1$ or $\xi^{(0)}_1$
when $b^{\pm}_1$ and $\xi^{(0)}_1$ are considered as some certain functions
of $\lambda_1$.

{\it Remark 3.4} ~ In this section, we have answered a question on
ordinary differential equations, i.e.,
how to obtained all possible existing solutions to the equations \eqref{cond-a'} and \eqref{cond-b'}
for any $N\times N$ constant matrix $A$. These solutions are in explicit form and
can be given according to the eigenvalues of $A$.
For example, let us see \eqref{mix-KdV}, \eqref{D-mix} and \eqref{eigenvalue-mix}.
If $A=T D_{mix}T^{-1}$ has eigenvalues given as \eqref{eigenvalue-mix},
then, based on Proposition \ref{Prop 3.8} and \ref{Prop 3.10}, the general solutions to
the equations \eqref{cond-a'} and \eqref{cond-b'} with such an $A$ is
$\phi=T\phi_{mix}$.

{\it Remark 3.5} ~ Solutions derived from {Case 1} and {Case 3} are called
{\it negatons}\cite{negaton-positon-96} of the KdV equation
due to the fact that $\phi^{-}_j$ in (\ref{entry-Case1}, \ref{entry-Case1-1})
is a solution to the stationary Schr\"{o}dinger equation \eqref{Lax-KdV-a}
with negative eigenvalues (and zero potential).
Similarly, solutions derived from {Case 2} and {Case 4} are called
{\it positons}\cite{Matveev-positon-kdv1,Matveev-positon-kdv2}.
Dynamics for some special positons and negatons of the KdV equation
have been discussed in \cite{Matveev-positon-kdv1,Matveev-positon-kdv2} and \cite{negaton-positon-96}.
Rational solutions in {Case 5} correspond to the case that the stationary
Schr\"{o}dinger equation \eqref{Lax-KdV-a}
has zero eigenvalues.

{\it Remark 3.6} ~ {\it Complexitons}\cite{Ma-complexiton} derived from {\it Case 6}
correspond to the case that the stationary Schr\"{o}dinger equation \eqref{Lax-KdV-a}
has complex eigenvalues,
and can also be called breathers of the KdV equation\cite{breather-KdV}.
They are real but have singularities due to the trigonometric functions involved.
The physical meanings of complexitons are still left to discuss.
Besides, the Wronskian $f(\phi^{\hbox{\it\tiny C}}_{\hbox{\tiny 2\it M}}[\lambda_1,\lambda_2,\cdots,\lambda_{M}])$
can also be expressed into Hirota's form \eqref{Hirota-KdV}.
That means complexitons in {\it Case 6} can also be obtained through the Hirota method by considering
$k_{2j-1}$ and $k_{2j}$ in
\eqref{Hirota-KdV} to be complex conjugate pairs,
as the well-known breathers of the mKdV equation.

{\it Remark 3.7} ~ An interesting question is the reality condition of square matrix $A$ in \eqref{cond-a'},
i.e., what conditions should be satisfied by $A$ so that it generates real solutions to the
KdV equation. We have shown that a real matrix $A$ can always generate a real solution
because in this case the eigenvalues of $A$ are either real or in conjugate pairs.
Now let $A$ be  complex  and we consider it
over the complex field $\mathbb{C}$. Hence we can always treat $A$ as a triangular matrix
because over the complex field
$\mathbb{C}$ any square matrix is similar to a triangular one.
In this case, $\{A_{jj}\}$ provide eigenvalues of $A$.
Then, besides complex eigenvalues appearing  in conjugate pairs,
what are the conditions for $A$ to generate real solutions to the KdV equation?

\vskip 16pt

\section {\normalsize Solutions in Casoratian form to the Toda lattice}

In this section,
we employ the Toda lattice to serve as a differential-difference example.
In fact, in terms of the relationship between Casoratian entry vectors and eigenvalues
of the coefficient matrix in the condition equations,
the Toda lattice has quite similar results to the KdV equation.

The Toda lattice is\cite{Toda lattice}
\begin{equation}
x_{n,tt}=e^{x_{n-1}-x_{n}}-e^{x_{n}-x_{n+1}},
\label{Toda}
\end{equation}
where $x_n\equiv x(n,t)$ is a function defined on $(\mathbb{Z},\mathbb{R})$.
By introducing $u_{n}=e^{x_{n-1}-x_{n}}$ and $v_{n}=-x_{n,t}$,
\eqref{Toda} reads
\begin{equation}
\begin{array}{l} u_{n,t}=u_{n}(v_{n}-v_{n-1}), \\
                   v_{n,t}=u_{n+1}-u_{n},
  \end{array}
\label{(3.2)}
\end{equation}
which is Lax integrable with the Lax pair
\begin{subequations}
\begin{eqnarray}
\Phi_{n+1}+u_{n}\Phi_{n-1}+v_{n}\Phi_{n}=\lambda \Phi_{n},
\label{(3.3a)}\\
\Phi_{n,t}=\Phi_{n}-u_{n}\Phi_{n-1}.
\label{(3.3b)}
\end{eqnarray}
\end{subequations}
Another form of the Toda lattice \eqref{Toda} is given as
\begin{equation}
[\ln(1+V_n)]_{tt}=V_{n+1}-2V_n +V_{n-1}
\label{(3.4)}
\end{equation}
where
\begin{equation}
V_n = e^{-y_{n}}-1, ~~~y_{n}=x_{n}-x_{n-1},
\end{equation}
Hirota\cite{Hirota-Toda lattice, Ablowitz-book-1981} gave the bilinear form of  \eqref{(3.4)} as
\begin{equation}
(D^2_t - 2\cosh D_n +1)f_n \cdot f_n=0,
\label{(3.6)}
\end{equation}
through the transformation
\begin{equation}
V_n=(\ln f_{n})_{tt},
\label{(3.7)}
\end{equation}
where $e^{\varepsilon D_n}$ is defined as\cite{Hirota-book}
\begin{equation}
e^{\varepsilon D_z}a(z)\cdot b(z)=e^{\varepsilon \partial_y}a(z+y) b(z-y)|_{y=0}=
a(z+\varepsilon) b(z-\varepsilon),
\label{(3.8)}
\end{equation}
no matter $z$ is continuous or discrete, where $\varepsilon$ is a parameter.

A very interesting fact is that if we choose a new transformation,
\begin{equation}
V_n=\frac{f_{n+1}f_{n-1}}{f^2_{n}}-1,
\label{trans-Toda}
\end{equation}
we can still reduce the Toda lattice \eqref{(3.4)} to the bilinear form \eqref{(3.6)}.
In fact, substituting \eqref{trans-Toda} into \eqref{(3.4)} yields
\begin{equation}
(\ln f_{n+1})_{tt}-2(\ln f_{n})_{tt}+(\ln f_{n-1})_{tt}=
\frac{f_{n+2}f_{n}}{f^2_{n+1}}-\frac{2f_{n+1}f_{n-1}}{f^2_{n}}
+\frac{f_{n}f_{n-1}}{f^2_{n-1}},
\end{equation}
which further reduces to
\eqref{(3.6)}.

\subsection{Casoratian solution}

A Casoratian is a discrete version of a Wronskian defined as
\begin{equation}
\begin{array}{rl}
Cas(\phi_1(n,t), \phi_2(n,t),\cdots,\phi_N(n,t)) &
=\left | \begin{array}{llll}
                \phi_{1}(n,t) & \phi_{1}(n+1,t) & \cdots  & \phi_{1}(n+N-1,t)\\
                \vdots   & \vdots  &  ~ & \vdots\\
                \phi_{N}(n,t) & \phi_{N}(n+1,t) & \cdots  & \phi_{N}(n+N-1,t)
                \end{array}
        \right |\\
~ & =|\phi(n,t), \phi(n+1,t),\cdots,\phi(n+N-1,t)|=|\widehat{N-1}|,
\end{array}
\label{(3.9)}
\end{equation}
where $\phi(n,t)=(\phi_1(n,t), \phi_2(n,t),\cdots,\phi_N(n,t))^T$.

Based on Proposition \ref{Prop 3.1} and \ref{Prop 3.2},
the Casoratian solution to the Toda lattice can be described as follows\cite{Zhang-Toda}.

\begin{Proposition}
\label{Prop 4.1}
{\it The following Casoratian $f_n$ solves the
bilinear Toda lattice \eqref{(3.6)}:
\begin{equation}
f_n={\rm Cas}(\phi(n,t))={\rm Cas}\left(\phi_1(n,t), \phi_2(n,t),\cdots, \phi_N(n,t)\right)
=|\widehat{N-1}|,
\label{(3.10)}
\end{equation}
where  $\phi(n,t)$ satisfies
\begin{equation}
\phi(n+1,t) +\phi(n-1,t)=A(t)\phi(n,t),
\label{(3.11a)}
\end{equation}
\begin{equation}
\pm\phi_{t}(n,t)=a\phi(n+1,t)  +c\phi(n-1,t)+ B(t) \phi(n,t),
\label{(3.11b)}
\end{equation}
$A(t)=(A_{ij})_{N\times N}$ and $B(t)=(B_{ij})_{N\times N}$ are two arbitrary $N\times N$  matrices of $t$ but
independent of $n$,
$B(t)$ satisfies
\begin{equation}
\Tr{B(t)}_t=0,
\label{B-cond-Toda}
\end{equation}
the constant pair $(a,c)$
is equal to $(\frac{1}{2},-\frac{1}{2})$ or $(1,0)$ or $(0,1)$.
Considering that \eqref{(3.11a)} and \eqref{(3.11b)} should be solvable, $A(t)$ and $B(t)$ must further satisfy
\begin{equation}
A(t)_t+[A(t),B(t)]=0.
\label{compat-Toda}
\end{equation}
} \hfill $\Box$
\end{Proposition}

In the proof for the above proposition, a key identity
\begin{equation*}
\Tr{A(t)}|\widehat{N-1}|
=|\widehat{N-2},N|+|-1,\widetilde{N-1}|,
\end{equation*}
which is the same as in the case that $A(t)$ is a diagonal\cite{Nimmo-Toda}
or triangular\cite{Zhang-Toda} constant matrix,
can be obtained
by taking $\Xi=|\W{N-1}|$ and $\Omega_{js}\equiv E+E^{-1}$ in Proposition \ref{Prop 3.1},
where $E$ is a shift operator defined as $E^jf(n)=f(n+j)$, and
 $\widetilde{N-j}$ indicates the set of consecutive columns $1,2,\cdots N-j$.
To get the condition \eqref{B-cond-Toda}, we have made use of Proposition \ref{Prop 3.2}.
\eqref{compat-Toda} comes from
the compatibility of \eqref{(3.11a)} and \eqref{(3.11b)}, as in Proposition \ref{Prop 3.5}.

The condition \eqref{B-cond-Toda} is necessary for the Casoratian solution \eqref{(3.10)} to the
bilinear Toda lattice \eqref{(3.6)}. However,
It is not necessary for the solution to the Toda lattice \eqref{(3.4)}.
We describe this fact in the following Proposition.

\begin{Proposition}
\label{Prop 4.2}
{\it The  Casoratian \eqref{(3.10)} provides
a solution to  the Toda lattice \eqref{(3.4)} through the transformation \eqref{trans-Toda},
if its entry vector $\phi(n,t)$ satisfies the equations \eqref{(3.11a)} and \eqref{(3.11b)},
where the constant pair $(a,c)$
equals to $(a,a\pm 1)$ in which $a$ is a real arbitrary number,
$A(t)$ and $B(t)$,
are two arbitrary $N\times N$  matrices of $t$ but
independent of $n$ and satisfy the compatible condition
\eqref{compat-Toda}.}
\end{Proposition}

{\it Proof:} ~ For any $n$-independent $N\times N$  matrix $B(t)$ in \eqref{(3.11b)}, if $\Tr{B(t)}\neq 0$, let
\begin{equation*}
B(t)=B_1(t) +B_2(t),
\end{equation*}
where
\begin{equation*}
B_2(t)=\frac{1}{N}\Tr{B(t)}I,~~~
B_1(t)=B(t)-\frac{1}{N}\Tr{B(t)}I,
\end{equation*}
and $I$ is the $N\times N$  unit matrix.
Defining a new vector $\psi(n,t)$ as
\begin{equation}
\psi(n,t)=e^{-\frac{1}{N}\int^{t}_{0}\Tr{B(t)}dt} \phi(n,t),
\label{psi-phi}
\end{equation}
it then follows from \eqref{(3.11a)} and \eqref{(3.11b)} that $\psi(n,t)$ satisfies
\begin{equation}
\psi(n+1,t) +\psi(n-1,t)=A(t)\psi(n,t),
\end{equation}
\begin{equation}
\pm\psi_{t}(n,t)=a\psi(n+1,t)  +c\psi(n-1,t)+ B_1(t) \psi(n,t).
\end{equation}
Due to $B_2(t)$ being a diagonal, \eqref{compat-Toda} suggests
\begin{equation*}
A_t(t)+[A(t),B_1(t)]=0.
\end{equation*}
Thus, in the light of Proposition \ref{Prop 4.1} and noticing $\Tr{B_1(t)}_t=0$,
we find the Casoratian ${\rm Cas}(\psi(n,t))$ solves the bilinear Toda lattice \eqref{(3.6)}.
Further, by noting that
\begin{equation*}
{\rm Cas}(\phi(n,t))=e^{\int^{t}_{0}\Tr{B(t)}dt}{\rm Cas}(\psi(n,t)),
\end{equation*}
${\rm Cas}(\phi(n,t))$ provides
a solution to  the Toda lattice \eqref{(3.4)} through the transformation \eqref{trans-Toda}.
Finally, with the arbitrariness of $B(t)$
in hand, we can substitute any $k$ times of \eqref{(3.11a)} into \eqref{(3.11b)},
and this provides the arbitrariness of $a$ in the pair $(a,a\pm 1)$.
Thus, the proof is completed.
\hfill $\Box$

In Ref.\cite{Zhang-Toda}, we listed some explicit choices of $\phi(n,t)$ meeting the conditions
\eqref{(3.11a)} and \eqref{(3.11b)},
and pointed out that some of them generate same solutions to the Toda lattice.
In this section, due to the arbitrariness of $B(t)$, we only need to discuss the case that $\phi(n,t)$  satisfies
\eqref{(3.11a)} and
\begin{equation*}
\phi_{t}(n,t)=\frac{1}{2}\phi(n+1,t) - \frac{1}{2} \phi(n-1,t)+ B(t) \phi(n,t).
\end{equation*}
By the similar discussions for the KdV equation in Sec.3,
the arbitrariness of $B(t)$ does not contribute new solutions to the Toda lattice through
the transformation \eqref{trans-Toda} in some sense; so we always take $B(t)=0$
in the following
and therefore $A(t)$ is independent of $t$. Thus we have
\begin{equation}
\phi(n+1,t) +\phi(n-1,t)=A \phi(n,t),
\label{cond-Toda-a'}
\end{equation}
\begin{equation}
2\phi_{t}(n,t)=\phi(n+1,t)  -\phi(n-1,t),
\label{cond-Toda-b'}
\end{equation}
where $A$ is an arbitrary $N\times N$ constant matrix.
Again based on the discussions in Sec.3, we can replace $A$ by its
any similar form $\Gamma$
and consequently consider the following Casoratian entry conditions
\begin{equation}
\phi(n+1,t) +\phi(n-1,t)=\Gamma \phi(n,t),
\label{cond-Toda-a}
\end{equation}
\begin{equation}
2\phi_{t}(n,t)=\phi(n+1,t)  -\phi(n-1,t).
\label{cond-Toda-b}
\end{equation}
\eqref{cond-Toda-a'} and \eqref{cond-Toda-b'}
or \eqref{cond-Toda-a} and \eqref{cond-Toda-b}
are called {\it Casoratian entry condition equations} of the Toda lattice.

\subsection{Solutions related to $\Gamma$}

In this subsection we skip over  detailed discussions and
directly list explicit expressions of all possible existing solutions to
the Casoratian entry condition equations \eqref{cond-Toda-a} and \eqref{cond-Toda-b}.
Relations between different kinds of solutions
will also be described.
We still use the same notations as in Sec.3.

{\it Case 1}
\begin{equation}
\Gamma=D^{-}_{N}[\lambda_1,\lambda_2,\cdots,\lambda_N]
={\rm Diag}(2\cosh k_1,2\cosh k_2,\cdots,2\cosh k_N),
\label{Toda-case1}
\end{equation}
where
\begin{equation}
\lambda_j=\omega_j+\frac{1}{\omega_j}=2\cosh k_j,~~~\omega_j=e^{k_j},
\end{equation}
each $|\lambda_j|>2$ and  all the $\{ k_j \}$ are arbitrary non-zero distinct real numbers.

In this case, $\phi\equiv\phi(n,t)$ in \eqref{cond-Toda-a} and \eqref{cond-Toda-b} is given as
\begin{equation}
\phi=\phi^{-}_{N}[\lambda_1,\lambda_2,\cdots,\lambda_N]
=(\phi^{-}_1,\phi^{-}_2,\cdots,\phi^{-}_N)^T,
\label{entry-Toda-I}
\end{equation}
in which
\begin{equation}
\phi^{-}_j=a^{+}_{j}\cosh {\xi_j} + a^{-}_{j}\sinh {\xi_j},
\label{entry-Toda-1a}
\end{equation}
or
\begin{equation}
\phi^{-}_j=b^{+}_{j}e^{\xi_j} + b^{-}_{j}e^{-\xi_j},
\label{entry-Toda-1b}
\end{equation}
where
\begin{equation}
\xi_j=k_j n +t \sinh k_j +\xi^{(0)}_j,~~j=1,2,\cdots N,
\label{xi-Toda}
\end{equation}
$a^{\pm}_{j}$, $b^{\pm}_{j}$ and $\xi^{(0)}_j$  are all real constants.

If we set $0<k_1 <k_2 < \cdots < k_N$ and take
$a^{\pm}_{j}=1\mp(-1)^{j}$, the corresponding Casoratian generates the normal $N$-soliton
solution and has the following Hirota's expression\cite{Zhang-Toda},
\begin{equation*}
f_n = 2^{-N} \biggl(\ds\prod^{N}_{1\leq j<l}
           \!2\sinh\ds\ds\frac{k_{l}-k_{j}}{2}\!\biggr)
       \exp\biggl\{\!\!-\ds\sum_{j=1}^{N}(\xi_{j} +\ds\frac{N-1}{2}k_j)\biggr\}
      \sum_{\mu=0,1}\!\exp \biggl\{\ds\sum_{j=1}^{N}2 \mu_{j}\eta_{j}+
           \sum^{N}_{1\leq j<l} \mu_{j}\mu_{l}A_{jl} \biggr\},
\end{equation*}
where the sum over $\mu=0,1$ refers to each of the $\mu_j=0,1$ for
$j=0,1,\cdots,N$,
and
\begin{equation*}
\eta_{j}=\xi_{j} +\ds\frac{N-1}{2}k_j -\frac{1}{4}\sum_{l=1,l\neq j}^{N}A_{jl},
~~~e^{A_{jl}}=\left(\frac{\sinh\frac{k_{l}-k_{j}}{2}}{\sinh\frac{k_{l}+k_{j}}{2}}\right)^2.
\end{equation*}

{\it Case 2}
\begin{equation}
\Gamma=D^{+}_{N}[\lambda_1,\lambda_2,\cdots,\lambda_N]
={\rm Diag}(2\cos k_1,2\cos k_2,\cdots,2\cos k_N),
\label{Toda-case2}
\end{equation}
where
\begin{equation}
\lambda_j=\omega_j+\frac{1}{\omega_j}=2\cos k_j,~~~\omega_j=e^{ik_j},
\end{equation}
each $|\lambda_j|<2$ and  all the $\{ k_j \}$ are arbitrary distinct real numbers
satisfying $\{k_j\neq s\pi,~ s\in \mathbb{Z}\}$.

The Casoratian entry  vector in this case is
\begin{equation}
\phi=\phi^{+}_{N}[\lambda_1,\lambda_2,\cdots,\lambda_N]
=(\phi^{+}_1,\phi^{+}_2,\cdots,\phi^{+}_N)^T,
\label{entry-Toda-II}
\end{equation}
in which
\begin{equation}
\phi^{+}_j=a^{+}_{j}\cos {\theta_j} + a^{-}_{j}\sin {\theta_j},
\label{entry-Toda-2a}
\end{equation}
or
\begin{equation}
\phi^{-}_j=b^{+}_{j}e^{i\theta_j} + b^{-}_{j}e^{-i\theta_j},
\label{entry-Toda-2b}
\end{equation}
where
\begin{equation}
\theta_j=k_j n + t \sin k_j+\theta^{(0)}_j,~~j=1,2,\cdots N,
\label{theta-Toda}
\end{equation}
$a^{\pm}_{j}$, $b^{\pm}_{j}$ and $\theta^{(0)}_j$  are all real constants.

%
%
%

{\it Case 3} ~
\begin{equation}
\Gamma=J^{-}_{N}[\lambda_1]
=\left(\begin{array}{cccccc}
2\cosh k_1 & 0    & 0   & \cdots & 0   & 0 \\
1   & 2\cosh k_1  & 0   & \cdots & 0   & 0 \\
\cdots &\cdots &\cdots &\cdots &\cdots &\cdots \\
0   & 0    & 0   & \cdots & 1   & 2\cosh k_1
\end{array}\right)_N,
\label{Toda-case3}
\end{equation}
where $ 2\cosh k_1=\lambda_1$ and $k_1$ is a nonzero real number.

The general solution to the condition equations \eqref{cond-Toda-a} and \eqref{cond-Toda-b} with
\eqref{Toda-case3} is given by
\begin{equation}
\phi^{J^{-}}_N[\lambda_1]=\mathcal{A}\mathcal{Q}^{+}_{0}+ \mathcal{B} \mathcal{Q}^{-}_{0},
~~\mathcal{A},\mathcal{B}\in \widetilde{G}_N,
\label{gen-sol-Toda-3}
\end{equation}
where
\begin{equation}
\mathcal{Q}^{\pm}_0=(\mathcal{Q}^{\pm}_{0,0},\mathcal{Q}^{\pm}_{0,1},\cdots,\mathcal{Q}^{\pm}_{0,N-1})^T,
~~~
\mathcal{Q}^{\pm}_{0,j}=\frac{1}{j!}\partial^{j}_{\lambda_1} b^{\pm}_1e^{\pm \xi_1},
\end{equation}
$\lambda_1=2\cosh k_1$ and $\xi_1$ is defined by \eqref{xi-Toda}.

The effective  form of \eqref{gen-sol-Toda-3} is
\begin{equation}
\phi=\mathcal{A}\mathcal{Q}^{+}_{0}+  \mathcal{Q}^{-}_{0},
~~ \mathcal{A} \in\widetilde{G}_N.
\end{equation}

This is the limit case of { Case 1} where
$\phi^{-}_1(\lambda_1,n,t)=b^{+}_{j}e^{\xi_j} + b^{-}_{j}e^{-\xi_j}$,
$\phi^{-}_j(\lambda_j,n,t)=\phi^{-}_1(\lambda_j,n,t)$, and
each $\lambda_j$ $(j=2,3\cdots, N)$ tends to $\lambda_1$,
as we have discussed for the KdV equation.

To calculate those derivatives $\partial^{j}_{\lambda_1} b^{\pm}_1e^{\pm \xi_1}$,
we have to rewrite $\xi_1$ as
\begin{equation}
\xi_1= n\ln \Bigl(\frac{\lambda_1}{2}+\frac{1}{2}\sqrt{\lambda_1^2-4}~\Bigr)
+\frac{t}{2}\sqrt{\lambda_1^2-4}+\xi^{(0)}_j,
\end{equation}
where we have made use of
\begin{equation}
k_1= \ln \Bigl(\frac{\lambda_1}{2}+\frac{1}{2}\sqrt{\lambda_1^2-4}~\Bigr),
~~|\lambda_1|>2.
\label{k-xi}
\end{equation}

Obviously, we prefer to calculate derivatives of $b^{\pm}_1e^{\pm \xi_1}$ with respect to $k_1$.
So we consider $\Gamma$ to be the following lower triangular $N\times N$ matrix defined as
\begin{equation}
\Gamma=\widehat{\Gamma}^{-}_N[k_1]=(\Gamma_{sj})_{N\times N},~~~
\Gamma_{sj}=\biggl\{\begin{array}{ll}
\frac{2}{(s-j)!}\partial^{s-j}_{k_1}\cosh k_1,& s\geq j,\\
0 & s<j.
\end{array}\biggr.
\label{Gamma-k-Toda}
\end{equation}
In this case a special solution to the condition equations \eqref{cond-Toda-a} and \eqref{cond-Toda-b}
is given as\cite{Zhang-Toda}
\begin{equation}
\widehat{\phi}=\widehat{\mathcal{Q}}^{+}_0+\widehat{\mathcal{Q}}^{-}_0,
\end{equation}
where
\begin{equation}
\widehat{\mathcal{Q}}^{\pm}_0=(\widehat{\mathcal{Q}}^{\pm}_{0,0},\widehat{\mathcal{Q}}^{\pm}_{0,1},\cdots,
\widehat{\mathcal{Q}}^{\pm}_{0,N-1})^T,~~~
\widehat{\mathcal{Q}}^{\pm}_{0,j}=\frac{1}{j!} \partial^{j}_{k_1} b^{\pm}_1 e^{\pm \xi_1}.
\end{equation}
The relation between $\widehat{\phi}$ and $\mathcal{Q}^{\pm}_0$
can be described as
\begin{equation}
\mathcal{Q}^{+}_0+\mathcal{Q}^{-}_0=M\widehat{\phi},
\end{equation}
where $M$ is an $N\times N$ lower triangular matrix and its entries $\{M_{js}\}$ are determined by
\begin{equation}
\Bigl(\frac{1}{2\sinh k_1}\partial_{k_1}\Bigr)^{j}=\sum^{j}_{s=0}M_{js}\partial^s_{k_1},~~~(j=0,1,\cdots, N-1).
\end{equation}
That means $\widehat{\phi}$ and $\mathcal{Q}^{+}_0+\mathcal{Q}^{-}_0$
lead to same solutions to the Toda lattice.

By noting that $\widehat{\Gamma}^{-}_N[k_1]$ also belongs to the Abelian semigroup $\widetilde{G}_N$,
the general solution to the condition equations \eqref{cond-Toda-a} and \eqref{cond-Toda-b}
with $\Gamma=\widehat{\Gamma}^{-}_N[k_1]$ can be given as
\begin{equation}
\widehat{\phi}^{J^{-}}_N[k_1]=\mathcal{A}\widehat{\mathcal{Q}}^{+}_{0}+ \mathcal{B} \widehat{\mathcal{Q}}^{-}_{0},
~~~~\mathcal{A},~\mathcal{B} \in \widetilde{G}_N,
\label{gen-sol-Toda-k-3}
\end{equation}
and its effective form is
\begin{equation}
\widehat{\phi}=\mathcal{A}\widehat{\mathcal{Q}}^{+}_{0}+ \widehat{\mathcal{Q}}^{-}_{0},
~~\mathcal{A}\in \widetilde{G}_N.
\label{e-gen-sol-Toda-k-3}
\end{equation}
Of course, \eqref{gen-sol-Toda-k-3} can be obtained through a limit procedure
from Case 1 where we rewrite
$\phi^{-}_1(k_1,n,t)=b^{+}_{1}e^{\xi_1} + b^{-}_{1}e^{-\xi_1}$,
$\phi^{-}_j(k_j,n,t)=\phi^{-}_1(k_j,n,t)$, and
let each $k_j$ $(j=2,3\cdots, N)$ tend to $k_1$.

We call the solutions generated from \eqref{gen-sol-Toda-3} or \eqref{Gamma-k-Toda} Jordan block solutions
of the Toda lattice.
They can also
be obtained through the IST
as  multi-pole solutions\cite{Wadati-82,Wadati-84}, or through a limit procedure
in Darboux transformation\cite{Matveev-positon-kdv1,Matveev-positon-kdv2},
or through a generalized Hirota's procedure
in Refs.\cite{Chen-02-JPSJ1,Chen-02-JPSJ2}.

{\it Case 4} ~
\begin{equation}
\Gamma=J^{+}_{N}[\lambda_1]
=\left(\begin{array}{cccccc}
2\cos k_1 & 0    & 0   & \cdots & 0   & 0 \\
1   & 2\cos k_1  & 0   & \cdots & 0   & 0 \\
\cdots &\cdots &\cdots &\cdots &\cdots &\cdots \\
0   & 0    & 0   & \cdots & 1   & 2\cos k_1
\end{array}\right)_N,
\label{Toda-case4}
\end{equation}
where $2\cos k_1= \lambda_1$ and $k_1\neq s\pi,~s\in \mathbb{Z}$.

The general solution to equations \eqref{cond-Toda-a} and \eqref{cond-Toda-b} with
\eqref{Toda-case4} is given by
\begin{equation}
\phi^{J^{+}}_{\hbox{\tiny{\it N}}}[\lambda_1]
=\mathcal{A}\mathcal{P}^{+}_{0}+ \mathcal{B} \mathcal{P}^{-}_{0},
~~\mathcal{A},\mathcal{B}\in \widetilde{G}_N,
\label{gen-sol-Toda-4}
\end{equation}
and the effective form of \eqref{gen-sol-Toda-4} is
\begin{equation}
\phi=\mathcal{A}\mathcal{P}^{+}_{0}+  \mathcal{P}^{-}_{0},
~~\mathcal{A}\in \widetilde{G}_N,
\label{egen-sol-Toda-4}
\end{equation}
where
\begin{equation}
\mathcal{P}^{\pm}_{0}=(\mathcal{P}^{\pm}_{0,0},\mathcal{P}^{\pm}_{0,1},\cdots,\mathcal{P}^{\pm}_{0,N-1})^T,
~~~
\mathcal{P}^{\pm}_{0,j}=\frac{1}{j!}\partial^{j}_{\lambda_1}
b^{\pm}_1e^{\pm i \theta_1},
\end{equation}
$\lambda_1=2\cos k_1$ and $\theta_1$ is defined by \eqref{theta-Toda}.

This is the limit case of {Case 2} where
$\phi^{+}_1(\lambda_1,n,t)=b^{+}_{1}e^{i\theta_1} + b^{-}_{1}e^{-i\theta_1}$,
$\phi^{+}_j(\lambda_j,n,t)=\phi^{+}_1(\lambda_j,n,t)$, and
each $\lambda_j$ $(j=2,3\cdots, N)$ tends to $\lambda_1$.

To calculate those derivatives $\partial^{j}_{\lambda_1} b^{\pm}_1e^{\pm \xi_1}$,
we have to make use of
\begin{equation}
k_1= \arccos \frac{\lambda_1}{2},
~~|\lambda_1|<2.
\label{k-xi-4}
\end{equation}
and rewrite
\begin{equation}
\theta_1= n \arccos \frac{\lambda_1}{2}
+\frac{t}{2}\sqrt{4- \lambda_1^2}+\theta^{(0)}_j.
\end{equation}

If we want to calculate derivatives of $b^{\pm}_1e^{\pm i\theta_1}$ with respect to $k_1$,
then we consider $\Gamma$ to be the following lower triangular $N\times N$ matrix defined as
\begin{equation}
\Gamma=\widehat{\Gamma}^{-}_N[k_1]=(\Gamma_{sj})_{N\times N},~~~
\Gamma_{sj}=\biggl\{\begin{array}{ll}
\frac{2}{(s-j)!}\partial^{s-j}_{k_1}\cos k_1,& s\geq j,\\
0 & s<j.
\end{array}\biggr.
\label{Gamma-k-Toda-4}
\end{equation}
In this case, the general solution
is
\begin{equation}
\widehat{\phi}^{J^{+}}_{\hbox{\tiny{\it N}}}[k_1]
=\mathcal{A}\widehat{\mathcal{P}}^{+}_{0}+ \mathcal{B} \widehat{\mathcal{P}}^{-}_{0},
~~~\mathcal{A}, \mathcal{B}\in \widetilde{G}_N,
\end{equation}
and its effective form is
\begin{equation}
\widehat{\phi}=\mathcal{A}\widehat{\mathcal{P}}^{+}_{0}+ \widehat{ \mathcal{P}}^{-}_{0},
~~~ \mathcal{A} \in \widetilde{G}_N,
\end{equation}
where
\begin{equation}
\widehat{\mathcal{P}}^{\pm}_{0}
=(\widehat{\mathcal{P}}^{\pm}_{0,0},\widehat{\mathcal{P}}^{\pm}_{0,1},\cdots,\widehat{\mathcal{P}}^{\pm}_{0,N-1})^T,
~~~\widehat{\mathcal{P}}^{\pm}_{0,j}=\frac{1}{j!}\partial^{j}_{k_1}
b^{\pm}_1 e^{\pm i \theta_1}.
\end{equation}
As a Casoratian entry vector, $\widehat{\mathcal{P}}^{+}_0+\widehat{\mathcal{P}}^{-}_0$
leads to the same solution to the bilinear Toda lattice \eqref{(3.6)} as
$\mathcal{P}^{+}_0+\mathcal{P}^{-}_0$ does, due to
\begin{equation*}
\mathcal{P}^{+}_0+\mathcal{P}^{-}_0=M(\widehat{\mathcal{P}}^{+}_0+\widehat{\mathcal{P}}^{-}_0),
\end{equation*}
where
$M$ is an $N\times N$ lower triangular matrix and its entries $\{M_{js}\}$ are determined by
\begin{equation*}
\Bigl(\frac{-1}{2\sin k_1}\partial_{k_1}\Bigr)^{j}=\sum^{j}_{s=0}M_{js}\partial^s_{k_1},~~~(j=0,1,\cdots, N-1).
\end{equation*}

{\it Case 5}
\begin{equation}
\Gamma=J^{0}_{N}
=\left(\begin{array}{cccccc}
2 & 0    & 0   & \cdots & 0   & 0 \\
1   & 2  & 0   & \cdots & 0   & 0 \\
\cdots &\cdots &\cdots &\cdots &\cdots &\cdots \\
0   & 0    & 0   & \cdots & 1   & 2
\end{array}\right)_N.
\label{(3.26)}
\end{equation}

This case corresponds to $k_1 \rightarrow 0$ in {Case 3} or {Case 4},
in this case we  get rational solutions to the Toda lattice.

However, it is not easy to directly calculate solutions to the condition equations
\eqref{cond-Toda-a} and \eqref{cond-Toda-b} with \eqref{(3.26)}, so we consider
the following alternative form of \eqref{(3.26)}, i.e.,
\begin{equation}
\Gamma=\widehat{\Gamma}^{0}_N=(\Gamma_{sj})_{N\times N},~~~
\Gamma_{sj}=\biggl\{\begin{array}{ll}
\frac{2}{[2(s-j)]!}& s\geq j,\\
0 & s<j,
\end{array}\biggr.
\label{Gamma-k-Toda-5}
\end{equation}
which is a lower triangular $N\times N$ matrix in the group $G_N$.
In this case, the general solution can be taken as
\begin{equation}
\phi^{0}_N=\mathcal{A}\mathcal{R}^{+}_{0}+ \mathcal{B} \mathcal{R}^{-}_{0},
~~\mathcal{A},\mathcal{B}\in \widetilde{G}_N,
\label{gen-sol-Toda-5}
\end{equation}
where
\begin{equation}
\mathcal{R}^{\pm}_0=(\mathcal{R}^{\pm}_{0,0},\mathcal{R}^{\pm}_{0,1},\cdots,\mathcal{R}^{\pm}_{0,N-1})^T,
\end{equation}
\begin{equation}
R^{+}_{0,j}=\frac{1}{(2j)!}\Bigl [\frac{\partial^{2j}}{{\partial k_1}^{2j}}\cosh \xi_1\Bigr ]_{k_1=0},~~~
R^{-}_{0,j}=\frac{1}{(2j+1)!}\Bigl [\frac{\partial^{2j+1}}{{\partial k_1}^{2j+1}}\sinh \xi_1\Bigr ]_{k_1=0},
\end{equation}
and $\xi_1$ is defined as \eqref{xi-Toda}. $ R^{+}_{0}$ and $ R^{-}_{0}$ are linearly independent.
When $\xi_1^{(0)}=0$ in $\xi_1$, $ R^{\pm}_{0,j}$ can be given as\cite{Zhang-Toda}
\begin{equation}
R^{+}_{0,j} =\sum^{2j}_{s=0}\frac{n^s}{s!}p_{2j-s}(\tilde{t}),~~~
R^{-}_{0,j}=\sum^{2j+1}_{s=0}\frac{n^s}{s!}p_{2j+1-s}(\tilde{t}),
\end{equation}
where
\begin{equation*}
p_{s}(\tilde{t})=\sum_{||\alpha||=s}\frac{\tilde{t}^\alpha}{\alpha!},
\end{equation*}
\begin{equation*}
\alpha=(\alpha_1,~\alpha_3,~\alpha_5,~\cdots),~~~~\alpha_j\geq 0,~~(j=1,3,5,\cdots),
\end{equation*}
\begin{equation*}
||\alpha||=\alpha_1+3\alpha_3 +5\alpha_5 +\cdots,~~~~
\alpha!=\alpha_1 !\alpha_3 ! \alpha_5 !\cdots,
\end{equation*}
\begin{equation*}
\tilde{t}=(t_1,~t_3,~t_5,~\cdots),~~~t_j=\frac{t}{j!},~~~
\tilde{t}^\alpha=t_1^{\alpha_1}t_3^{\alpha_3}t_5^{\alpha_5}\cdots.
\end{equation*}

The effective form of \eqref{gen-sol-Toda-5} is
\begin{equation}
\phi=\mathcal{A}\mathcal{R}^{+}_{0}+ \mathcal{R}^{-}_{0},
~~\mathcal{A}\in \widetilde{G}_N.
\label{egen-sol-Toda-5}
\end{equation}

We note that \eqref{gen-sol-Toda-5} can also be obtained by following the procedure
in Ref.\cite{Zhang-Toda rational,rational-1};
and $\cos \theta_1$ and $\sin \theta_1$
generate the same rational solutions as \eqref{gen-sol-Toda-5}\cite{Zhang-Toda}.

{\it Case 6}
\begin{equation}
\begin{array}{rl}
\Gamma=D^{\hbox{\it\tiny C}}_{\hbox{\tiny 2\it M}}[\lambda_1,\lambda_2,\cdots,\lambda_M]
  & ={\rm Diag}(\cosh k_1,\cosh k^{*}_1,\cosh k_2, \cosh k^{*}_2,\cdots,\cosh k_M,\cosh k^{*}_M)\\
~ & ={\rm Diag}(\lambda_1,\lambda^{*}_1,\lambda_2, \lambda^{*}_2,\cdots,\lambda_M,\lambda^{*}_M),
\end{array}
\label{Gamma-Toda-6-1}
\end{equation}
where \{$\lambda_j=\cosh k_j\}$ are $M$ distinct complex numbers, and $*$ still means complex conjugate.
Besides, $|\lambda_j|\neq 2$ and we set
\begin{equation}
\lambda_j=\lambda_{j1}+i\lambda_{j2} ,~~
k_j=k_{j1}+i k_{j2}, ~~ k_{j1} k_{j2} \neq 0, ~~~ j=1,2,\cdots,M.
\end{equation}
In fact, if $k_{j1}=0$ or $k_{j2}=0$, then $\lambda_j\in \mathbb{R}$;
if both, then $\lambda_j=2$.
If we consider $\Gamma$ to be a canonical form of $A$ in \eqref{cond-Toda-a'}, then this case corresponds to
the real matrix $A$ having $N=2M$ distinct complex eigenvalues which appear in conjugate pairs.
We also note that it is not necessary to consider the case \{$\lambda_j=\cos k_j\}$
due to $\cosh k_j=\cos ik_j$.

The relations between $\lambda_{js}$ and $k_{js}$ can be described as follows,
\begin{equation}
\lambda_{j1}=2\cos k_{j2} \cosh k_{j1},~~~
\lambda_{j2}=2\sin k_{j2} \sinh k_{j1},
\end{equation}
and
\begin{equation}
k_{j1}= \ln |\lambda_j+\sqrt{\lambda_j^2-4}|-\ln 2,
~~~k_{j2}=\arg (\lambda_j+\sqrt{\lambda_j^2-4}),
\label{k-xi-Toda}
\end{equation}
where $\arg$ is the argument principle value function and we have taken one branch of $k_j$.

In this case, we take Casoratian vector as
\begin{equation}
\phi^{\hbox{\it\tiny C}}_{\hbox{\tiny 2\it M}}[\lambda_1,\lambda_2,\cdots,\lambda_{M}]
=(\phi^{\hbox{\it\tiny C}}_1, \phi^{\hbox{\it\tiny C}*}_1,
\phi^{{\hbox{\it\tiny C}}}_2, \phi^{{\hbox{\it\tiny C}}*}_2,\cdots,
\phi^{\hbox{\it\tiny C}}_{\hbox{\it\tiny M}}, \phi^{\hbox{\it\tiny C}*}_{\hbox{\it\tiny M}})^T,
\label{entry-Toda-Case6-0}
\end{equation}
where
\begin{equation}
\phi^{\hbox{\it\tiny C}}_j=a^{+}_{j}\cosh {\xi_j} + a^{-}_{j}\sinh {\xi_j},
\end{equation}
or
\begin{equation}
\phi^{\hbox{\it\tiny C}}_j=b^{+}_{j}e^{\xi_j} + b^{-}_{j}e^{-\xi_j},
\label{entry-Toda-Case6-1}
\end{equation}
with
\begin{equation}
\xi_j=k_j n +t \sinh k_j +\xi^{(0)}_j,~~j=1,2,\cdots M.
\label{xi-Tota-6}
\end{equation}
Here
$a^{\pm}_{j}$, $b^{\pm}_{j}$ and $\xi^{(0)}_j$  are all complex constants.
\eqref{entry-Toda-Case6-0} guarantees
${\rm{Cas}}(\phi^{\hbox{\it\tiny C}}_{\hbox{\tiny 2\it M}}[\lambda_1,\lambda_2,\cdots,\lambda_{M}])$
always generates a real solution to the Toda lattice through the transformation \eqref{(3.7)} or
\eqref{trans-Toda}.

We can also consider the real version of \eqref{Gamma-Toda-6-1},
i.e.,
\begin{equation}
\Gamma={\rm Diag}(\widetilde{\Lambda}_1,\widetilde{\Lambda}_2,\cdots,\widetilde{\Lambda}_M),
~~\widetilde{\Lambda}_j=\biggl(\begin{matrix}\lambda_{j1} & -\lambda_{j2}\\
                            \lambda_{j2} & \lambda_{j1}
              \end{matrix}\biggr),~~j=1,2,\cdots M,
\label{Gamma-Toda-6-2}
\end{equation}
which is connected with \eqref{Gamma-Toda-6-1} through
\begin{equation}
{\rm Diag}(\widetilde{\Lambda}_1,\widetilde{\Lambda}_2,\cdots,\widetilde{\Lambda}_M)=
U^{-1}D^{c}_{2M}[\lambda_1,\lambda_2,\cdots,\lambda_M]U
\end{equation}
where $U$ is defined by \eqref{(2.50)}.

Similar to the same case for the KdV equation,
the solution to the condition equations \eqref{cond-Toda-a} and \eqref{cond-Toda-b}
with \eqref{Gamma-Toda-6-2} can be taken as
\begin{equation}
\phi^{\hbox{\it\tiny R}}_{\hbox{\tiny 2\it M}}[\lambda_1,\lambda_2,\cdots,\lambda_{M}]
=(\phi_{11},\phi_{12},\phi_{21},\phi_{22},\cdots,\phi_{\hbox{\tiny {\it M}1}},\phi_{\hbox{\tiny{\it M}2}})^T
=U^{-1}\phi^{\hbox{\it\tiny C}}_{\hbox{\tiny 2\it M}}[\lambda_1,\lambda_2,\cdots,\lambda_{M}]
\label{phi-R-C-Toda}
\end{equation}
where, if $\phi^{\hbox{\it\tiny C}}_j$ is defined by \eqref{entry-Toda-Case6-1},
\begin{equation}
\begin{array}{rl}
\phi_{j1}= & \Bigl\{ b^{+}_{j1}\cos [k_{j2}(n-\nu_j t)+\xi^{(0)}_{j2}]
-b^{+}_{j2} \sin [k_{j2}(n-\nu_j t)+\xi^{(0)}_{j2}]\Bigr\}
e^{[k_{j1}(n-\mu_j t)+\xi^{(0)}_{j1}]}\\
~& +\Bigl\{ b^{-}_{j1}\cos [k_{j2}(n-\nu_j t)+\xi^{(0)}_{j2}]
+b^{-}_{j2} \sin [k_{j2}(n-\nu_j t)+\xi^{(0)}_{j2}]\Bigr\}
e^{-[k_{j1}(n-\mu_j t)+\xi^{(0)}_{j1}]},
\end{array}
\label{entry-Toda-Case6-R-a}
\end{equation}
\begin{equation}
\begin{array}{rl}
\phi_{j2} = & \Bigl\{ b^{+}_{j2}\cos [k_{j2}(n-\nu_j t)+\xi^{(0)}_{j2}]
+b^{+}_{j1} \sin [k_{j2}(n-\nu_j t)+\xi^{(0)}_{j2}]\Bigr\}
e^{[k_{j1}(n-\mu_j t)+\xi^{(0)}_{j1}]}\\
~& +\Bigl\{ b^{-}_{j2}\cos [k_{j2}(n-\nu_j t)+\xi^{(0)}_{j2}]
-b^{-}_{j1} \sin [k_{j2}(n-\nu_j t)+\xi^{(0)}_{j2}]\Bigr\}
e^{-[k_{j1}(n-\mu_j t)+\xi^{(0)}_{j1}]},
\end{array}
\label{entry-Toda-Case6-R-b}
\end{equation}
and
\begin{equation*}
\mu_j=-\frac{\cos k_{j2} \sinh k_{j1}}{k_{j1}},~~ \nu_j=-\frac{\sin k_{j2} \cosh k_{j1}}{k_{j2}}.
\end{equation*}

$\phi^{\hbox{\it\tiny R}}_{\hbox{\tiny 2\it M}}[\lambda_1,\lambda_2,\cdots,\lambda_{M}]$
and $\phi^{\hbox{\it\tiny C}}_{\hbox{\tiny 2\it M}}[\lambda_1,\lambda_2,\cdots,\lambda_{M}]$
lead to  same solutions to the Toda equation due to the relation \eqref{phi-R-C-Toda}.

%

{\it Case 7} ~
\begin{equation}
\Gamma= J^{\hbox{\it\tiny C}}_{2M}[\lambda_1]=\biggl(\begin{matrix}\Lambda & 0\\
                                 0  & \Lambda^* \end{matrix}\biggr),
~~\Lambda
=\left(\begin{array}{ccccc}
2\cosh k_1 & 0    &  \cdots & 0   & 0 \\
1   & 2\cosh k_1 &  \cdots & 0   & 0 \\
\cdots &\cdots &\cdots &\cdots  &\cdots \\
0   & 0    &  \cdots & 1   & 2\cosh k_1
\end{array}\right)_{2M},
\label{Gamma-Toda-7-1}
\end{equation}
i.e.,  $A$ has $M=N/2$ same complex conjugate eigenvalue pairs, where we set $2\cosh k_1=\lambda_1$.
%

The general solution
to the condition equations \eqref{cond-Toda-a} and \eqref{cond-Toda-b} in this case can be taken as
\begin{equation}
\phi
=\Biggl(\begin{array}{cc}
\mathcal{A}& 0\\
0& \mathcal{A}^*
\end{array}\Biggr)\Biggl(
\begin{array}{c}
\mathcal{Q}^{+}_{0}\\
{\mathcal{Q}^{+}_{0}}^*
\end{array}\Biggr)
+\Biggl(\begin{array}{cc}
\mathcal{B}& 0\\
0& \mathcal{B}^*
\end{array}\Biggr)\Biggl(
\begin{array}{c}
\mathcal{Q}^{-}_{0}\\
{\mathcal{Q}^{-}_{0}}^*
\end{array}\Biggr),
~~ \mathcal{A}, \mathcal{B} \in \widetilde{G}_M,
\label{gen-sol-Toda-7-1}
\end{equation}
where
\begin{equation}
\mathcal{Q}^{\pm}_0=(\mathcal{Q}^{\pm}_{0,0},\mathcal{Q}^{\pm}_{0,1},\cdots,\mathcal{Q}^{\pm}_{0,M-1})^T,
~~
\mathcal{Q}^{\pm}_{0,j}=\frac{(-1)^j}{j!}\partial^{j}_{\lambda_1} b^{\pm}_1e^{\pm \xi_1},
\end{equation}
and the effective form of \eqref{gen-sol-KdV-c} can be taken as
\begin{equation}
\phi=\Biggl(\begin{array}{cc}
\mathcal{A}& 0\\
0& \mathcal{A}^*
\end{array}\Biggr)\Biggl(
\begin{array}{c}
\mathcal{Q}^{+}_{0}\\
{\mathcal{Q}^{+}_{0}}^*
\end{array}\Biggr)
+\Biggl (
\begin{array}{c}
\mathcal{Q}^{-}_{0}\\
{\mathcal{Q}^{-}_{0}}^*
\end{array}\Biggr),
~~ \mathcal{A} \in \widetilde{G}_M.
\end{equation}
Similar to the same case for the KdV equation,
$\mathcal{A}^*$ and $\mathcal{B}^*$ in \eqref{gen-sol-Toda-7-1} can be substituted by arbitrary matrices
$\mathcal{C}$ and $\mathcal{D}$ in $\widetilde{G}_M$, but such a $\phi$ does not
guarantee to generate a real solution to the Toda lattice.

If we take
\begin{equation}
\Gamma=J^{\hbox{\it\tiny C}}_{2M}[\Lambda_1]
=\left(\begin{array}{ccccc}
\Lambda_1 & 0    &  \cdots & 0   & 0 \\
I_1   & \Lambda_1  &  \cdots & 0   & 0 \\
\cdots &\cdots &\cdots &\cdots  &\cdots \\
0   & 0    &  \cdots & I_1   & \Lambda_1
\end{array}\right)_{2M}
\label{Gamma-Toda-7-2}
\end{equation}
instead of \eqref{Gamma-Toda-7-1},
where
\begin{equation}
I_1 = \biggl(\begin{matrix}1 & 0\\
                                 0  & 1 \end{matrix}\biggr), ~~
\Lambda_1\equiv \biggl(\begin{matrix}2\cosh k_1 & 0\\
                                 0 & 2\cosh k_1^*
                       \end{matrix}\biggr),
\end{equation}
then the general solution to the condition equations \eqref{cond-KdV-a} and \eqref{cond-KdV-b} can be taken as
\begin{equation}
\phi^{J_C}_{\hbox{\tiny 2\it M}}[\lambda_1]
=\mathcal{A}^B\widetilde{\mathcal{Q}}^{+}_{0}+
\mathcal{B}^B \widetilde{\mathcal{Q}}^{-}_{0},
~~~
\mathcal{A}^B, \mathcal{B}^B \in \widetilde{G}^{B}_{2M},
\label{gen-sol-Toda-7-2}
\end{equation}
where
\begin{equation}
\widetilde{\mathcal{Q}}^{\pm}_0
=\widetilde{{\rm Diag}}_{2M}[{\mathcal{C}}_{1,\lambda_1}]\Phi^{\pm}_{\hbox{\tiny C}},
~~~
{\mathcal{C}}_{1,\lambda_1}=\biggl(\begin{matrix} \partial_{\lambda_1} & 0\\
                                 0 & \partial_{\lambda^*_1} \end{matrix}\biggr),
\end{equation}
$\widetilde{{\rm Diag}}_{2M}[\cdot]$ is defined by
\begin{equation}
\widetilde{{\rm Diag}}_{2M}[P]={\rm Diag}\Bigl(I_1, \frac{1}{1!}P,
\frac{1}{2!}P^2,\cdots, \frac{1}{(M-1)!}P^{M-1}\Bigr),
\label{diag-P-Toda}
\end{equation}
$P$ is a $2\times 2$ matrix,
$\Phi^{\pm}_{\hbox{\tiny C}}$ is a $2M$-order column vector defined as
\begin{equation}
\Phi^{\pm}_{\hbox{\tiny C}}=(\mathcal{Q}^{\pm}_{0,0},{\mathcal{Q}^{\pm *}_{0,0}},
\mathcal{Q}^{\pm}_{0,0},{\mathcal{Q}^{\pm *}_{0,0}},\cdots,
\mathcal{Q}^{\pm}_{0,0},{\mathcal{Q}^{\pm *}_{0,0}})^T.
\label{Phi-C-Toda}
\end{equation}
%
%
$A_j$ and $B_j$
in $\mathcal{A}^B$ and $\mathcal{B}^B$  are always taken as
\begin{equation}
A_j = \biggl(\begin{matrix} a_{j1} & 0\\
                                 0  & a^*_{j1} \end{matrix}\biggr),~~~
B_j = \biggl(\begin{matrix} b_{j1} & 0\\
                                 0  & b^*_{j1} \end{matrix}\biggr)
\label{entry-AB}
\end{equation}
so as to generate real solutions to the Toda lattice.
\eqref{gen-sol-Toda-7-2} has the following effective form
\begin{equation}
\phi=\mathcal{A}^B\widetilde{\mathcal{Q}}^{+}_{0}+
\widetilde{\mathcal{Q}}^{-}_{0},
~~\mathcal{A}^B \in \widetilde{G}^{B}_N.
\end{equation}

Besides, \eqref{gen-sol-Toda-7-1} and \eqref{gen-sol-Toda-7-2}
generate same solutions to the Toda lattice.

If we replace \eqref{Gamma-Toda-7-2} by its real version, i.e.,
\begin{equation}
\Gamma=\widetilde{J}^{\hbox{\it\tiny C}}_{2M}[\widetilde{\Lambda}_1]
=\left(\begin{array}{ccccc}
\widetilde{\Lambda}_1 & 0    &  \cdots & 0   & 0 \\
I_1   & \widetilde{\Lambda}_1  &  \cdots & 0   & 0 \\
\cdots &\cdots &\cdots &\cdots  &\cdots \\
0   & 0    &  \cdots & I_1   & \widetilde{\Lambda}_1
\end{array}\right)_{2M},
\end{equation}
where $\widetilde{\Lambda}_1$ is defined by \eqref{Gamma-Toda-6-2}, i.e.,
$\widetilde{\Lambda}_1 \equiv \biggl(\begin{matrix}\lambda_{11} & -\lambda_{12}\\
                                 \lambda_{12} & \lambda_{11} \end{matrix}\biggr)$,
then the corresponding general solution to the condition equations \eqref{cond-Toda-a}
and \eqref{cond-Toda-b} can be given as
\begin{equation}
\phi^{J_R}_{\hbox{\tiny 2\it M}}[\lambda_1]
=\mathcal{A}^{\widetilde{B}} \bar{\mathcal{Q}}^{+}_{0} + \mathcal{B}^{\widetilde{B}}
\bar{\mathcal{Q}}^{-}_{0}, ~~~
\mathcal{A}^{\widetilde{B}},\mathcal{B}^{\widetilde{B}}\in \widetilde{G}_{2M}^{\widetilde{B}},
\label{gen-sol-Toda-7-3}
\end{equation}
and the effective form  is
\begin{equation}
\phi
=\mathcal{A}^{\widetilde{B}}\bar{\mathcal{Q}}^{+}_{0} +
\bar{\mathcal{Q}}^{-}_{0},
~~~
\mathcal{A}^{\widetilde{B}}\in \widetilde{G}_{2M}^{\widetilde{B}},
\end{equation}
where
\begin{equation}
\bar{\mathcal{Q}}^{\pm}_0
=\widetilde{{\rm Diag}}_{2M}[{\mathcal{R}}_{1,\lambda_1}]\Phi^{\pm}_{\hbox{\tiny R}},~~~
\mathcal{R}_{1,\lambda_1}=U_1^{-1}{\mathcal{C}}_{1,\lambda_1} U_1,
\label{R-C-Toda}
\end{equation}
\begin{equation}
\Phi^{\pm}_{\hbox{\tiny R}}=U^{-1}\Phi^{\pm}_{\hbox{\tiny C}}
=(\mathcal{Q}^{R\pm}_{0,0},{\mathcal{Q}^{I\pm }_{0,0}},
\mathcal{Q}^{R\pm}_{0,0},{\mathcal{Q}^{I\pm }_{0,0}},\cdots,
\mathcal{Q}^{R\pm}_{0,0},{\mathcal{Q}^{I\pm }_{0,0}})^T,
\label{Phi-R-Toda}
\end{equation}
\begin{equation*}
\begin{array}{rl}
\mathcal{Q}^{R\pm}_{0,0}= & \Big\{ b^{\pm}_{11}\cos [k_{12}(n-\nu_1 t)+\xi^{(0)}_{12}] \mp
b^{\pm}_{12} \sin [k_{12}(n-\nu_1 t)+\xi^{(0)}_{12}]\Bigr\}
e^{\pm [k_{11}(n-\mu_1 t)+\xi^{(0)}_{11}]},\\
\mathcal{Q}^{I\pm}_{0,0} = & \Big\{b^{\pm}_{12} \cos [k_{12}(n-\nu_1 t)+\xi^{(0)}_{12}]\pm
b^{\pm}_{11} \sin [k_{12}(n-\nu_1 t)+\xi^{(0)}_{12}]\Bigr\}e^{\pm [k_{11}(n-\mu_1 t)+\xi^{(0)}_{11}]},
\end{array}
\end{equation*}
and here $k_{1j}$, $\mu_1$ and $\nu_1$ are defined as in Case 6.
To get the above general solution, we have made use of
the following relations,
\begin{equation}
\widetilde{J}^{\hbox{\it\tiny C}}_{2M}[\widetilde{\Lambda}_1]=
U^{-1}J^{\hbox{\it\tiny C}}_{2M}[\Lambda_1]U,~~~
\phi^{J_R}_{\hbox{\tiny 2\it M}}[\lambda_1]
=U^{-1}\phi^{J_C}_{\hbox{\tiny 2\it M}}[\lambda_1],~~~
\mathcal{A}^{\widetilde{B}}=U^{-1}\mathcal{A}^{B}U,~~~\mathcal{B}^{\widetilde{B}}=U^{-1}\mathcal{B}^{B}U.
\end{equation}
Obviously, $\phi^{J_R}_{\hbox{\tiny 2\it M}}[\lambda_1]$ leads to the same solution to
the Toda lattice as $\phi^{J_C}_{\hbox{\tiny 2\it M}}[\lambda_1]$ does.

Similar to the same case for the KdV equation, we have the following results.

\begin{Proposition}
\label{Prop 4.3}
{\it
a). $\mathcal{R}_{1,\lambda_1}$ in \eqref{R-C-Toda} can be taken one of the following forms,
\begin{equation}
\mathcal{R}_{1,\lambda_1}=
\mathcal{R}_{1,\lambda_{11}}\equiv \partial_{\lambda_{11}}I_1,
\end{equation}
or
\begin{equation}
\mathcal{R}_{1,\lambda_1}=
\sigma_1\mathcal{R}_{1,\lambda_{12}}\equiv\partial_{\lambda_{12}}\sigma_1, ~~
\sigma_1=\biggl(\begin{matrix} 0 & -1\\
                               1 & 0 \end{matrix}\biggr).
\end{equation}
b). The following vectors
\begin{equation}
\phi^{\lambda_{11}}_{_{2M}}
=\widetilde{{\rm Diag}}_{2M}[{\mathcal{R}}_{1,\lambda_{11}}]
(\Phi^{+}_{\hbox{\tiny R}}+\Phi^{-}_{\hbox{\tiny R}}),
\end{equation}
\begin{equation}
\phi^{\lambda_{12}}_{_{2M}}
=\widetilde{{\rm Diag}}_{2M}[\sigma_1{\mathcal{R}}_{1,\lambda_{12}}]
(\Phi^{+}_{\hbox{\tiny R}}+\Phi^{-}_{\hbox{\tiny R}})
\end{equation}
and
\begin{equation}
\widetilde{\phi}^{\lambda_{12}}_{_{2M}}
=\widetilde{{\rm Diag}}_{2M}[{{\mathcal{R}}}_{1,\lambda_{12}}]
(\Phi^{+}_{\hbox{\tiny R}}+\Phi^{-}_{\hbox{\tiny R}}),
\end{equation}
as Casoratian entries,
lead to the same solutions to the Toda lattice as $\phi=\mathcal{Q}^{+}_{0}+\mathcal{Q}^{-}_{0}$ does.
In addition, the following Casoratian entry vector
\begin{equation}
\phi=\left((\phi^{\lambda_{11}}_{_{2M_1}})^T,
({\phi}^{\lambda_{12}}_{_{2M_2}})^T\right)^T~~
{\rm or}~~
\phi=\left((\phi^{\lambda_{11}}_{_{2M_1}})^T,
(\widetilde{\phi}^{\lambda_{12}}_{_{2M_2}})^T\right)^T,
~~M_1 +M_2=M
\end{equation}
always generates a zero solution to the KdV equation.}
\hfill $\Box$
\end{Proposition}

As we mentioned in Case 3, it is not convenient to calculate derivatives
with respect to $\lambda$, so we consider another form of $\Gamma$.

Let
\begin{equation}
\Gamma=\check{J}^{\hbox{\it\tiny C}}_{2M}[\Gamma_0,\Gamma_1,\cdots,\Gamma_{M-1}]
=\left(\begin{array}{ccccccc}
\Gamma_0 & 0   & 0  &  \cdots & 0   & 0 & 0 \\
\Gamma_1  & \Gamma_0  & 0& \cdots & 0   & 0 & 0 \\
\Gamma_2  & \Gamma_1 & \Gamma_0 & \cdots & 0   & 0 & 0 \\
\cdots &\cdots &\cdots & \cdots &\cdots  &\cdots & \cdots\\
\Gamma_{M-1}  & \Gamma_{M-2}  & \Gamma_{M-3}  &  \cdots & \Gamma_2 & \Gamma_1   & \Gamma_0
\end{array}\right)_{2M},
\label{Gamma-Toda-7-4-a}
\end{equation}
where
\begin{equation}
\Gamma_{j}=\frac{1}{j!}\biggl(\begin{matrix}e^{k_1} +(-1)^j e^{-k_1} & 0\\
                                 0  & e^{k^*_1} +(-1)^j e^{-k^*_1} \end{matrix}\biggr),
~~~j=0,1,\cdots,M-1.
\label{Gamma-KdV-7-4-b}
\end{equation}

Thus we can give a general solution to the condition equations \eqref{cond-Toda-a} and
\eqref{cond-Toda-b} by calculating derivatives with respect to $k_1$ instead of $\lambda_1$.
The corresponding general solution is
\begin{equation}
\check{\phi}^{J_C}_{\hbox{\tiny 2\it M}}[k_1]
=\mathcal{A}^B\check{\mathcal{Q}}^{+}_{0}+
\mathcal{B}^B \check{\mathcal{Q}}^{-}_{0},
~~~
\mathcal{A}^B, \mathcal{B}^B \in \widetilde{G}^{B}_{2M},
\label{gen-sol-Toda-7-4}
\end{equation}
where
\begin{equation}
{\check{\mathcal{Q}}}^{\pm}_0
=\widetilde{{\rm Diag}}_{2M}[{\mathcal{C}}_{1,k_1}]{\Phi}^{\pm}_{\hbox{\tiny C}},
~~~
{\mathcal{C}}_{1,k_1}=\biggl(\begin{matrix} \partial_{k_1} & 0\\
                                 0 & \partial_{k^*_1} \end{matrix}\biggr),
\label{entry-Toda-7-4}
\end{equation}
${\Phi}^{\pm}_{\hbox{\tiny C}}$ is given by \eqref{Phi-C-Toda},
and in order to get real solutions to the Toda lattice, we still take entries $A_j$ and $B_j$
in $\mathcal{A}^B$ and $\mathcal{B}^B$  as \eqref{entry-AB}.
The effective form of \eqref{gen-sol-Toda-7-4} is
\begin{equation}
\check{\phi}=\mathcal{A}^B\check{\mathcal{Q}}^{+}_{0}+
\check{\mathcal{Q}}^{-}_{0},
~~\mathcal{A}^B \in \widetilde{G}^{B}_{2M}.
\end{equation}

We note that ${\mathcal{C}}_{1,k_1}$ in \eqref{entry-Toda-7-4}
can be simplified to
\begin{equation}
\biggl(\begin{matrix} \partial_{k_{11}} & 0\\
                                 0 & \partial_{k_{11}} \end{matrix}\biggr)~~
{\rm or}~~
\biggl(\begin{matrix} -i \partial_{k_{12}} & 0\\
                                 0 & i \partial_{k_{12}} \end{matrix}\biggr)
\label{simp-C}
\end{equation}
which means, as a Casoratian entry vector,
$\widetilde{{\rm Diag}}_{2M}[P](\Phi^{+}_{\hbox{\tiny C}}+\Phi^{-}_{\hbox{\tiny C}})$
generates same solutions to the Toda lattice when $P$ is taken as ${\mathcal{C}}_{1,k_1}$,
$\partial_{k_{11}}I_1$ or $\partial_{k_{12}}I_1$.

We can also consider the real version of \eqref{Gamma-Toda-7-4-a}, i.e.,
\begin{equation}
\Gamma=\widehat{J}^{\hbox{\it\tiny C}}_{2M}
[\widehat{\Gamma}_1,\widehat{\Gamma}_2,\cdots,\widehat{\Gamma}_{M-1}]
=\left(\begin{array}{ccccccc}
\widehat{\Gamma}_0 & 0   & 0  &  \cdots & 0   & 0 & 0 \\
\widehat{\Gamma}_1  & \widehat{\Gamma}_0  & 0& \cdots & 0   & 0 & 0 \\
\widehat{\Gamma}_2  & \widehat{\Gamma}_1 & \widehat{\Gamma}_0 & \cdots & 0   & 0 & 0 \\
\cdots &\cdots &\cdots & \cdots &\cdots  &\cdots & \cdots\\
\widehat{\Gamma}_{M-1}  & \widehat{\Gamma}_{M-2}  & \widehat{\Gamma}_{M-3}  &
\cdots & \widehat{\Gamma}_2 & \widehat{\Gamma}_1   & \widehat{\Gamma}_0
\end{array}\right)_{2M},
\label{Gamma-Toda-7-5-a}
\end{equation}
which is connected with $\check{J}^{\hbox{\it\tiny C}}_{2M}[\Gamma_1,\Gamma_2,\cdots,\Gamma_{M-1}]$ by
\begin{equation}
\widehat{J}^{\hbox{\it\tiny C}}_{2M}
[\widehat{\Gamma}_1,\widehat{\Gamma}_2,\cdots,\widehat{\Gamma}_{M-1}]=
U\check{J}^{\hbox{\it\tiny C}}_{2M}[\Gamma_1,\Gamma_2,\cdots,\Gamma_{M-1}]U^{-1},
\end{equation}
%
where $U$ is defined by \eqref{(2.50)}, and
\begin{equation}
\widehat{\Gamma}_{j}=\frac{1}{j!}
\biggl(\begin{matrix}\cos k_{12}[e^{k_{11}} +(-1)^j e^{-k_{11}}] & -\sin k_{12}[e^{k_{11}}-(-1)^j e^{-k_{11}}]\\
   \sin k_{12}[e^{k_{11}}-(-1)^j e^{-k_{11}}] & \cos k_{12}[e^{k_{11}} +(-1)^j e^{-k_{11}}]
\end{matrix}\biggr),
~~~j=0,1,\cdots,M-1.
\label{Gamma-KdV-7-5-b}
\end{equation}

In this case, the general solution to the condition
equations \eqref{cond-Toda-a} and \eqref{cond-Toda-b} can be taken as
\begin{equation}
\widehat{\phi}^{J_R}_{\hbox{\tiny 2\it M}}[k_1]
=U^{-1}\check{\phi}^{J_C}_{\hbox{\tiny 2\it M}}[k_1]
=\mathcal{A}^{\widetilde{B}}\widehat{\mathcal{Q}}^{+}_{0}+
\mathcal{B}^{\widetilde{B}} \widehat{\mathcal{Q}}^{-}_{0},~~~
\mathcal{A}^{\widetilde{B}},\mathcal{B}^{\widetilde{B}}\in \widetilde{G}_{2M}^{\widetilde{B}},
\label{gen-sol-Toda-7-5}
\end{equation}
and its effective form is
\begin{equation}
\widehat{\phi}^{J_R}_{\hbox{\tiny 2\it M}}[k_1]
=\mathcal{A}^{\widetilde{B}}\widehat{\mathcal{Q}}^{+}_{0}+
 \widehat{\mathcal{Q}}^{-}_{0},~~~
\mathcal{A}^{\widetilde{B}}\in \widetilde{G}_{2M}^{\widetilde{B}},
\end{equation}
where
\begin{equation}
\widehat{\mathcal{Q}}^{\pm}_{0}=U^{-1}\check{\mathcal{Q}}^{\pm}_{0}
=\widetilde{{\rm Diag}}_{2M}[\widehat{\mathcal{R}}_{1,k_{1}}]\Phi^{\pm}_{\hbox{\tiny R}},~~~
\widehat{\mathcal{R}}_{1,k_{1}}=U_1^{-1}{\mathcal{C}}_{1,k_1}U_1,
\end{equation}
and $\Phi^{\pm}_{\hbox{\tiny R}}$ is \eqref{Phi-R-Toda}.
By noticing  the simplified forms \eqref{simp-C} of ${\mathcal{C}}_{1,k_1}$,
$\widehat{\mathcal{R}}_{1,k_{1}}$ can be taken
as $\partial_{k_{11}}I_1$ or $\partial_{k_{12}}\sigma_1$.
Consequently, we have the following result similar to Proposition \ref{Prop 4.3}.

\begin{Proposition}
\label{Prop 4.4}
{\it
As a Casoratian entry vector, $\widetilde{{\rm Diag}}_{2M}[P](\Phi^{+}_{\hbox{\tiny R}}+\Phi^{-}_{\hbox{\tiny R}})$
generates same solutions to the Toda lattice no matter  $P$ is taken as
$\partial_{k_{11}}I_1$, $\partial_{k_{12}}\sigma_1$ or $\partial_{k_{12}}I_1$.
In addition, the following vector
\begin{equation}
\phi={\rm Diag}(\widetilde{{\rm Diag}}_{2M_1}[\partial_{k_{11}}I_1],
\widetilde{{\rm Diag}}_{2M_2}[\partial_{k_{12}}I_1])(\Phi^{+}_{\hbox{\tiny R}}+\Phi^{-}_{\hbox{\tiny R}}),
~~M_1 +M_2=M
\end{equation}
always generates a zero solution to the KdV equation.}
\hfill $\Box$
\end{Proposition}

All the Casoratians obtained in this case are  Jordan block solutions and can be obtained
through certain limit procedures.

\vskip 10pt
We end this section by the following remarks.

{\it Remark 4.1} ~ In this section we first discussed the Casoratian conditions
for the Toda lattices. Then
we  discussed general solutions to the condition equations
\eqref{cond-Toda-a} and \eqref{cond-Toda-b} according to
$\Gamma$ taking different canonical forms or similar forms of $A$ in equations
\eqref{cond-Toda-a'} and \eqref{cond-Toda-b'},
particularly, we gave explicit forms of general Jordan-block solutions,
reduced them to their effective forms, and explained the  relations
between Jordan-block solutions and diagonal cases through exact limit procedures.
In addition, for those complexiton\cite{Ma-Maruno-PA} solutions, we gave
several different choices for a Casoratian entry vector while
they generate same solutions to the Toda lattice.
All  results are similar to those for the KdV equation in Sec.3.

{\it Remark 4.2} ~
A mixed solution to the Toda lattice is generated from a Casoratian entry vector
which is composed by arbitrarily picking up entries from
the above 7 cases.
For example,
\begin{equation}
\phi=\left((\phi^{+}_{\rho_1}[\lambda_{l_1},\cdots, \lambda_{l_{\rho_1}}])^T,
(\phi^{J^{-}}_{\rho_2}[\lambda_{g_1}])^T ,(\phi^{J^{-}}_{\rho_3}[\lambda_{g_2}])^T,
\right)^T,
\label{mix-Toda}
\end{equation}
where $\sum^{3}_{s=1}\rho_s=N$,
$\phi^{+}_{\rho_1}[\lambda_{l_1},\cdots, \lambda_{l_{\rho_1}}]$
and $\phi^{J^{-}}_{\rho_j}[\lambda_{g_s}]$
are respectively defined as \eqref{entry-Toda-II} and \eqref{gen-sol-Toda-3}.
Of course, we can substitute \eqref{gen-sol-Toda-k-3} $\widehat{\phi}^{J^{-}}_{\rho_j}[k_{g_s}]$
for $\phi^{J^{-}}_{\rho_j}[\lambda_{g_s}]$.
\eqref{mix-Toda} is a solution of the condition equations \eqref{cond-KdV-a} and \eqref{cond-KdV-b} when
$\Gamma$ is the following block diagonal matrix
\begin{equation}
{\rm Diag} (D^{+}_{\rho_1}[\lambda_{l_1},\cdots, \lambda_{l_{\rho_1}}],
J^{-}_{\rho_2}[\lambda_{g_1}], J^{-}_{\rho_3}[\lambda_{g_2}]).
\end{equation}
This case corresponds to $A$ having the following eigenvalues
\begin{equation*}
\lambda_{l_1},\cdots, \lambda_{l_{\rho_1}},
\overbrace{\lambda_{g_1},\cdots,\lambda_{g_1}}^{\rho_2},
\overbrace{\lambda_{g_2},\cdots,\lambda_{g_2}}^{\rho_3},
\end{equation*}
where $\sum^{3}_{s=1}\rho_s=N$.
In order to get a nonzero solution to the Toda lattice, we always let all the $\{\lambda_j\}$ be distinct.
Any mixed solution can be obtained from a diagonal case through
certain local limit procedure.

{\it Remark 4.3} ~
Algebraic properties of the lower triangular Toeplitz matrices, i.e., the matrices
commuting with a Jordan block,
are used to easily construct a general Jordan block solution
from a set of special solutions and further give its effective form.
These properties also tell us how to generate the arbitrary constants in  general Jordan-block solution
through a limit procedure.

{\it Remark 4.4} ~ Consider the spectral problem \eqref{(3.3a)}
\begin{equation}
\Phi_{n+1}+ \Phi_{n-1}=(\omega+\frac{1}{\omega}) \Phi_{n},
\label{spect-Toda}
\end{equation}
where we have taken $(u_n, v_n)=(1,0)$ and $\lambda=\omega+\omega^{-1}$.
\eqref{entry-Toda-1a} and \eqref{entry-Toda-1b}
can be considered as  solutions to \eqref{spect-Toda} when $\omega=e^{k_j}$ and $k_j$ is a non-zero real number,
i.e., $|\lambda|>2$; and in this case, the classical solitons are generated.
On the other hand, when $\omega=e^{ik_j}$ where $k_j\in \mathbb{R}$ and $\{k_j\neq s\pi,~s\in \mathbb{Z}\}$,
i.e., $|\lambda|<2$,
\eqref{spect-Toda} has a solution as \eqref{entry-Toda-2a} or \eqref{entry-Toda-2b},
which leads to singular solutions to the Toda lattice.
Corresponding to the solution classification of the KdV equation\cite{Matveev-positon-kdv1,Matveev-positon-kdv2},
Matveev {\it et.al.}\cite{Matveev-positon-Toda} named
the solutions  generated from {Case 2} and {Case 4} positons
of the Toda lattice. And consequently, solutions  generated from {Case 1} and {Case 3}
are called negatons.
Dynamics for these  solutions has been discussed in \cite{Zhang-Toda}.
When $|\lambda|=2$, rational solutions are obtained in Case 5.
Solutions derived from {Case 6} and {7} are called complexitons\cite{Ma-Maruno-PA} to the Toda lattices
due to the fact that they correspond to complex eigenvalues in \eqref{spect-Toda}.
complexitons can also be derived from the Hirota method by taking $k_1$ to be complex and
$k_2$ the conjugate of $k_1$, and so on.

{\it Remark 4.5} ~ The reality condition of square matrix $A$ in \eqref{cond-Toda-a'}
is an interesting question.

\vskip 16pt

\section {\normalsize Solutions in Wronskian form to the KP equation}

In this section, as an (1+2)-dimensional example,
we will investigate a kind of generalization of Wronskian solutions to the Kadomtsev-Petviashvili (KP)
equation. Two arbitrary matrices will be introduced in
our generalization and the verification will easily be achieved by virtue of Proposition \ref{Prop 3.2}.

The KP equation is
\begin{equation}
(u_t + 6uu_{x} + u_{xxx})_x + 3 u_{yy}=0.
\label{KP}
\end{equation}
Through the transformation\cite{Hirota-1971}
\begin{equation}
u=2(\ln f_{n})_{xx},
\label{trans-KP}
\end{equation}
the bilinear form of \eqref{KP} is given as
\begin{equation}
(D_tD_x + D^4_x +3D^2_y) f \cdot f =0,
\label{bilinear-KP}
\end{equation}
which admits the Wronskian solution\cite{Freeman-Nimmo-KP}
\begin{equation}
f=|\widehat{N-1}|,
\label{wrons-KP}
\end{equation}
with the $N$-order entry vector $\phi=\phi(t,x,y)=(\phi_1,\phi_2,\cdots,\phi_N)^T$ satisfying
\begin{subequations}
\begin{eqnarray}
\phi_{j,y}=\phi_{j,xx},
\label{cond-y-KP}\\
\phi_{j,t}=-4\phi_{j,xxx},
\label{cond-t-KP}
\end{eqnarray}
\end{subequations}
for $j=1,2,\cdots, N$.
The function $\phi_j$ meeting the above conditions can be taken as\cite{Freeman-Nimmo-KP}
\begin{equation}
\phi_{j}(t,x,y)=a^{+}_j e^{k_jx +k_j^2 y -4k_j^3 t}
+a^{-}_j e^{h_jx +h_j^2 y -4h_j^3 t},
\label{wrons-entry-KP}
\end{equation}
Where $a^{\pm}_j$ and $k_j \neq h_j$ are parameters independent of $t,~x$ and $y$.
An obvious generalization is that,
$\frac{\partial^m}{{\partial \varrho_j}^m}\phi_j$ still satisfies
\eqref{cond-y-KP} and \eqref{cond-t-KP} for any $m=0,1,2,\cdots$,
and therefore can be a new Wronskian entry,
if considering $a^{\pm}_j$, $k_j$ and $h_j$ are some functions of
$\varrho_j$;
or, alternatively, instead of \eqref{wrons-entry-KP},
\begin{equation}
\phi_{j}(t,x,y)=\frac{\partial^m}{{\partial k_j}^m}\Bigl(a^{+}_j(k_j) e^{k_jx +k_j^2 y -4k_j^3 t}\Bigr)
+\frac{\partial^n}{{\partial h_j}^n}\Bigl(a^{-}_j(h_j) e^{h_jx +h_j^2 y -4h_j^3 t}\Bigr)
\label{wrons-entry-KP-alt}
\end{equation}
with arbitrary $m,n=0,1,2,\cdots$.

\subsection{Wronskian solutions}

By virtue of Proposition \ref{Prop 3.2}, it is easy to get the following generalized Wronskian solution
to the bilinear KP equation \eqref{bilinear-KP}.

\begin{Proposition}
\label{Prop 5.1}
{\it The  Wronskian \eqref{wrons-KP} solves
the bilinear KP equation \eqref{bilinear-KP} if  $\phi$ satisfies
\begin{subequations}
\begin{eqnarray}
\phi_{y}=\phi_{xx}+A(t,y)\phi,
\label{cond-y-new-KP}\\
\phi_{t}=-4\phi_{xxx}+B(t,y)\phi,
\label{cond-t-new-KP}
\end{eqnarray}
\end{subequations}
where $A(t,y)$ and $B(t,y)$ are two arbitrary $N\times N$ matrix functions
of $t$ and $y$ but independent of $x$, satisfying
\begin{equation}
[{\rm Tr}(A(t,y))]_y  =0
\label{trace-KP}
\end{equation}
and the compatible condition
\begin{equation}
A_t(t,y)-B_y(t,y)+ [A(t,y),B(t,y)]=0.
\label{compat-KP}
\end{equation}
}
\hfill $\Box$
\end{Proposition}

The Wronskian satisfying the above proposition solves the bilinear KP equation \eqref{bilinear-KP}.
If we consider solutions of the KP equation \eqref{KP}, which are recovered through the transformation
\eqref{trans-KP}, then the condition \eqref{trace-KP} in Proposition \ref{Prop 5.1}  can be neglected.
Let us prove this in the following.

\begin{Proposition}
\label{Prop 5.2}
~{\eqref{trans-KP}\it provides a solution to the KP equation \eqref{KP}
if $f$ is the Wronskian \eqref{wrons-KP} with  $\phi$ satisfying
the Wronskian conditions \eqref{cond-y-new-KP} and  \eqref{cond-t-new-KP}
where $A(t,y)$ and $B(t,y)$ are two arbitrary $N\times N$ matrix functions
of $t$ and $y$ but independent of $x$, satisfying
the compatible condition \eqref{compat-KP}.}
\end{Proposition}

{\it Proof:} The proof is quite similar to the one for Proposition \ref{Prop 4.2}.
We introduce a Wronskian vector $\psi$ defined as
\begin{equation}
\psi=e^{-\frac{1}{N}\int^{y}_{0}\Tr{A(t,y)}dy}\phi.
\label{psi}
\end{equation}
It then follows from \eqref{cond-y-new-KP} and  \eqref{cond-t-new-KP} that
\begin{subequations}
\begin{eqnarray}
\psi_{y}=\psi_{xx}+\widetilde{A}(t,y)\psi,
\label{cond-y-psi-KP}\\
\psi_{t}=-4\psi_{xxx}+\widetilde{B}(t,y)\psi,
\label{cond-t-psi-KP}
\end{eqnarray}
\end{subequations}
where
\begin{equation}
\widetilde{A}(t,y)=A(t,y)-\frac{1}{N} \Tr{A(t,y)} I
\end{equation}
and
\begin{equation}
\widetilde{B}(t,y)=B(t,y)-\frac{1}{N} \int^y_0 [\Tr{A(t,y)}]_t dy
= B(t,y)-\frac{1}{N} \Tr{B(t,y)}I + \frac{1}{N} \Tr{B(t,0)}I.
\label{B-tilde}
\end{equation}
In \eqref{B-tilde} we have made use of the equality
$[{\rm Tr}(A(t,y))]_t  =[{\rm Tr}(B(t,y))]_y$ implied from \eqref{compat-KP}.
Obviously,
$\widetilde{A}(t,y)$ and $\widetilde{B}(t,y)$ satisfy
\begin{equation}
[{\rm Tr}(\widetilde{A}(t,y))]_y  =0
\end{equation}
and
\begin{equation}
\widetilde{A}_t(t,y)-\widetilde{B}_y(t,y)+ [\widetilde{A}(t,y),\widetilde{B}(t,y)]=0.
\end{equation}
Thus, in the light of Proposition \ref{Prop 5.1}
and by noting that the Wronskian relation
$$f(\phi)=e^{\int^{y}_{0}\Tr{A(t,y)}dy}f(\psi),$$
this proposition holds.
\hfill $\Box$

\subsection {Further discussion}

Although we introduced two new matrices $A(t,y)$ and $B(t,y)$ in the Wronskian
conditions  \eqref{cond-y-new-KP} and  \eqref{cond-t-new-KP},
we now argue that this generalization is trivial for deriving new solutions in some sense.

In fact, if
$A(t,y)$ and $B(t,y)$ belong to ${\rm C}[a,b;c,d]$ ($a$, $b$, $c$ and $d$ can be infinite),
then on the basis of Proposition \ref{Prop 3.6}, there exists a non-singular $N \times N$ matrix $H(t,y)$ solving
\begin{equation}
H_t(t,y)=-H(t,y)B(t,y).
\end{equation}
By introducing
\begin{equation}
\widetilde{\phi}=H(t,y)\phi,
\label{phi-tilde-KP}
\end{equation}
we can rewrite  \eqref{cond-y-new-KP} and \eqref{cond-t-new-KP} as
\begin{subequations}
\begin{eqnarray}
\widetilde{\phi}_y=\widetilde{\phi}_{xx} +\widetilde{A}(t,y)\widetilde{\phi} ,
\label{cond-y-new-3}\\
\widetilde{\phi}_t=-4\widetilde{\phi}_{xxx},
\label{cond-t-new-3}
\end{eqnarray}
\end{subequations}
where
\begin{equation}
\widetilde{A}(t,y)=(H_y(t,y)+H(t,y)A(t,y))H^{-1}(t,y).
\label{A-tilde-KP}
\end{equation}
In addition, noting that $\phi_{ty}=\phi_{yt}$, from \eqref{phi-tilde-KP}
we have
$\widetilde{\phi}_{ty}=\widetilde{\phi}_{yt}$
which implies
\begin{equation}
\widetilde{A}_t(t,y)=0,
\end{equation}
i.e., $\widetilde{A}(t,y)=\overline{A}(y)$.
Without lose of generality, let $\overline{A}(y) \in {\rm C}[c,d]$.
Next, we further introduce
\begin{equation}
\overline{\phi}=G(y)\widetilde{\phi},
\label{phi-bar}
\end{equation}
where $G(y)$ is an non-singular $N \times N$ matrix solving
\begin{equation}
G_y(y)=-G(y)\overline{A}(y).
\end{equation}
It then follows from \eqref{cond-y-new-3} and \eqref{cond-t-new-3} that
\begin{subequations}
\begin{eqnarray}
\overline{\phi}_y=\overline{\phi}_{xx},
\label{cond-y-new-4}\\
\overline{\phi}_t=-4\overline{\phi}_{xxx}.
\label{cond-t-new-4}
\end{eqnarray}
\end{subequations}
Thus, noting that the Wronskians $f(\overline{\phi})$ and $f(\phi)$ follow $f(\overline{\phi})=|G(y)H(t,y)|f(\phi)$,
we conclude that $\overline{\phi}$ and $\phi$ lead to same solutions
to the KP equation \eqref{KP} through the transformation \eqref{trans-KP}.
That means,  $A(t,y)$ and $B(t,y)$ in \eqref{cond-y-new-KP} and \eqref{cond-t-new-KP}
do not generate any new solutions for the KP equation in the case that $B(t,y)$ and $\widetilde{A}(t,y)$
defined by \eqref{A-tilde-KP} have mutual continuous area.

\vskip 10pt

{\it Remark 5.1} ~ We have given a kind of generalization for
the Wronskian solutions to the KP equation,
and this generalization is easily obtained by virtue of
the property given in Proposition \ref{Prop 3.2}.
Although we further argued that the generalization is trivial for generating new solutions
in some sense, our discussion is still meaningful for the study of
Wronskian technique.

{\it Remark 5.2} ~ Such generalizations also hold for some other  (1+2)-dimensional cases
such as the Casoratian solutions to the 2-dimensional Toda lattice\cite{Hirota-O-S-1988}
and Casoratian solutions to the differential-difference KP equation\cite{Tamizhmani-JPA},
even for the Grammian solution ${\rm Det}\Bigl(\int^{x}f_i g_j dx\Bigr)_{1\leq i,j\leq N}$
to the KP equation\cite{Nakamura-Gram,Hirota-book}.


\vskip 16pt

\leftline{\bf Conclusions:} ~
Now Let us formulate the Wronskian technique as the following four steps.
The first step, which is the most  basic and important one, is to find Wronskian/Casoratian entry condition equations.
The second step is to solve the condition equations and try to get all their
possible existing  solutions.
The third step is to describe relations between different kinds of solutions,
and the final step is to discuss  parameter effects and dynamics of the solutions obtained.

Let us see what we have achieved  for soliton equations with KdV-type bilinear forms in this paper.
For the first step, we have given more general condition equations for the KdV equation,
the Toda lattice and the KP equation, and shown the arbitrariness of some matrices does not contribute
any new solutions in some sense.
For the second step, we have proposed an easy approach to obtain all possible existing solutions
of the condition equations for any constant coefficient matrix $A$ for
the KdV equation and the Toda lattice.
These solutions are in explicit form and
can be given according to the eigenvalues of $A$.
Obviously, Proposition \ref{Prop 3.8}, \ref{Prop 3.9} and Remark 3.4
which are related to the KdV equation are still valid for
the Toda lattice.
For the third step, we have explained the limit relations between Jordan-block
and diagonal solutions.
For the final step,
we have given the effective forms of the Wronskian/Casoratian
entry vectors for Jordan-block solutions in which the number of effective parameters
are reduced to the least.
These will be helpful to the study of dynamics of the obtained solutions.
Although for the KP equation the generalization is proved to be trivial for generating new solutions
in some sense, our discussion is still meaningful for the study of
Wronskian technique.

The method used in this paper is general and can apply to other soliton equations
with  Wronskian/Casoratian solutions, for example,
the nonlinear Schr\"{o}dinger equation\cite{Nimmo-NLS,Zhang-Hietarinta} .
We have also investigated the mKdV equation
and found it does not have complexitons and rational solutions in single Wronskian form\cite{Zhang-personal report}.
However, the mKdV equation admits rational solutions in double-Wronskian form\cite{Yin-2005}.
How to get its more solutions in (double-)Wronskian form is an interesting question.
For nonisospectral equations, the coefficient matrices in the condition equations have to
depend on time. In this case, Proposition \ref{Prop 3.2} and \ref{Prop 3.6}
will help us to do similar discussions
and this will be considered elsewhere.

\vskip 16 pt
\leftline {\bf Acknowledgments}

I sincerely thank Prof. J. Hietarinta for his  discussions and hospitality when I
visited University of Turku.
This project is supported by the National Natural Science
Foundation of China (10371070), the  Foundation of Shanghai
Education Committee for Shanghai Prospective
Excellent Young Teachers.


\end{document}